\documentclass[reprint,prd,onecolumn,11pt,notitlepage,nofootinbib,superscriptaddress]{revtex4-1}
\usepackage[english]{babel}
\usepackage{amsmath,amssymb,graphicx,bm,lipsum,float}
\usepackage[colorlinks=true,citecolor=blue,linkcolor=blue,urlcolor=blue]{hyperref} 
\usepackage[font={small},flushleft,indent]{caption}
\usepackage{setspace}
\usepackage{subcaption}
\usepackage{placeins}
\usepackage{overpic}

\begin{document}
\title{Exploring the statistical anisotropy of primordial curvature perturbations with pulsar timing arrays}
\author{Fengting Xie}
\email{xiefengting@stu.cqu.edu.cn} 
\affiliation{School of Physics, Chongqing University, Chongqing 401331, China} 
\author{Zhi-Chao Zhao}
\email{zhaozc@cau.edu.cn}
	\affiliation{Department of Applied Physics, College of Science, China Agricultural University, Qinghua East Road, Beijing 100083, China}
\author{Qing-Hua Zhu}
\email{zhuqh@cqu.edu.cn (corresponding author)} 
\affiliation{School of Physics, Chongqing University, Chongqing 401331, China} 
\author{Xin Li}
\email{lixin1981@cqu.edu.cn (corresponding author)} 
\affiliation{School of Physics, Chongqing University, Chongqing 401331, China} 

\begin{abstract}
The recent detection of a stochastic gravitational wave background by pulsar timing arrays has opened a new window in understanding supermassive black hole binaries and in probing the universe at the early time. Recently, pulsar timing array (PTA) collaborations have been further paving the way to probe anisotropies in the stochastic gravitational wave background. This study investigates dipole-type statistical anisotropy in the primordial power spectrum within a phenomenological framework.
We demonstrate that the primordial dipole induces both dipolar and quadrupolar anisotropies in the energy density spectrum of scalar-induced gravitational waves (SIGWs), without generating extra polarization modes. 
Based on this anisotropic spectrum, we derive the corresponding PTA overlap reduction functions (ORFs), which exhibit frequency dependence, with the anisotropies enhanced on small scales. Furthermore, owing to the non-uniform distribution of millisecond pulsars over the sky in current PTA dataset, the ORFs exhibit a morphology that explicitly depends on the preferred direction of the anisotropy.
However, our bayesian analysis of the NANOGrav 15-year dataset still yields no significant evidence for a preferred direction and a weak upper limit on anisotropy amplitude $(g\lesssim0.5)$. 
This result arises because the observational frequency band lies below the spectral peak, where our models predict suppressed anisotropic contributions. 
This limitation highlights the potential of future PTA observations. Specifically, datasets with broader frequency coverage are expected to tighten constraints on dipole-type anisotropy.

\end{abstract}
\maketitle  
\section{Introduction}

The detection of the stochastic gravitational wave background (SGWB) by pulsar timing arrays (PTAs) \cite{NANOGrav:2023gor,EPTA:2023fyk,Reardon:2023gzh,Xu:2023wog} has opened a new era both in understanding supermassive black hole binaries (SMBHBs) \cite{NANOGrav:2023hfp,NANOGrav:2023pdq} and in probing the universe at the early time  \cite{NANOGrav:2023tcn,EPTA:2023xxk}. Currently, the most plausible source of the signal is believed to be the cosmic population of supermassive black hole binaries (SMBHBs) in galactic centers, which appears preferred to the cosmological interpretations \cite{Agazie:2024kdi,Agazie:2026tui}. In contrast to cosmological models, a distinguishing feature of the SMBHB-dominated SGWB is signal anisotropy arising from the intrinsic statistical properties of the source population \cite{Mingarelli:2017fbe,Sato-Polito:2023spo,Grimm:2024lfj,Lemke:2024cdu,Yang:2024mqz}. Since cosmological models have not been fully ruled out \cite{Lemke:2024cdu}, it remains valuable to explore anisotropy features by comparing them against extended cosmological scenarios, such as anisotropy from the cosmic string \cite{Olmez:2011cg,Kuroyanagi:2016ugi,Jenkins:2018nty}, phase transition gravitational waves \cite{Geller:2018mwu,Liu:2020mru,Li:2021iva}, and scalar-induced gravitational wave \cite{Ota:2020vfn,Zhang:2022dgx,Chen:2022qec,Kuang:2023urj,Li:2023qua,Yu:2023jrs,Mukherjee:2025dcv,Luo:2025lgr,Li:2025met,Li:2024zwx,Yu:2023jrs}.     
 

The SGWB from cosmological origins encodes information about Universe the at the early time. The gravitational waves can be emitted by oscillating loops and cusp or kink features on cosmic strings, producing a nearly scale-invariant spectrum \cite{Vilenkin:1981bx,Hogan:1984is,Vachaspati:1984gt,Damour:2004kw,Blanco-Pillado:2017oxo,Ringeval:2017eww,Baeza-Ballesteros:2023say,Fu:2023nrn,Qiu:2023wbs,Li:2024iua,Hu:2025sxv}. First-order phase transitions generate gravitational waves via bubble collisions, sound waves, and turbulence, at frequencies set by the transition temperature \cite{Hogan:1986dsh,Barni:2025gnm,Chatrchyan:2025wop,Inomata:2024rkt,Shao:2024dxt,Tian:2024ysd,Niemi:2024vzw,Jiang:2023nkj,Cai:2023guc,Liu:2022lvz}. Scalar-induced gravitational waves (SIGWs) are generated by primordial curvature perturbations re-entering the horizon in the early universe and are enhanced by the formation of the primordial black holes (PBHs) \cite{Ananda:2006af,Baumann:2007zm,Espinosa:2018eve,Kohri:2018awv,Sasaki:2018dmp,Domenech:2021ztg,Yuan:2021qgz}.
We restrict our attention to the SIGWs, which potentially offer a compelling interpretation for the recent PTA observation \cite{NANOGrav:2023tcn,EPTA:2023xxk,Domenech:2024rks,Wang:2023sij,Chen:2024fir,Chen:2019xse,Yuan:2019udt,Chen:2024fir}. 
If detected, it would serve as a powerful probe of the early universe on small scales, complementary to observations of the cosmic microwave background \cite{Planck:2018vyg}. Therefore, although recent studies suggest that the parameter space of SIGWs has been constrained to a narrow region due to the PBH overproduction \cite{Nakama:2015nea,NANOGrav:2023hvm}, theoretical efforts to alleviate this situation persist \cite{Domenech:2024rks,Zhu:2023gmx,Harigaya:2023pmw,DeLuca:2023tun,Choudhury:2023fwk,Wang:2024nmd,Choudhury:2024kjj,Yu:2024xmz}. An important aspect to consider is the unconstrained nature of primordial curvature perturbations on small scales, which have not yet been completely probed by observations. Various inflationary models yet to be experimentally verified can generate primordial power spectra with distinct shapes \cite{Suyama:2011pu,Cai:2018tuh,Cai:2019jah,Cai:2019bmk,Ozsoy:2019lyy,Cai:2021yvq}, exhibiting non-Gaussianity \cite{Maldacena:2002vr,Byrnes:2010em,Dimastrogiovanni:2010sm,Adshead:2011jq,Cai:2018dig,Franciolini:2023pbf,Gorji:2023sil,Zhao:2024gan,Li:2023xtl,Wang:2023ost} or statistical anisotropy \cite{Ackerman:2007nb,Yokoyama:2008xw,Dulaney:2010sq}. In this study, we adopt a phenomenological approach to investigate anisotropy in primordial power spectrum. Such anisotropies can arise from the presence of gauge fields \cite{Maleknejad:2012fw,Dimastrogiovanni:2010sm,Soda:2012zm,Chen:2025qyv} or inflation within anisotropic spacetimes \cite{Chang:2013xwa,Chang:2013lxa,Jain:2014cpa,Li:2015sja,Rath:2014cka,Chang:2018msh}. 

The universe is homogeneous and isotropic on large scales, which has served as the foundational assumption underlying standard cosmology, known as the cosmological principle. However, precise observations of the CMB have revealed subtle anisotropies in temperature and polarization, providing evidence for departures from statistical isotropy \cite{Bennett:2010jb,Planck:2015igc,Planck:2019evm}. This picture is complemented by large-scale structure surveys, which map the distribution of galaxies and reveal anisotropic clustering on Gpc scales \cite{SDSS:2005xqv,eBOSS:2020lta,eBOSS:2020gbb,DESI:2025qqy,DESI:2025zgx}. With the advent of gravitational-wave astronomy, the KAGRA-Virgo-LIGO collaboration has searched for anisotropies from unresolved sources of the gravitational waves \cite{KAGRA:2021mth,KAGRA:2021kbb,LIGOScientific:2019vic}.
Recently, PTA collaborations reported their search for anisotropy in the SGWBs, placing an upper limit of $C_l/C_0 < 27\%$, and indicated that observational sensitivity is approaching the regime where constraints on dipolar anisotropy ($l=1$) become feasible \cite{NANOGrav:2023tcn}. In PTAs, the model-independent framework used for anisotropy searches assumes that the SGWB power spectrum can be factorized into a frequency-dependent part and a direction-dependent part, i.e., $P(f,\hat{k})\equiv H(f)\Omega(\hat{k})$ \cite{Mingarelli:2013dsa,Taylor:2013esa,Gair:2014rwa,Romano:2016dpx,Pol:2022sjn,Wang:2023div,Chen:2026mid}.
It would be well-suited for searching the anisotropy from SMBHBs \cite{Mingarelli:2017fbe}. However, it might not encompass all possibility for modeling the anisotropy. 
A counterexample is kinematic anisotropy \cite{Cusin:2022cbb,Chowdhury:2022pnv,Tasinato:2023zcg,Heisenberg:2024var,Mentasti:2025ywl}, which arises from the relative motion of the observer with respect to the SGWB rest frame. Notably, the corresponding overlap reduction function (ORF) exhibits a characteristic frequency dependence. A natural question arises whether cosmological models exhibit such analogous signatures. Addressing this inquiry is one of the motivations for our study.


In this study, we explore the statistical anisotropy of primordial power spectrum based on the PTA response to SIGWs. At leading order, we phenomenologically consider dipolar anisotropy of the primordial power spectrum, which could arise from the inflation models with parity violation \cite{Chang:2013xwa,Chang:2013lxa,Jain:2014cpa,Li:2015sja,Rath:2014cka,Chang:2018msh}. We utilize analytical expressions for ORFs to perform parameter estimation within a Bayesian framework. Notably, the imprint of statistical anisotropy in the primordial power spectrum on ORFs has often been overlooked in previous studies \cite{Chen:2022qec,Kuang:2023urj}. In addition to standard SIGW parameters (amplitude and reference frequency), our work aims to constrain the anisotropic amplitude and the preferred direction of the dipole.

The rest of this paper is organized as follows. In Sec.~\ref{II}, we derive the SIGW spectrum generated by the primordial power spectrum with statistical anisotropy. In Sec.~\ref{III}, we present ORFs for anisotropic SIGWs and analyze their characteristics. In Sec.~\ref{IV}, we perform parameter estimation using the NANOGrav 15-year dataset. Finally, Section~\ref{V} summarizes our conclusions and discussions.

\section{Scalar-induced gravitational waves with statistical primordial anisotropy\label{II}}

The SIGWs are generated when the primordial curvature perturbation re-enters the horizon in the radiation domination era. The motion of equation for the SIGWs, denoted as $h_{ij}$, is derived from Einstein field equations to the secondary order, namely, \cite{Baumann:2007zm,Espinosa:2018eve,Kohri:2018awv}
\begin{eqnarray}
  h_{i  j}'' + 2\mathcal{H}  h_{i  j}' - \Delta h_{i
   j} & = & - 4 \Lambda^{a  b}_{i  j} \mathcal{S}_{a
   b} ~, \label{eq:h}
\end{eqnarray}
where $\Delta\equiv \delta^{ij}\partial_i\partial_j$, $\mathcal{H}$ is conformal Hubble parameter, $\Lambda_{ij}^{ab}$ is the transverse-traceless operator, and the source term $\mathcal{S}_{ab}$ on the rhs. of Eq.~(\ref{eq:h}) is
\begin{eqnarray}
  \mathcal{S}_{a  b} & = & 3 \partial_{a  } \psi \partial_b
  \psi + \frac{1}{\mathcal{H}} (\partial_a \psi \partial_b \psi' + \partial_a
  \psi' \partial_b \psi') + \frac{1}{\mathcal{H}^2} \partial_a \psi'
  \partial_b \psi' ~.
\end{eqnarray}
The source term is quadratic in the curvature perturbation $\psi$, which represents the scalar component of the metric perturbation. 
The Fourier mode of the $\psi$ can be factorized as two parts in the form of $\psi_{\textbf{k}} = \Psi_{\textbf{k}} T (k \eta)$, where $\eta$ is the conformal time. The initial value $\Psi_{\textbf{k}}$ can be related to primordial curvature perturbation $\zeta_{\textbf{k}}$, namely, $\Psi_{\textbf{k}} = (2 / 3) \zeta_{\textbf{k}}$ \cite{Maggiore:2018sht}. The transfer function $T (k \eta)[=T (x) = 9 ( \sqrt{3} \sin ( x / \sqrt{3}) / x - \cos( x / \sqrt{3} ) ) / x^2]$ describes the evolution of the curvature perturbation and can be obtained by solving the scalar component of the Einstein field equation at the first order \cite{Baumann:2007zm,Espinosa:2018eve,Kohri:2018awv}. 

Based on motion of equation in Eq.~(\ref{eq:h}), the gravitational wave $h_{ij}$ in Fourier space takes the form of
\begin{eqnarray}
  h^{\lambda}_{\textbf{k}} & = & k^{- 2} e^{\lambda, a  b} (\hat{k})
  \int \frac{\textrm{d}^3 p}{(2 \pi)^3} \left\{ \zeta_{\textbf{k} - \textbf{p}}
  \zeta_{\textbf{p}} p_a p_b I_h \left( \left| \textbf{k} - \textbf{p}
  \right|, p, k, \eta \right) \right\} ~, \label{3}
\end{eqnarray}
where $e^{\lambda,ab}$ is the polarization tensor, $h^\lambda_{\textbf{k}}[=e^{\lambda,ab} h_{ab}]$ is polarization component of the gravitational wave $h_{ij}$,  $I_h \left( \left| \textbf{k} - \textbf{p} \right|, p, k, \eta \right)$ denotes the kernel function obtained by solving Eq.~(\ref{eq:h}) via the Green's function method. The analytic expression for the kernel function, which is well-established in the literature on SIGWs \cite{Kohri:2018awv}, is given by
\begin{eqnarray}
  I_h \left( \left| \textbf{k} - \textbf{p} \right|, p, k, \eta \right)^2 & =   & \frac{1}{2 k \eta} \left( \frac{27 \left( 3 k^2 - \left| \textbf{k} -   \textbf{p} \right|^2 - p^2 \right)}{4 \left| \textbf{k} - \textbf{p}  \right|^3 p^3} \right)^2 \nonumber \\ && \times \Bigg( \pi^2 \left( 3 k^2 - \left| \textbf{k} -  \textbf{p} \right|^2 - p^2 \right)^2 \Theta \left( \left| \textbf{k} -  \textbf{p} \right| + p - \sqrt{3} k \right) \nonumber \\ &&  + \left( 4 \left| \textbf{k} -  \textbf{p} \right| p + \left( 3 k^2 - \left| \textbf{k} - \textbf{p}  \right|^2 - p^2 \right) \ln \left| \frac{3 k^2 + \left( \left| \textbf{k} -  \textbf{p} \right| + p \right)^2}{3 k^2 + \left( \left| \textbf{k} -  \textbf{p} \right| - p \right)^2} \right| \right)^2 \Bigg) ~.
\end{eqnarray}

SIGWs serve as a promising candidate for the SGWB of cosmological origin in PTAs. Their stochastic nature originates from the statistical properties of the primordial curvature perturbations. Specifically, the two-point correlation function of the primordial curvature perturbations takes the form of 
\begin{eqnarray}
  \langle \zeta_{\textbf{k}} \zeta_{\textbf{k}'} \rangle & = & (2
  \pi)^3 \delta \left( \textbf{k} + \textbf{k}' \right) P_{\zeta} \left(
  \textbf{k} \right) ~, \label{4}
\end{eqnarray}
where $P_\zeta(\textbf{k})$ is the primordial curvature power spectrum. In this work, we do not restrict our study to the standard assumption of an isotropic power spectrum. Given that PTA observations are becoming sensitive enough to constrain dipolar anisotropy in the future \cite{NANOGrav:2023tcn}, we adopt a parameterization incorporating dipole terms, namely,
\begin{eqnarray}
  P_{\zeta} (\textbf{k})  & = & P_{\zeta} (k) (1 + g (\hat{d} \cdot \hat{k})) ~, \label{6}
\end{eqnarray}
where unit spatial vector $\hat{d}$ denotes the preferred direction of statistical anisotropy in the primordial power spectrum and $g(\in [0,1])$ represents the anisotropy amplitude. Such dipolar anisotropy can arise in inflation models involving parity violation \cite{Chang:2013xwa,Chang:2013lxa,Jain:2014cpa,Li:2015sja,Rath:2014cka,Chang:2018msh}. In contrast, statistical anisotropy originating from gauge fields during the inflationary epoch typically manifests as a quadrupolar anisotropy \cite{Maleknejad:2012fw,Dimastrogiovanni:2010sm,Soda:2012zm,Chen:2025qyv}. However, the quadrupolar effect is expected to be subdominant in PTA observations relative to the dipolar anisotropy.

By making use of Eqs.~(\ref{3}) and (\ref{4}), we obtain the power spectrum of SIGWs, namely,
\begin{eqnarray}
  P^{\lambda \lambda'}_h \left( \textbf{k} \right) & \equiv & \frac{2}{k^4}
  e^{\lambda, a  b} (\hat{k}) e^{\lambda', i  j} (\hat{k})
  \int \frac{\textrm{d}^3 p}{(2 \pi)^3} \left\{ p_a p_b p_i p_j I_h^2 \left(
  \left| \textbf{k} - \textbf{p} \right|, p, k, \eta \right) P_{\zeta} \left(
  \textbf{k} - \textbf{p} \right) P_{\zeta} \left( \textbf{p} \right) \right\}~, \label{7}
\end{eqnarray}
where power spectrum $P^{\lambda \lambda'}_h\left( \textbf{k} \right)$ is defined with the two-point correlation of the gravitational waves $h_{ij}$, namely, $\langle h_{\textbf{k}}^{\lambda} h_{\textbf{k}'}^{\lambda'}  \rangle  =  (2 \pi)^3 \delta ( \textbf{k} + \textbf{k}' )   P_h^{\lambda \lambda'} ( \textbf{k} )$. Using the primordial power spectrum in Eq.~(\ref{6}), we evaluate the $P^{\lambda \lambda'}_h \left( \textbf{k} \right)$ in Eq.~(\ref{7}) in the form of
\begin{eqnarray}
  P^{\lambda \lambda'}_h \left( \textbf{k} \right) & = & \delta^{\lambda
  \lambda'} (H_0 (k) + g (\hat{d} \cdot \hat{k}) H_1 (k) + g^2 (\hat{d} \cdot \hat{k})^2
  H_2 (k))~, \label{8}
\end{eqnarray}
where
\begin{subequations}
  \begin{flalign}
  H_0 (k)  = \hspace{0.1cm}& \frac{1}{2} \int \frac{\textrm{d}^3 p}{(2 \pi)^3} \left\{\left( 1 - \frac{g^2 p \sin^2  \theta}{2 \left| \textbf{k} - \textbf{p} \right|} \right)  \left(  \frac{p\sin \theta}{k} \right)^4   I_h^2 \left(  \left| \textbf{k} - \textbf{p} \right|, p, k, \eta \right) P_{\zeta} \left(  \textbf{k} - \textbf{p} \right) P_{\zeta} \left( \textbf{p} \right) \right\}~,  &
  \end{flalign}
      \vspace{-1.25cm} 
  \begin{flalign}
    \vspace{-1cm}
  H_1 (k)  = \hspace{0.1cm}& \frac{1}{2} \int \frac{\textrm{d}^3 p}{(2 \pi)^3} \Bigg\{ \left( \frac{k}{\left| \textbf{k} -  \textbf{p} \right|} + \left( 1 - \frac{p}{\left| \textbf{k} - \textbf{p}  \right|} \right) \cos \theta \right) \nonumber \\ & \times \left(  \frac{p \sin \theta}{k} \right)^4  I_h^2 \left( \left| \textbf{k} -  \textbf{p} \right|, p, k, \eta \right) P_{\zeta} \left( \left| \textbf{k} -  \textbf{p} \right| \right) P_{\zeta} (p) \Bigg\}~, &
    \end{flalign}
        \vspace{-1.25cm} 
  \begin{flalign}
  H_2 (k)  = \hspace{0.1cm}& \frac{1}{2} \int \frac{\textrm{d}^3 p}{(2 \pi)^3} \Bigg\{ \left( \frac{ k    }{\left| \textbf{k} - \textbf{p} \right|} \cos \theta + \frac{p}{2  \left| \textbf{k} - \textbf{p} \right|} (1 - 3 \cos^2 \theta) \right) \nonumber &\\ & \times \left(  \frac{p \sin \theta}{k} \right)^4    I_h^2   \left( \left| \textbf{k} - \textbf{p} \right|, p, k, \eta \right) P_{\zeta}   \left( \left| \textbf{k} - \textbf{p} \right| \right) P_{\zeta} (p) \Bigg\}~.  &
\end{flalign}\label{9}
\end{subequations}
\hspace{-0.225cm} It is found that that the dipolar anisotropy in the primordial power spectrum [Eq.~(\ref{6})] induces both dipolar and quadrupolar anisotropies in the power spectrum of SIGWs. The limit $g=0$ recovers its isotropic case. While the angular dependence in Eq.~(\ref{8}) resembles that of kinematic anisotropies \cite{Cusin:2022cbb,Chowdhury:2022pnv,Tasinato:2023zcg,Heisenberg:2024var,Mentasti:2025ywl}, the spectral functions $H_1(k)$ and $H_2(k)$ are of cosmological origin. Detailed derivations of the polarization term $\delta_{\lambda\lambda'}$ in Eq.~(\ref{8}) are provided in Appendix~\ref{A}. Here, we demonstrate that the primordial dipolar anisotropy does not induce additional polarization modes in the SIGWs.

On small scales, the primordial curvature power spectrum might be significantly enhanced to facilitate the production of PBHs \cite{Ananda:2006af,Baumann:2007zm,Espinosa:2018eve,Kohri:2018awv,Sasaki:2018dmp,Domenech:2021ztg,Yuan:2021qgz}. Specifically, the parametric resonance mechanism can generate such enhancements, leading to a power spectrum of the form of \cite{Cai:2018tuh,Cai:2019jah,Cai:2019bmk}
\begin{eqnarray}
  P_{\zeta} \left( \textbf{k} \right) & = & A_{\zeta}  k_{\ast} \delta
  (k - k_{\ast})~, \label{10}
\end{eqnarray}
where $A_\zeta$ is the spectral amplitude, and $k_\ast$ the reference scale of the PBH production. 
Using Eqs.~(\ref{9}) and (\ref{10}), we explicitly obtain
\begin{subequations}
  \begin{eqnarray}
  H_0 (k) & = & \left( 1 - \frac{g^2}{2} \left( 1 - \frac{k^2}{4 k_{\ast}^2}
  \right) \right) \left( \frac{2 k_{\ast}}{k} + \frac{k}{2 k_{\ast}} \right)^2
  I_h (k_{\ast}, k_{\ast}, k, \eta)^2 \Theta (2 k_{\ast} - k)~,\\
  H_1 (k) & = & \left( \frac{k}{k_{\ast}} \right) \left( \frac{2 k_{\ast}}{k}
  + \frac{k}{2 k_{\ast}} \right)^2 I_h (k_{\ast}, k_{\ast}, k, \eta)^2 \Theta
  (2 k_{\ast} - k)~,\\
  H_2 (k) & = & \left( \frac{k^2}{2 k_{\ast}^2} + \frac{1}{2} \left( 1 -
  \frac{3 k^2}{4 k_{\ast}^2} \right) \right) \left( \frac{2 k_{\ast}}{k} +
  \frac{k}{2 k_{\ast}} \right)^2 I_h (k_{\ast}, k_{\ast}, k, \eta)^2 \Theta (2
  k_{\ast} - k)~.
\end{eqnarray}
\end{subequations}
In the large-scale limit ($k \to 0$), we find that $H_1(k) \propto k H_0(k)$ while $H_2(k) \propto H_0(k)$. The dipolar anisotropy in SIGWs is significantly suppressed on large scales. 

\begin{figure}
  \includegraphics[width=1\linewidth]{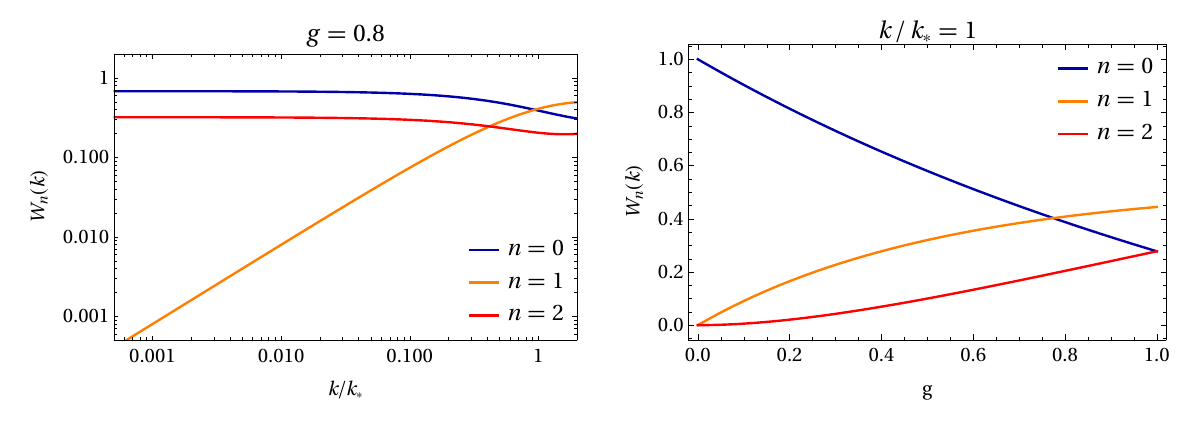}
  \caption{The scale-weighted function $W_n(k)$ as function of scale $k$ (left panel) and anisotropy magnitude $g$ (right panel). The $n=0$, $1$, and $2$ represent the isotropic part, dipole anisotropic part and quadrupole anisotropic part, respectively. \label{F1}}
\end{figure}

To facilitate comparison with existing analyses of anisotropic SGWBs in PTAs \cite{Mingarelli:2013dsa,Taylor:2013esa,Gair:2014rwa}, we define an effective power spectrum based on Eq.~(\ref{8}) as
\begin{eqnarray}
  \mathcal{P}^{\lambda \lambda'}_h \left( \textbf{k} \right) & = & 
  \delta^{\lambda \lambda'} \mathcal{P}_{\text{eff}} (k) \Omega (k, \hat{k},
  d)~, \label{12}
\end{eqnarray}
where dimensionless power spectrum is given by $\mathcal{P}^{\lambda \lambda}_h 
(k) = (k^3/2 \pi^2) P_h^{\lambda \lambda}(k)$, and
\begin{eqnarray}
  \mathcal{P}_{\text{eff}} (k) & = & \frac{k^3}{2 \pi^2} \sum_{m = 0}^2 g^m
  H_m (k)~,\\
  \Omega (k, \hat{k}, d) & = & 
  \sum_{n =
  0}^2 (\hat{d} \cdot \hat{k})^n W_n (k)~.
\end{eqnarray}
The scale-dependent weight functions $W_n(k)[\equiv {g^n H_n (k)}/{\sum_{m = 0}^2 g^m H_m (k)}]$ satisfy the normalization condition $\sum_{n=0}^{2} W_n(k) = 1$ for all $k$. Fig.~\ref{F1} illustrates their dependence on the wavenumber $k$ and anisotropy amplitude $g$. As anticipated, the dipole weight $W_1(k)$ is suppressed on the large scale, but dominates over $W_0(k)$ near the reference scale $k_\ast$ of PBH production. In contrast, the quadrupole weight $W_2(k)$ remains subdominant to $W_0(k)$ across all scales, satisfying $W_2(k) \leq W_0(k)$.

\begin{figure}
  \includegraphics[width=0.7\linewidth]{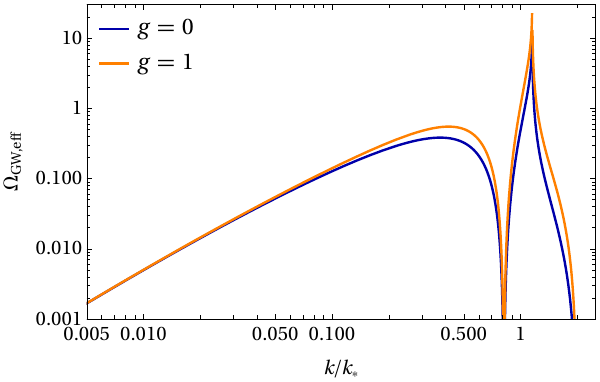}
  \caption{Effective energy density fraction of SIGWs in the presence of statistical anisotropy of primordial power spectrum in the form of Eq.~(\ref{10}). \label{F2}}
\end{figure}

The energy density fraction spectrum of SIGWs is related to the power spectrum by $\Omega_{\rm GW,eff}(k) = (1/24)(k/H)^2 \mathcal{P}_{\rm eff}(k)$ \cite{Kohri:2018awv}. In the presence of the statistical anisotropy of the primordial power spectrum, the effective energy density fraction takes the form of
\begin{eqnarray}
  \Omega_{\text{GW}, \text{eff}} (k) & = & \left(1+\frac{g k}{2k_\ast}\right)^2 \Omega_\text{GW,iso}(k)~, \label{15}
\end{eqnarray}
where $\Omega_{\rm GW,iso} \equiv \Omega_{\rm GW,eff}|_{g=0}$ denotes the isotropic SIGW energy density fraction in the standard literatures \cite{Ananda:2006af,Baumann:2007zm,Espinosa:2018eve,Kohri:2018awv,Sasaki:2018dmp,Domenech:2021ztg,Yuan:2021qgz}. Our analysis reveals that $\Omega_{\rm GW,eff}(k) \leq 2\,\Omega_{\rm GW,iso}(k)$, placing an upper bound on the anisotropic enhancement. We shows the effective energy density fraction spectrum $\Omega_{\text{GW}, \text{eff}} (k)$ in Fig.~\ref{F2}. 
Here, the $\Omega_\text{GW,eff}$ is a phenomenological parameterization designed for anisotropy searches and differs from the standard cosmological density parameter, which quantifies the fractional contribution of gravitational waves to the universe's energy budget.

The weak dependence of the spectrum in Eq.~(\ref{15}) on the anisotropic amplitude $g$ results in a challenge for constraining $g$ using current datasets. However, since anisotropic SIGWs can imprint distinctive signatures on the observables in PTAs, analyzing these imprints might represent a promising complementary approach to constraining anisotropies. In the subsequent section, we will derive the angular correlation in PTAs originating from the anisotropic SIGWs in Eq.~(\ref{8}).


\section{The deformation of Hellings-Downs curves \label{III}}

Due to spacetime fluctuations, such as gravitational waves, which can affect the propagation of radio signals from a pulsar to Earth, the corresponding effect can be encoded in pulsar timing \cite{Detweiler:1979wn}. In this framework, the timing residuals originating from the SGWB can be expressed as follows \cite{Maggiore:2018sht},
\begin{eqnarray}
  R_\text{A} (t) & = & \sum_{\lambda}^{\times, +} \int_0^t \textrm{d} t \int \frac{\textrm{d}^3 k}{(2
  \pi)^3} \left\{ h^{\lambda}_{\textbf{k}} F^{\lambda}_\text{A} (\hat{k}) e^{- i
    k  t} (1 - e^{i  k  t (1 + \hat{n}_\text{A}
  \cdot \hat{k})}) \right\}~,
\end{eqnarray}
where the response function takes the form of $
  F^{\lambda}_\text{A} (\hat{k})  =  {\hat{n}_\text{A}^i \hat{n}_\text{A}^j e^{\lambda}_{i
   j} (\hat{k})}/{(2 (1 + \hat{n}_\text{A} \cdot \hat{k}))}
$ and the direction vector $\hat{n}_\text{A}$ represent the location of a pulsar, denoted as $\text{A}$. 
The correlation of timing residuals between a pulsar pair (labeled A and B) encodes the information of the SGWB \cite{Hellings:1983fr}, namely, $\langle R_\text{A}(t) R_\text{B}(t) \rangle \propto \int df\, \mathcal{P}_{\rm eff}(2\pi f)\, \Gamma(\theta_\text{AB})$. Here, $\Gamma(\theta_\text{AB})$ denotes the overlap reduction function (ORF), which quantifies the angular correlation of the signals and is commonly known as the Hellings-Downs curve \cite{Hellings:1983fr}. This relation indicates that the correlation amplitude is modulated by the angular separation between pulsar pairs. The primary objective of this study is to investigate how the anisotropies of the SIGWs given in Eq.~(\ref{8}) deforms the Hellings-Downs curve. 

Signatures of anisotropic SGWBs in PTAs have been widely investigated \cite{Mingarelli:2013dsa,Taylor:2013esa,Gair:2014rwa,Romano:2016dpx,Pol:2022sjn,Chen:2026mid}. Following this formalism, our ORFs can be obtained by integrating the product of the pulsar pair response functions and the direction-dependent term $\Omega(k,\hat{k},d)$ defined in Eq.~(\ref{12}) over the celestial sphere, namely,
\begin{eqnarray}
  \Gamma (k, \hat{n}_\text{A}, \hat{n}_\text{B}, d)  =  \int \frac{\textrm{d}^2 \hat{k}}{4 \pi}
  \{ F^{\lambda}_\text{A} (\hat{k}) F^{\lambda}_\text{B} (\hat{k}) \Omega (k, \hat{n}_\text{A}, \hat{n}_\text{B}, d)
  \}
   =  \sum^2_{n = 0} W_n (k) \Gamma_n (\hat{n}_\text{A}, \hat{n}_\text{B}, d)~,\label{ORF}
\end{eqnarray}
where $k=2\pi f$ for the SIGWs and
\begin{eqnarray}
  \Gamma_n (\hat{n}_\text{A}, \hat{n}_\text{B}, d) & = & \int \frac{\textrm{d}^2 \hat{k}}{4 \pi}
  \sum_{\lambda, \lambda'} F^{\lambda} (\hat{n}_\text{A}) F^{\lambda'} (\hat{n}_\text{B}) (\hat{d}
  \cdot \hat{k})^n~. \label{18}
\end{eqnarray}
In contrast to ORFs derived for anisotropic SGWBs in previous studies \cite{Mingarelli:2013dsa,Taylor:2013esa,Gair:2014rwa,Romano:2016dpx,Pol:2022sjn,Chen:2026mid}, the ORFs presented in Eq.~(\ref{ORF}) exhibit frequency dependence, analogous to kinematic anisotropy \cite{Mingarelli:2013dsa,Taylor:2013esa,Gair:2014rwa,Romano:2016dpx,Pol:2022sjn,Chen:2026mid}. To facilitate efficient data analysis, we derive analytic expressions for the ORFs by adopting a specific coordinate system, $\hat{n}_\text{A} = (0,0,1)$ and $\hat{n}_\text{B} = (\sin\theta_\text{AB}, 0, \cos\theta_\text{AB})$, where the anisotropy direction vector $\hat{d}$ is parameterized as
\begin{eqnarray}
  \hat{d} = \big(\csc\theta_\text{AB}\cos\theta_\text{BD}-\cot\theta_\text{AB}\cos\theta_\text{AD},\hspace{0.2cm}((\hat{n}_\text{A}\times \hat{n}_\text{B})\cdot \hat{d})\csc\theta_\text{AB},\hspace{0.2cm}\cos\theta_\text{AD}\big)~,
\end{eqnarray}
where $\theta_\text{AB}\equiv \arccos(\hat{n}_\text{A} \cdot \hat{n}_\text{B})$, $\theta_\text{AD}\equiv \arccos(\hat{n}_\text{A} \cdot d)$ and $\theta_\text{BD}\equiv \arccos(\hat{n}_\text{B} \cdot d)$. By making use of these coordinate choices, the ORFs in Eqs.~(\ref{18}) can be given by
\begin{subequations}
  \begin{flalign}
  \Gamma_0 (\hat{n}_\text{A}, \hat{n}_\text{B}, d)  = \hspace{0.1cm}& \frac{1}{4} + \frac{1}{12} \cos
  \theta_{\text{AB}} + (1 - \cos \theta_{\text{AB}}) \ln \left( \sin \left(
  \frac{\theta_{\text{AB}}}{2} \right) \right)~, &
  \end{flalign}
      \vspace{-1.25cm} 
  \begin{flalign}
  \Gamma_1 (\hat{n}_\text{A}, \hat{n}_\text{B}, d) = \hspace{0.1cm}& \frac{1}{48} (\cos \theta_{\text{AD}}
  + \cos \theta_{\text{BD}}) \Bigg( - 5 - 2 \cos \theta_{\text{AB}} + 3 \cos
  (2 \theta_{\text{AB}}) \nonumber &\\ & - 24 (1 - \cos \theta_{\text{AB}}) \ln \left( \sin 
  \left( \frac{\theta_{\text{AB}}}{2} \right) \right) \Bigg) \sec \left(
  \frac{\theta_{\text{AB}}}{2} \right)^2~, & \label{20b}
  \end{flalign}
      \vspace{-1.25cm} 
  \begin{flalign}
  \Gamma_2 (\hat{n}_\text{A}, \hat{n}_\text{B}, d)  = \hspace{0.1cm} & \frac{1}{480} \cos\theta_\text{AD} \sec^2\left(\left(\frac{\theta_\text{AB}}{2}\right)\right) \Bigg( 
    54 \cos\theta_\text{AD} + 28 \cos\theta_\text{AB} \cos\theta_\text{AD} - 26 \cos(2\theta_\text{AB}) \cos\theta_\text{AD} \nonumber  &\\
    & + 44 \cos\theta_\text{BD} + 9 \cos\theta_\text{AB} \cos\theta_\text{BD} - 40 \cos(2\theta_\text{AB}) \cos\theta_\text{BD} - 5 \cos(3\theta_\text{AB}) \cos\theta_\text{BD} \nonumber \\
    & - 240 (\cos\theta_\text{AB} - 1) (\cos\theta_\text{AD} + \cos\theta_\text{BD}) \ln\left(\sin\left(\left(\frac{\theta_\text{AB}}{2}\right)\right)\right) \Bigg) \nonumber\\
    & + \frac{1}{240} (\hat{d}\cdot(\hat{n}_\text{A}\times \hat{n}_\text{B}))^2 \csc^2\theta_\text{AB} \sec^2\left(\left(\frac{\theta_\text{AB}}{2}\right)\right) \nonumber \\ & \times \Bigg( 
    -19 + 12 \cos\theta_\text{AB} + 31 \cos(2\theta_\text{AB}) - 480 \ln\left(\sin\left(\left(\frac{\theta_\text{AB}}{2}\right)\right)\right) \sin^4\left(\left(\frac{\theta_\text{AB}}{2}\right)\right) \Bigg) \nonumber\\
    & + \left(\cos\theta_\text{BD} \csc\theta_\text{AB}  -\cos\theta_\text{AD} \cot\theta_\text{AB} \right) \nonumber \\ & \times \Bigg( 
    \frac{1}{60} \cos\theta_\text{AD} \csc\theta_\text{AB}\left( 25 - 35 \cos\theta_\text{AB} + 4 \cos(2\theta_\text{AB}) \right)  \nonumber\\
    & + \frac{1}{240} \cos\theta_\text{BD} \left( -121 \cot\theta_\text{AB} + 5 \csc\theta_\text{AB} \left( 22 + 6 \cos(2\theta_\text{AB}) + \cos(3\theta_\text{AB}) \right)  \right) \nonumber\\
    & + (\cos\theta_\text{AD} + \cos\theta_\text{BD}) \ln\left(\sin\left(\left(\frac{\theta_\text{AB}}{2}\right)\right)\right) \tan^3\left(\left(\frac{\theta_\text{AB}}{2}\right)\right) \Bigg)\Bigg)~. \label{20c}
\end{flalign} \label{20}
\end{subequations}
Detailed derivations of the aforementioned ORFs are provided in Appendix~\ref{B}. The analytic expressions in Eqs.~(\ref{20}) are consistent with results derived via different approaches in Refs.~\cite{Tasinato:2023zcg,Mingarelli:2013dsa}. Here, the cosmological preferred direction $\hat{d}$ is independent of the specific locations of pulsar pairs $(\hat{n}_\text{A}, \hat{n}_\text{B})$.

\begin{figure}
  \includegraphics[width=1\linewidth]{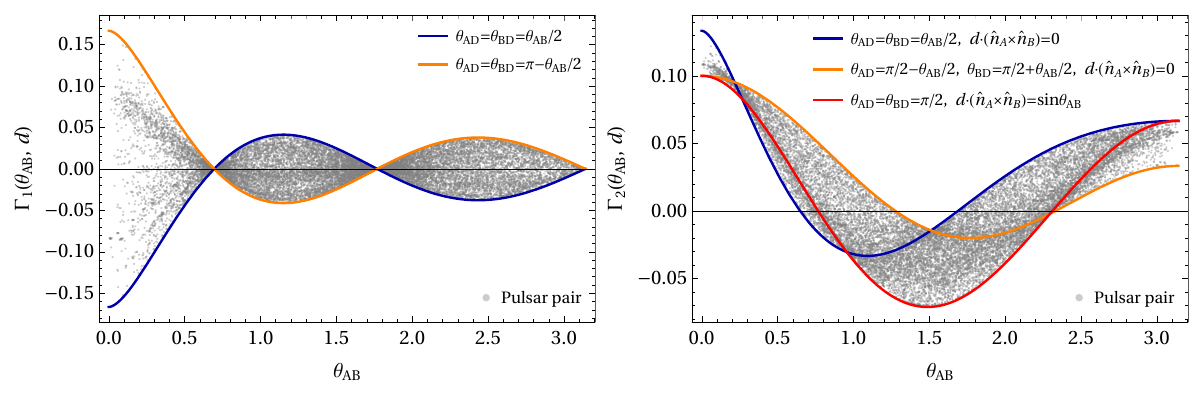}
  \caption{Overlap reduction functions for dipole (left panel) and quadrupole (right panel) anisotropies of the SIGWs. The scatter points are given by pairs of simulated 200 pulsars over the sky. The colored curves determine the envelope curves of all the scatter points. \label{F3}}
\end{figure}

\begin{figure}  
  \centering
  \begin{subfigure}[t]{0.49\textwidth}
    \caption{Dipole \label{F3a}}
  \includegraphics[width=0.8\linewidth]{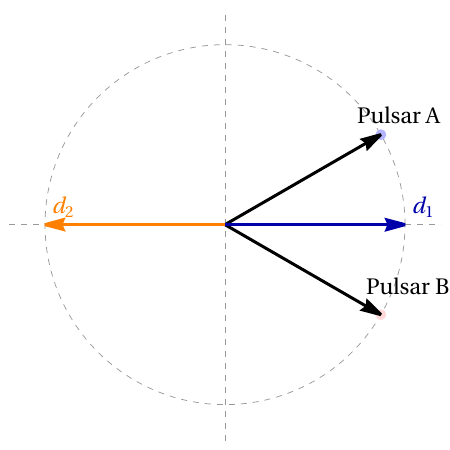}
    \end{subfigure} 
   \begin{subfigure}[t]{0.49\textwidth} 
    \caption{Quadrupole\label{F3b}}
    \includegraphics[width=0.95\linewidth]{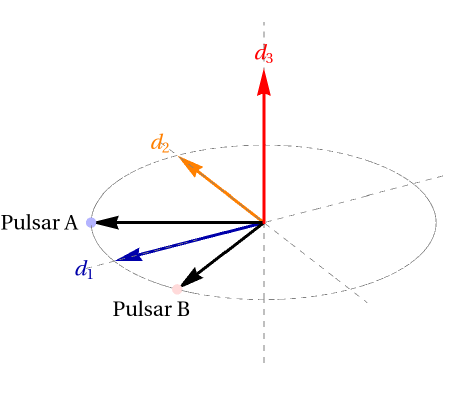}
      \end{subfigure}
  \caption{The typical configurations of the pulsar pairs relative to the preferred direction $\hat{d}$ for the envelope curves illustrated in Fig.~\ref{F3}.  Left panels: the configurations for dipolar anisotropies with $d_1$ and $d_2$ corresponding the solid curves with the same colors. Right panels: the configurations for quadrupolar anisotropies with $d_1$, $d_2$ and $d_3$ corresponding the solid curves with the same colors. For the colored curves in Fig.~\ref{F3} as functions of $\theta_\text{AB}$, the $\hat{d}_1\cdot\hat{n}_\text{A}=\hat{d}_1\cdot\hat{n}_\text{B}$ is always maintained. \label{F4}} 
\end{figure}

Fig.~\ref{F3} shows the dipole and quadrupole anisotropic parts of the ORFs given in Eq.~(\ref{20b}) and (\ref{20c}), respectively. The gray scatter points represent the value of ORFs for the pulsar pair at fixed locations $\hat{n}_\text{A}$ and $\hat{n}_\text{B}$. These scatter points are distributed within the colored envelope curves, rather than lying on a single curve as in the standard Hellings-Downs curve for an isotropic SGWB. This scatter arise because the ORFs not only depend on the angular separation of the pulsar pair but also on their individual locations relative to the preferred direction $\hat{d}$. Remarkably, these envelope curves are universal. They are independent of the preferred direction $\hat{d}$, and thus can be determined by fixing $\hat{d}\cdot\hat{n}_\text{A}$, $\hat{d}\cdot\hat{n}_\text{B}$ and $\hat{d}\cdot(\hat{n}_\text{A}\times \hat{n}_\text{B})$. For dipole part $\Gamma_1(\hat{n}_\text{A},\hat{n}_\text{B},\hat{d})$, as shown in the left panel of Fig.~\ref{F3}, it exhibits symmetry about the zero axis. 
The envelope curves intersect at zero for specific angular separations  $\theta_\text{AB}\approx0.7$ and $1.8$, indicating that the dispersion of the scatter points vanishes at these angles. The quadrupole part $\Gamma_2(\hat{n}_\text{A},\hat{n}_\text{B},\hat{d})$ yields positive correlations for pulsar pairs at angular separations $\theta_\text{AB}=0$ and $\pi$ as shown in the right panel of Fig.~\ref{F3}, in contrast with to $\Gamma_1(\hat{n}_\text{A},\hat{n}_\text{B},\hat{d})$. It is found that the envelope curves  of $\Gamma_2(\hat{n}_\text{A},\hat{n}_\text{B},\hat{d})$ are determined by three curves with fixed $\hat{d}\cdot\hat{n}_\text{A}$, $\hat{d}\cdot\hat{n}_\text{B}$ and $\hat{d}\cdot(\hat{n}_\text{A}\times \hat{n}_\text{B})$, whereas those of $\Gamma_2(\hat{n}_\text{A},\hat{n}_\text{B},\hat{d})$ are determined by two curves with fixed $\hat{d}\cdot\hat{n}_\text{A}$ and $\hat{d}\cdot\hat{n}_\text{B}$.
In Fig.~\ref{F4}, we characterize these envelope curves by illustrating their geometric configurations of the pulsar pairs relative to $\hat{d}$. The configurations of pulsar pairs for $\Gamma_1(\hat{n}_\text{A},\hat{n}_\text{B},\hat{d})$ are all planar, while those for $\Gamma_2(\hat{n}_\text{A},\hat{n}_\text{B},\hat{d})$ consist of two planar and one three-dimensional configuration.



\begin{figure}
  \includegraphics[width=1\linewidth]{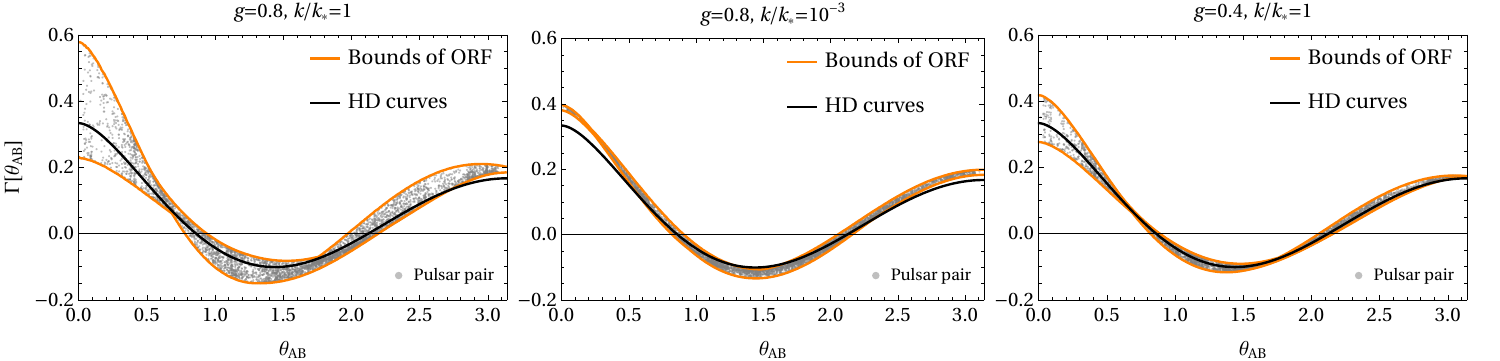}
  \caption{The deformed Hellings-Downs curves due to the dipolar anisotropy in primordial power spectrum. The scatter points represent the ORFs as function of the angular separation of pulsar pairs, which consist of 100 simulated pulsars cross the sky. The deformation of the ORFs varies with the scale $k/k_\ast$ and the anisotropy magnitude $g$. \label{F5}}
\end{figure}

In Fig.~\ref{F5}, we show the total ORFs $\Gamma(k,\hat{n}_\text{A},\hat{n}_\text{B},\hat{d})$ in the presence of the statistical anisotropy. Due to non-vanishing values of $\Gamma_2(\hat{n}_\text{A},\hat{n}_\text{B},\hat{d})$ at specific $\theta_\text{AB}$, the standard Hellings--Downs curve is not necessarily contained within the envelope of the anisotropic ORFs at all $\theta_\text{AB}$. In the large-scale limit ($k\to 0$), the dispersion of the ORF scatter points reduces, indicating that effects of dipole-type primordial anisotropy on SIGWs are suppressed on large scales. Therefore, to constrain the anisotropies of SIGWs, a promising strategy is to probe deformation of the Hellings-Downs curves on the small scale $k/k_\ast \simeq \mathcal{O}(1)$. 

\section{Data analysis\label{IV}}



We analyze the NANOGrav 15-year dataset \cite{NANOGrav:2023gor} for our anisotropic SIGW model described in Sec.~\ref{II} via performing Bayesian parameter estimation. To incorporate deformed Hellings-Downs curves as presented in Sec.~\ref{III}, we employ a modified version of the next-generation PTA data-analysis package \texttt{discovery} developed by NANOGrav \cite{Vallisneri_nanograv_discovery_2025}. This package leverages automatic differentiation and GPU acceleration to perform efficient Bayesian inference on PTA dataset. 
Here, our spectrum of SGWBs is derived from the SIGWs, $\Omega_{\rm SIGW}(f) \simeq \Omega_{r,0}\, \Omega_{\rm GW,eff}(k=2\pi f)$, where $\Omega_{\rm GW,eff}$ is given by Eq.~(\ref{15}) and the present-day radiation density parameter is $h^2 \Omega_{r,0} \simeq 4.2\times 10^{-5}$ \cite{Planck:2018vyg}. This spectrum is then converted to signal spectral density. We employ the standard ORF normalization $C(\hat{n}_\text{A},\hat{n}_\text{B},\hat{d}) \equiv (3/2)\,\Gamma(\hat{n}_\text{A},\hat{n}_\text{B},\hat{d})$, with the locations of the millisecond pulsars \cite{NANOGrav:2023hde} explicitly accounted for in the correlation calculations.

\begin{figure}
  \includegraphics[width=1\linewidth]{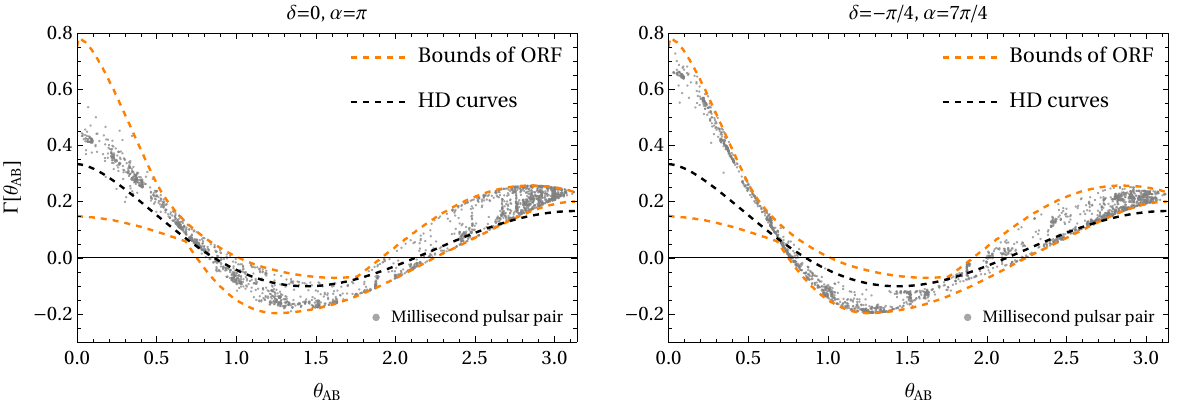}
  \caption{The deformed Hellings-Downs curves with $g=1$ on the scale of $k/k_\ast=1.5$ for selected direction of $\hat{d}$. The scatter points represent the  ORFs as function of the angular separation of pulsar pairs, which consist of 67 millisecond pulsars used by NANOGrav \cite{NANOGrav:2023hde}. \label{F6}}
\end{figure}

Our Bayesian inference spans the parameter space $\{\log_{10} A_\zeta, \log_{10}(f_*/{\rm Hz}), g, \hat{d}\}$, where the preferred direction $\hat{d}$ is parameterized in equatorial coordinates, namely,
\begin{eqnarray}
  \hat{d} = (\cos\alpha \cos\delta, \sin\alpha \cos\delta, \sin\delta)~,
\end{eqnarray}
with right ascension $\alpha \in [0,2\pi]$ and declination $\delta \in [-\pi/2,\pi/2]$. Due to the non-uniform distribution of the millisecond pulsars on the sky \cite{NANOGrav:2023hde}, the ORF scatter points for pulsar pairs exhibits $\hat{d}$-dependent morphology, as shown in Fig.~\ref{F6}, providing, in principle, sensitivity to the anisotropy direction.

\begin{figure}
  \includegraphics[width=0.85\linewidth]{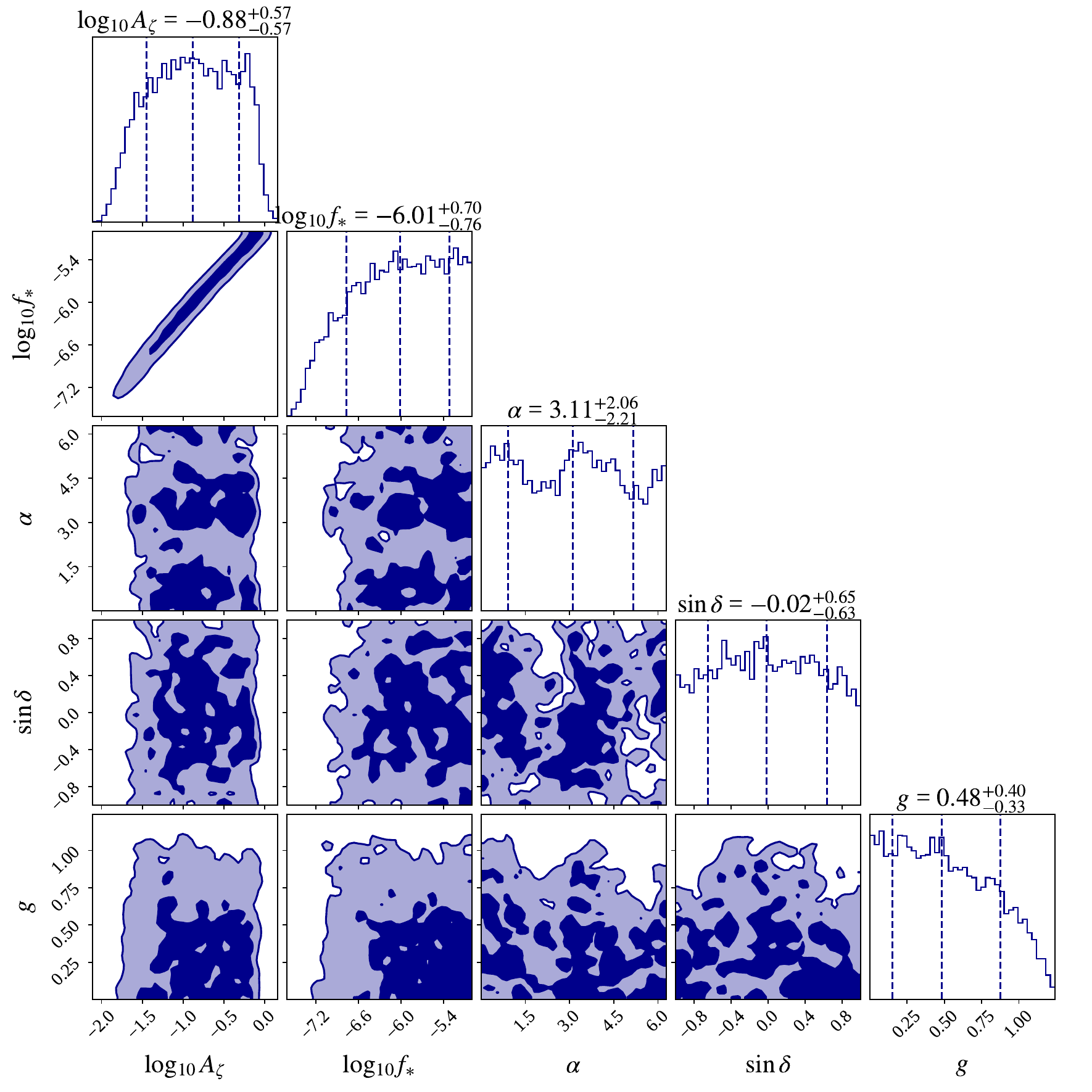}
  \caption{The posterior distributions of our model parameters of SIGWs with the dipole statistical anisotropy in primordial power spectrum. \label{F7}} 
\end{figure}

\begin{figure}
  \includegraphics[width=0.65\linewidth]{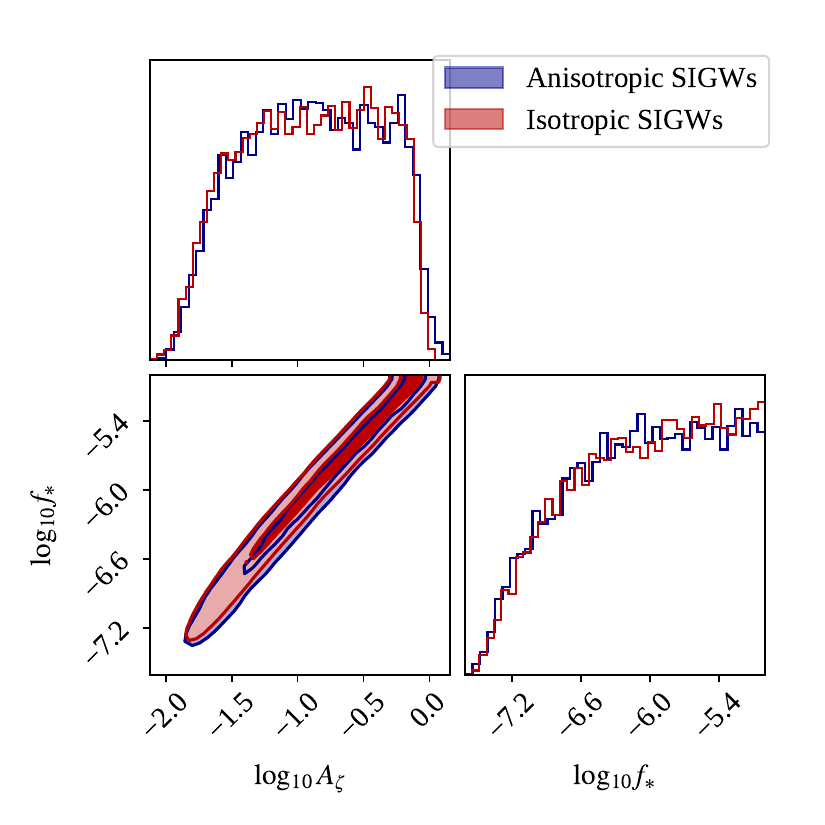}
  \caption{Comparison of the posterior distributions of the anisotropic SIGWs and the isotropic SIGWs \label{F8}} 
\end{figure}

We perform Bayesian parameter estimation with uniform priors for all parameters. The posterior distributions of our model parameters are presented in Fig.~\ref{F7}, and the results are summarized in Tab.~\ref{T1}.  The posterior for $\hat{d}$ is essentially uniform, indicating no significant constraint on the preferred direction. The anisotropy amplitude $g$ exhibits a broad posterior peaking at small values ($g \lesssim 0.1$), consistent with the absence of detectable statistical anisotropy in the current dataset. The spectral parameters $A_\zeta$ and $f_\ast$ are nearly unaffected by the anisotropy in Eq.~(\ref{6}), with posteriors nearly identical to those of isotropic SIGWs, shown in Fig.~\ref{F8}.
\begin{table}[ht]
    \centering
        \caption{Priors and posterior median for our model parameters.}
        \begin{ruledtabular}
    \begin{tabular}{clc}
        Parameter &Priors & Posterior median  \\
    \hline
        $\log_{10}A_\zeta$  &$\mathcal{U}[-14,6]$ & $-0.88^{+0.57}_{-0.57}$  \\
        $\log_{10}f_\ast$  &$\mathcal{U}[-7.8,-4]$ & $-6.01^{+0.70}_{-0.76}$\\
        $\alpha$  &$\mathcal{U}[0,2\pi]$& $3.11^{+2.06}_{-2.21}$ \\
        $\sin\delta$   &$\mathcal{U}[-1,1]$ & $-0.02^{+0.65}_{-0.63}$  \\
        $g$ & $\mathcal{U}[0,1]$ & $0.48^{+0.40}_{-0.33}$ \\
    \end{tabular}
  \end{ruledtabular}
    \label{T1}
\end{table}

Our results shows that high values of $g(\approx1)$ are largely ruled out, while the constraint weakens significantly for $g\lesssim0.5$. This trend arises because moderate values of $g$ induce only subtle deviations from the standard Hellings-Downs curve, as illustrated in the right panel of Fig.~\ref{F5}. For the moderate $g$, the constraint on $\hat{d}$ degrades significantly. In this regime, the ORFs in Eq.~(\ref{ORF}) converge towards the Hellings-Downs curve, rendering the effect of the non-uniform pulsar distribution insufficient to resolve the direction $\hat{d}$. Furthermore, the agreement between the constraints on $A_\zeta$ and $f_\ast$ and those from the isotropic SIGWs suggests that the observational frequency band lies below the spectral peak ($f_{\text{data}} \ll f_*$), as illustrated in Fig.~4 of Ref.~\cite{NANOGrav:2023hvm}. Consequently, anisotropies of SIGWs is suppressed in this low-frequency regime, which is indicated in middle panel of Fig.~\ref{F5}. All these factors collectively determine that the anisotropy of SIGWs is weakly constrained in this study.

\section{Conclusions and discussions\label{V}}

This study investigated dipole-type statistical anisotropy in the primordial power spectrum through the PTA response to anisotropic SIGWs. Theoretically, such anisotropy induces both dipolar and quadrupolar anisotropies in the SIGW energy density spectrum without generating additional polarization modes. It is found that the dipole component is suppressed on  the large scale but dominates the isotropic component on small scales, whereas the quadrupole component remains subdominant across all scales. Notably, the ORFs for anisotropic SIGWs exhibit frequency dependence, analogous to kinematic anisotropy. Owing to the non-uniform distribution of millisecond pulsars over the sky \cite{NANOGrav:2023hde}, the ORFs exhibit a morphology that explicitly depends on preferred direction $\hat{d}$ (Fig.~\ref{F6}). This directional dependence provides a possibility to constrain the preferred direction of the anisotropies of SIGWs.

Via Bayesian parameter estimation using the NANOGrav 15-year dataset, we assessed constraints on $\hat{d}$ and the anisotropy amplitude $g$. The data provide no significant constraint on $\hat{d}$ and yield only a weak upper limit on $g$ ($g \lesssim 0.5$), with the posterior peaking near zero. The weak constraints likely stems from the fact that current PTA observations probe frequencies $f_{\text{data}} \ll f_*$, where anisotropic parts of our SIGWs are suppressed (Fig.~\ref{F5}). Future PTA datasets with broader frequency coverage and improved sensitivity are expected to enhance constraints on these primordial anisotropies.

The anisotropic SIGW in this study employs a simplified model, because additional parameters of SGWBs, such as the peak width of the primordial power spectrum \cite{Pi:2020otn} and the spectrum from SMBHBs \cite{NANOGrav:2023hvm}, were not considered. We adopted a simplified strategy for searching for statistical anisotropy in the PTA dataset, assuming the signals are all interpreted as SIGWs with peaked primordial spectrum. However, the nearly uninformative results might lower our expectations for the existence of anisotropy in the primordial power spectrum.

\smallskip
{\it Acknowledgments}:  This work has been supported by the National Natural
Science Fund of China Grants (No. 12305073, No. 12347101, and No. 12275034) and the National Key Research and Development Program of China Grant No. 2021YFC2203001. The authors Thank Prof. Hai-Nan Lin for useful discussions.

\appendix

\section{Identities for deriving polarization components in Eq.~(\ref{8})\label{A}}
With the primordial anisotropies in Eq.~(\ref{6}), the polarization components of power spectrum of SIGWs can be given by
\begin{eqnarray}
  P^{\lambda \lambda'}_h \left( \textbf{k} \right) & = & \frac{2}{k^4}
  e^{\lambda, a  b} (\hat{k}) e^{\lambda', i  j} (\hat{k})
  \int \frac{\textrm{d}^3 p}{(2 \pi)^3} \Bigg\{ p_a p_b p_i p_j I_h^2 \left(
  \left| \textbf{k} - \textbf{p} \right|, p, k, \eta \right) P_{\zeta} \left(
  \textbf{k} - \textbf{p} \right) P_{\zeta} \left( \textbf{p} \right) \Bigg\} \nonumber \\
  & = & \frac{2}{k^4}
  e^{\lambda, a  b} (\hat{k}) e^{\lambda', i  j} (\hat{k})
  \int \frac{\textrm{d}^3 p}{(2 \pi)^3} \Bigg\{ p_a p_b p_i p_j I_h^2 \left(
  \left| \textbf{k} - \textbf{p} \right|, p, k, \eta \right) P_{\zeta} \left(
  |\textbf{k} - \textbf{p}| \right) P_{\zeta} \left( |\textbf{p}| \right) \nonumber \\
  && \times \Bigg( 1+g\left( \frac{k}{|\textbf{k}-\textbf{p}|}(\hat{d}\cdot\hat{k}) +\left(1-\frac{p}{|\textbf{k}-\textbf{p}|}\right)(\hat{d}\cdot\hat{p}) \right) \nonumber \\ && +g^2\left( \frac{k}{|\textbf{k}-\textbf{p}|} (\hat{d}\cdot \hat{k})(\hat{d}\cdot\hat{p})- \frac{p}{|\textbf{k}-\textbf{p}|} (\hat{d}\cdot\hat{p})^2\right) \Bigg)  \Bigg\}~, \label{A1}
\end{eqnarray}
where $\hat{p}_i\equiv p_i/p$.
The key to obtaining the polarization component is to evaluate the contraction between the polarization tensor and the multiple momentum $p_i$, which can be formally given by the identities as follows,
\begin{subequations}
  \begin{flalign}
\quad &e^{\lambda, a   b} (\hat{k}) e^{\lambda', i   j} (\hat{k})
  \int \frac{\mathrm{d}^3 p}{(2 \pi)^3} p_a p_b p_i p_j f \left( \left| \textbf{k}
  - \textbf{p} \right|, p \right) \nonumber & \\
&\quad =   \delta^{\lambda \lambda'} \int
  \frac{\mathrm{d}^3 p}{(2 \pi)^3} \left\{ \frac{p^4 \sin^4\theta}{4} f \left(
   \left| \textbf{k} - \textbf{p} \right|, p \right) \right\}~,&
\end{flalign}
\vspace{-1.25cm} 
\begin{flalign}
\quad &e^{\lambda, a   b} (\hat{k}) e^{\lambda', c   d} (\hat{k}) d^i
  \int \frac{\mathrm{d}^3 p}{(2 \pi)^3} \left\{ p_a p_b p_c p_d p_i   f
  \left( \left| \textbf{k} - \textbf{p} \right|, p \right) \right\} \nonumber & \\
&\quad = 
  \frac{1}{4} \delta^{\lambda \lambda'} (\hat{d} \cdot \hat{k}) \int \frac{\mathrm{d}^3
  p}{(2 \pi)^3} \Big\{  p^5 \cos \theta  \sin^4\theta   f \left( \left|
  \textbf{k} - \textbf{p} \right|, p \right) \Big\}~,&
\end{flalign}
\vspace{-1.25cm} 
  \begin{flalign}
  \quad& e^{\lambda, a   b} (\hat{k}) e^{\lambda', c   d} (\hat{k}) d^i
  d^j \int \frac{\mathrm{d}^3 p}{(2 \pi)^3} \left\{ p_a p_b p_c p_d p_i p_j
    f \left( \left| \textbf{k} - \textbf{p} \right|, p \right) \right\}
  \nonumber & \\  
  & \quad=  \delta^{\lambda \lambda'} \int \frac{\mathrm{d}^3 p}{(2 \pi)^3} \left\{
  p^6 \left( \frac{1}{4} \left( \cos^2 \theta  \sin^4\theta - \frac{1}{2}
  \sin^6 \theta \right) (\hat{d} \cdot \hat{k})^2 + \frac{1}{8} \sin^6 \theta
  \right)   f \left( \left| \textbf{k} - \textbf{p} \right|, p \right)
  \right\}~, &
\end{flalign}\label{A2}
\end{subequations}
where we have used the results for integrating over the multiple $p_i$, namely,
\begin{subequations}
\begin{flalign}
   \int \frac{\mathrm{d} \phi}{2 \pi} p_a p_b p_c p_d  = &  p^4 \Bigg( \left(  \cos^2 \theta - 3 \cos^2 \theta \sin^2 \theta + \frac{3}{8}  \sin^4\theta  \right) \hat{k}_a \hat{k}_b \hat{k}_c \hat{k}_d  
  \nonumber & \\ 
  & + \left( 3 \cos^2 \theta  \sin^2 \theta - \frac{3}{4}  \sin^4\theta \right) \delta_{  (a    b} \hat{k}_c \hat{k}_{  d)} \nonumber 
  \\ &  + \frac{3}{8}  \sin^4\theta  \delta_{  (a   b} \delta_{  c   d)} \Bigg)~, &
\end{flalign}
\vspace{-1.25cm} 
\begin{flalign}
  \int \frac{\mathrm{d} \phi}{2 \pi} p_a p_b p_c p_d p_i  = & p^5 \Bigg( \left(  \cos^5 \theta - 5 \cos^3 \theta \sin^2 \theta + \frac{15}{8} \cos \theta   \sin^4\theta \right) \hat{k}_a \hat{k}_b \hat{k}_c \hat{k}_d \hat{k}_i 
  \nonumber &\\ &  +
  \left( 5 \cos^3 \theta \sin^2 \theta - \frac{15}{4} \cos \theta \sin^4  \theta \right) \delta_{  (a   b} \hat{k}_c \hat{k}_d  \hat{k}_{  i)} 
  \nonumber &\\ & + \frac{15}{8} \cos \theta  \sin^4\theta  \delta_{  ((a   b  } \delta_{  c 
  d)} \hat{k}_{  i)} \Bigg)~, &
\end{flalign}
\vspace{-1.25cm}
\begin{flalign}
    \int \frac{\mathrm{d} \phi}{2 \pi} p_a p_b p_c p_d p_i p_j  = & p^6 \Bigg(
  \left( \cos^6 \theta - \frac{15}{2} \cos^4 \theta \sin^2 \theta +
  \frac{45}{8} \cos^2 \theta  \sin^4\theta - \frac{5}{16} \sin^6 \theta
  \right) \hat{k}_a \hat{k}_b \hat{k}_c \hat{k}_d \hat{k}_i \hat{k}_j 
  \nonumber &\\ &  + \left(
  \frac{15}{2} \cos^4 \theta \sin^2 \theta - \frac{45}{4} \cos^2 \theta \sin^4
  \theta + \frac{15}{16} \sin^6 \theta \right)   \delta_{  (a
    b} \hat{k}_c \hat{k}_d \hat{k}_i \hat{k}_{  j)} \nonumber &\\ &  + \left(
  \frac{45}{8} \cos^2 \theta  \sin^4\theta - \frac{15}{16} \sin^6 \theta
  \right) \delta_{  ((a   b  } \delta_{  c
    d)} \hat{k}_i \hat{k}_{  j)} \nonumber &\\ &  + \frac{5}{16} \sin^6 \theta
  \delta_{  (a   b} \delta_{c   d} \delta_{  i
    j)} \Bigg)  ~. &
\end{flalign}
\end{subequations}
Substituting Eqs.~(\ref{A2}) into Eq.~(\ref{A1}), it is not difficult to obtain Eq.~(\ref{8}).

\section{Analytic derivation of overlap reduction functions \label{B}}
In Eqs.~(\ref{18}), the vector $\hat{d}$ can be separated from the integration in the ORFs, Namely, we have
$\Gamma_1 (\hat{n}_\text{A}, \hat{n}_\text{B}, d)=: d^i\Gamma_i (\theta_\text{AB})$ and  $\Gamma_2 (\hat{n}_\text{A}, \hat{n}_\text{B}, d)=: d^j d^i\Gamma_{ij} (\theta_\text{AB})$, where
\begin{eqnarray}
  \Gamma_i (\theta_\text{AB}) & \equiv & \int \frac{\textrm{d}^2 \hat{k}}{4 \pi}
  \sum_{\lambda, \lambda'} F^{\lambda} (\hat{n}_\text{A}) F^{\lambda'} (\hat{n}_\text{B}) \hat{k}_i~. \\
  \Gamma_{ij} (\theta_\text{AB}) & \equiv & \int \frac{\textrm{d}^2 \hat{k}}{4 \pi}
  \sum_{\lambda, \lambda'} F^{\lambda} (\hat{n}_\text{A}) F^{\lambda'} (\hat{n}_\text{B}) \hat{k}_i \hat{k}_j~.\label{B2} 
\end{eqnarray}
In Cartesian coordinates, the $k_i$ can be rewritten as
\begin{eqnarray}
  \hat{k}_x =\sin\theta\cos\phi,~ \hat{k}_y=\sin\theta\sin\phi, ~\hat{k}_z=\cos\theta~,
\end{eqnarray}
and thus $\mathrm{d}^2\hat{k}=\sin\theta\mathrm{d}\theta\mathrm{d}\phi$. The coordinate choice of $\hat{n}_\text{A}$ and $\hat{n}_\text{B}$ given mentioned in Sec.~\ref{III}.
Based on the coordinate choice, the analytic expressions of the ORFs in Eqs.~(\ref{B2}) can be obtained, namely,
\begin{subequations}
\begin{eqnarray}
\Gamma_x (\theta_\text{AB}) &=&
\frac{1}{24} \Bigg( -5 + 3 \cos(2 \theta_\text{AB}) - 24 \log\left(\sin\left(\frac{\theta_\text{AB}}{2}\right)\right) \nonumber \\ && + \cos(\theta_\text{AB}) \left(-2 + 24 \log\left(\sin\left(\frac{\theta_\text{AB}}{2}\right)\right)\right) \Bigg) \tan\left(\frac{\theta_\text{AB}}{2}\right)~, \\
\Gamma_y (\theta_\text{AB})&=& 0~, \\
\Gamma_z (\theta_\text{AB}) &=& \Gamma_x (\hat{n}_\text{A}, \hat{n}_\text{B}, d) \cot\left(\frac{\theta_\text{AB}}{2}\right) ~.
\end{eqnarray}
\end{subequations}
and
\begin{subequations}
  \begin{flalign}
\Gamma_{xx} (\theta_\text{AB})= & \frac{1}{960} \Bigg( 99 + 40 \cos(3 \theta_\text{AB}) + 5 \cos(4 \theta_\text{AB}) + \cos(\theta_\text{AB}) \left( 8 - 960 \log\left(\sin\left(\frac{\theta_\text{AB}}{2}\right)\right) \right) \nonumber &\\ &  + 720 \log\left(\sin\left(\frac{\theta_\text{AB}}{2}\right)\right) + 8 \cos(2 \theta_\text{AB}) \left( -7 + 30 \log\left(\sin\left(\frac{\theta_\text{AB}}{2}\right)\right) \right) \Bigg) \nonumber &\\ & \times \sec^2\left(\frac{\theta_\text{AB}}{2}\right) ~, 
\end{flalign}
      \vspace{-1.4cm} 
  \begin{flalign}
\Gamma_{xy} (\theta_\text{AB})  = &  0 ~,& 
\end{flalign}
    \vspace{-1.4cm} 
\begin{flalign}
\Gamma_{xz} (\theta_\text{AB})  = &  -\frac{1}{240} \Bigg( -44 + 40 \cos(2 \theta_\text{AB}) + 5 \cos(3 \theta_\text{AB}) - 240 \log\left(\sin\left(\frac{\theta_\text{AB}}{2}\right)\right) \nonumber &\\ &  + 3 \cos(\theta_\text{AB}) \left( -3 + 80 \log\left(\sin\left(\frac{\theta_\text{AB}}{2}\right)\right) \right) \Bigg) \tan\left(\frac{\theta_\text{AB}}{2}\right)~, 
\end{flalign}
      \vspace{-1.4cm} 
  \begin{flalign}
\Gamma_{yy} (\theta_\text{AB})  = &  \frac{1}{240} \Bigg( -19 + \cos(2 \theta_\text{AB}) \left( 31 - 60 \log\left(\sin\left(\frac{\theta_\text{AB}}{2}\right)\right) \right) - 180 \log\left(\sin\left(\frac{\theta_\text{AB}}{2}\right)\right) \nonumber &\\ &  + 12 \cos(\theta_\text{AB}) \left( 1 + 20 \log\left(\sin\left(\frac{\theta_\text{AB}}{2}\right)\right) \right) \Bigg) \sec^2\left(\frac{\theta_\text{AB}}{2}\right)~, 
\end{flalign}
      \vspace{-1.4cm} 
  \begin{flalign}
\Gamma_{yz} (\theta_\text{AB})  = &  0~ &, 
\end{flalign}
      \vspace{-1.4cm} 
  \begin{flalign}
\Gamma_{zz} (\theta_\text{AB})  = &  \frac{5}{24} - \frac{1}{8} \cos(2 \theta_\text{AB}) - \frac{1}{48} \cos(3 \theta_\text{AB}) \nonumber &\\ &  + \cos(\theta_\text{AB}) \left( \frac{17}{240} - \log\left(\sin\left(\frac{\theta_\text{AB}}{2}\right)\right) \right) + \log\left(\sin\left(\frac{\theta_\text{AB}}{2}\right)\right) ~,
\end{flalign}
\end{subequations}

\bibliography{ref}

\begin{thebibliography}{131}%
\makeatletter
\providecommand \@ifxundefined [1]{%
 \@ifx{#1\undefined}
}%
\providecommand \@ifnum [1]{%
 \ifnum #1\expandafter \@firstoftwo
 \else \expandafter \@secondoftwo
 \fi
}%
\providecommand \@ifx [1]{%
 \ifx #1\expandafter \@firstoftwo
 \else \expandafter \@secondoftwo
 \fi
}%
\providecommand \natexlab [1]{#1}%
\providecommand \enquote  [1]{``#1''}%
\providecommand \bibnamefont  [1]{#1}%
\providecommand \bibfnamefont [1]{#1}%
\providecommand \citenamefont [1]{#1}%
\providecommand \href@noop [0]{\@secondoftwo}%
\providecommand \href [0]{\begingroup \@sanitize@url \@href}%
\providecommand \@href[1]{\@@startlink{#1}\@@href}%
\providecommand \@@href[1]{\endgroup#1\@@endlink}%
\providecommand \@sanitize@url [0]{\catcode `\\12\catcode `\$12\catcode `\&12\catcode `\#12\catcode `\^12\catcode `\_12\catcode `\%12\relax}%
\providecommand \@@startlink[1]{}%
\providecommand \@@endlink[0]{}%
\providecommand \url  [0]{\begingroup\@sanitize@url \@url }%
\providecommand \@url [1]{\endgroup\@href {#1}{\urlprefix }}%
\providecommand \urlprefix  [0]{URL }%
\providecommand \Eprint [0]{\href }%
\providecommand \doibase [0]{http://dx.doi.org/}%
\providecommand \selectlanguage [0]{\@gobble}%
\providecommand \bibinfo  [0]{\@secondoftwo}%
\providecommand \bibfield  [0]{\@secondoftwo}%
\providecommand \translation [1]{[#1]}%
\providecommand \BibitemOpen [0]{}%
\providecommand \bibitemStop [0]{}%
\providecommand \bibitemNoStop [0]{.\EOS\space}%
\providecommand \EOS [0]{\spacefactor3000\relax}%
\providecommand \BibitemShut  [1]{\csname bibitem#1\endcsname}%
\let\auto@bib@innerbib\@empty
\bibitem [{\citenamefont {Agazie}\ \emph {et~al.}(2023{\natexlab{a}})\citenamefont {Agazie} \emph {et~al.}}]{NANOGrav:2023gor}%
  \BibitemOpen
  \bibfield  {author} {\bibinfo {author} {\bibfnamefont {G.}~\bibnamefont {Agazie}} \emph {et~al.} (\bibinfo {collaboration} {NANOGrav}),\ }\href {\doibase 10.3847/2041-8213/acdac6} {\bibfield  {journal} {\bibinfo  {journal} {Astrophys. J. Lett.}\ }\textbf {\bibinfo {volume} {951}} (\bibinfo {year} {2023}{\natexlab{a}}),\ 10.3847/2041-8213/acdac6},\ \Eprint {http://arxiv.org/abs/2306.16213} {arXiv:2306.16213 [astro-ph.HE]} \BibitemShut {NoStop}%
\bibitem [{\citenamefont {Antoniadis}\ \emph {et~al.}(2023)\citenamefont {Antoniadis} \emph {et~al.}}]{EPTA:2023fyk}%
  \BibitemOpen
  \bibfield  {author} {\bibinfo {author} {\bibfnamefont {J.}~\bibnamefont {Antoniadis}} \emph {et~al.} (\bibinfo {collaboration} {EPTA, InPTA:}),\ }\href {\doibase 10.1051/0004-6361/202346844} {\bibfield  {journal} {\bibinfo  {journal} {Astron. Astrophys.}\ }\textbf {\bibinfo {volume} {678}},\ \bibinfo {pages} {A50} (\bibinfo {year} {2023})},\ \Eprint {http://arxiv.org/abs/2306.16214} {arXiv:2306.16214 [astro-ph.HE]} \BibitemShut {NoStop}%
\bibitem [{\citenamefont {Reardon}\ \emph {et~al.}(2023)\citenamefont {Reardon} \emph {et~al.}}]{Reardon:2023gzh}%
  \BibitemOpen
  \bibfield  {author} {\bibinfo {author} {\bibfnamefont {D.~J.}\ \bibnamefont {Reardon}} \emph {et~al.},\ }\href {\doibase 10.3847/2041-8213/acdd02} {\bibfield  {journal} {\bibinfo  {journal} {Astrophys. J. Lett.}\ }\textbf {\bibinfo {volume} {951}},\ \bibinfo {pages} {L6} (\bibinfo {year} {2023})},\ \Eprint {http://arxiv.org/abs/2306.16215} {arXiv:2306.16215 [astro-ph.HE]} \BibitemShut {NoStop}%
\bibitem [{\citenamefont {Xu}\ \emph {et~al.}(2023)\citenamefont {Xu} \emph {et~al.}}]{Xu:2023wog}%
  \BibitemOpen
  \bibfield  {author} {\bibinfo {author} {\bibfnamefont {H.}~\bibnamefont {Xu}} \emph {et~al.},\ }\href {\doibase 10.1088/1674-4527/acdfa5} {\bibfield  {journal} {\bibinfo  {journal} {Res. Astron. Astrophys.}\ }\textbf {\bibinfo {volume} {23}},\ \bibinfo {pages} {075024} (\bibinfo {year} {2023})},\ \Eprint {http://arxiv.org/abs/2306.16216} {arXiv:2306.16216 [astro-ph.HE]} \BibitemShut {NoStop}%
\bibitem [{\citenamefont {Agazie}\ \emph {et~al.}(2023{\natexlab{b}})\citenamefont {Agazie} \emph {et~al.}}]{NANOGrav:2023hfp}%
  \BibitemOpen
  \bibfield  {author} {\bibinfo {author} {\bibfnamefont {G.}~\bibnamefont {Agazie}} \emph {et~al.} (\bibinfo {collaboration} {NANOGrav}),\ }\href {\doibase 10.3847/2041-8213/ace18b} {\bibfield  {journal} {\bibinfo  {journal} {Astrophys. J. Lett.}\ }\textbf {\bibinfo {volume} {952}},\ \bibinfo {pages} {L37} (\bibinfo {year} {2023}{\natexlab{b}})},\ \Eprint {http://arxiv.org/abs/2306.16220} {arXiv:2306.16220 [astro-ph.HE]} \BibitemShut {NoStop}%
\bibitem [{\citenamefont {Agazie}\ \emph {et~al.}(2023{\natexlab{c}})\citenamefont {Agazie} \emph {et~al.}}]{NANOGrav:2023pdq}%
  \BibitemOpen
  \bibfield  {author} {\bibinfo {author} {\bibfnamefont {G.}~\bibnamefont {Agazie}} \emph {et~al.} (\bibinfo {collaboration} {NANOGrav}),\ }\href {\doibase 10.3847/2041-8213/ace18a} {\bibfield  {journal} {\bibinfo  {journal} {Astrophys. J. Lett.}\ }\textbf {\bibinfo {volume} {951}},\ \bibinfo {pages} {L50} (\bibinfo {year} {2023}{\natexlab{c}})},\ \Eprint {http://arxiv.org/abs/2306.16222} {arXiv:2306.16222 [astro-ph.HE]} \BibitemShut {NoStop}%
\bibitem [{\citenamefont {Agazie}\ \emph {et~al.}(2023{\natexlab{d}})\citenamefont {Agazie} \emph {et~al.}}]{NANOGrav:2023tcn}%
  \BibitemOpen
  \bibfield  {author} {\bibinfo {author} {\bibfnamefont {G.}~\bibnamefont {Agazie}} \emph {et~al.} (\bibinfo {collaboration} {NANOGrav}),\ }\href {\doibase 10.3847/2041-8213/acf4fd} {\bibfield  {journal} {\bibinfo  {journal} {Astrophys. J. Lett.}\ }\textbf {\bibinfo {volume} {956}},\ \bibinfo {pages} {L3} (\bibinfo {year} {2023}{\natexlab{d}})},\ \Eprint {http://arxiv.org/abs/2306.16221} {arXiv:2306.16221 [astro-ph.HE]} \BibitemShut {NoStop}%
\bibitem [{\citenamefont {Antoniadis}\ \emph {et~al.}(2024)\citenamefont {Antoniadis} \emph {et~al.}}]{EPTA:2023xxk}%
  \BibitemOpen
  \bibfield  {author} {\bibinfo {author} {\bibfnamefont {J.}~\bibnamefont {Antoniadis}} \emph {et~al.} (\bibinfo {collaboration} {EPTA, InPTA}),\ }\href {\doibase 10.1051/0004-6361/202347433} {\bibfield  {journal} {\bibinfo  {journal} {Astron. Astrophys.}\ }\textbf {\bibinfo {volume} {685}},\ \bibinfo {pages} {A94} (\bibinfo {year} {2024})},\ \Eprint {http://arxiv.org/abs/2306.16227} {arXiv:2306.16227 [astro-ph.CO]} \BibitemShut {NoStop}%
\bibitem [{\citenamefont {Agazie}\ \emph {et~al.}(2025)\citenamefont {Agazie} \emph {et~al.}}]{Agazie:2024kdi}%
  \BibitemOpen
  \bibfield  {author} {\bibinfo {author} {\bibfnamefont {G.}~\bibnamefont {Agazie}} \emph {et~al.},\ }\href {\doibase 10.3847/2041-8213/ad99d3} {\bibfield  {journal} {\bibinfo  {journal} {Astrophys. J. Lett.}\ }\textbf {\bibinfo {volume} {978}},\ \bibinfo {pages} {L29} (\bibinfo {year} {2025})},\ \Eprint {http://arxiv.org/abs/2408.10166} {arXiv:2408.10166 [astro-ph.HE]} \BibitemShut {NoStop}%
\bibitem [{\citenamefont {Agazie}\ \emph {et~al.}(2026)\citenamefont {Agazie} \emph {et~al.}}]{Agazie:2026tui}%
  \BibitemOpen
  \bibfield  {author} {\bibinfo {author} {\bibfnamefont {G.}~\bibnamefont {Agazie}} \emph {et~al.},\ }\href@noop {} {\  (\bibinfo {year} {2026})},\ \Eprint {http://arxiv.org/abs/2601.09481} {arXiv:2601.09481 [astro-ph.HE]} \BibitemShut {NoStop}%
\bibitem [{\citenamefont {Mingarelli}\ \emph {et~al.}(2017)\citenamefont {Mingarelli}, \citenamefont {Lazio}, \citenamefont {Sesana}, \citenamefont {Greene}, \citenamefont {Ellis}, \citenamefont {Ma}, \citenamefont {Croft}, \citenamefont {Burke-Spolaor},\ and\ \citenamefont {Taylor}}]{Mingarelli:2017fbe}%
  \BibitemOpen
  \bibfield  {author} {\bibinfo {author} {\bibfnamefont {C.~M.~F.}\ \bibnamefont {Mingarelli}}, \bibinfo {author} {\bibfnamefont {T.~J.~W.}\ \bibnamefont {Lazio}}, \bibinfo {author} {\bibfnamefont {A.}~\bibnamefont {Sesana}}, \bibinfo {author} {\bibfnamefont {J.~E.}\ \bibnamefont {Greene}}, \bibinfo {author} {\bibfnamefont {J.~A.}\ \bibnamefont {Ellis}}, \bibinfo {author} {\bibfnamefont {C.-P.}\ \bibnamefont {Ma}}, \bibinfo {author} {\bibfnamefont {S.}~\bibnamefont {Croft}}, \bibinfo {author} {\bibfnamefont {S.}~\bibnamefont {Burke-Spolaor}}, \ and\ \bibinfo {author} {\bibfnamefont {S.~R.}\ \bibnamefont {Taylor}},\ }\href {\doibase 10.1038/s41550-017-0299-6} {\bibfield  {journal} {\bibinfo  {journal} {Nature Astron.}\ }\textbf {\bibinfo {volume} {1}},\ \bibinfo {pages} {886} (\bibinfo {year} {2017})},\ \Eprint {http://arxiv.org/abs/1708.03491} {arXiv:1708.03491 [astro-ph.GA]} \BibitemShut {NoStop}%
\bibitem [{\citenamefont {Sato-Polito}\ and\ \citenamefont {Kamionkowski}(2024)}]{Sato-Polito:2023spo}%
  \BibitemOpen
  \bibfield  {author} {\bibinfo {author} {\bibfnamefont {G.}~\bibnamefont {Sato-Polito}}\ and\ \bibinfo {author} {\bibfnamefont {M.}~\bibnamefont {Kamionkowski}},\ }\href {\doibase 10.1103/PhysRevD.109.123544} {\bibfield  {journal} {\bibinfo  {journal} {Phys. Rev. D}\ }\textbf {\bibinfo {volume} {109}},\ \bibinfo {pages} {123544} (\bibinfo {year} {2024})},\ \Eprint {http://arxiv.org/abs/2305.05690} {arXiv:2305.05690 [astro-ph.CO]} \BibitemShut {NoStop}%
\bibitem [{\citenamefont {Grimm}\ \emph {et~al.}(2025)\citenamefont {Grimm}, \citenamefont {Pijnenburg}, \citenamefont {Cusin},\ and\ \citenamefont {Bonvin}}]{Grimm:2024lfj}%
  \BibitemOpen
  \bibfield  {author} {\bibinfo {author} {\bibfnamefont {N.}~\bibnamefont {Grimm}}, \bibinfo {author} {\bibfnamefont {M.}~\bibnamefont {Pijnenburg}}, \bibinfo {author} {\bibfnamefont {G.}~\bibnamefont {Cusin}}, \ and\ \bibinfo {author} {\bibfnamefont {C.}~\bibnamefont {Bonvin}},\ }\href {\doibase 10.1088/1475-7516/2025/03/011} {\bibfield  {journal} {\bibinfo  {journal} {JCAP}\ }\textbf {\bibinfo {volume} {03}},\ \bibinfo {pages} {011} (\bibinfo {year} {2025})},\ \Eprint {http://arxiv.org/abs/2404.05670} {arXiv:2404.05670 [astro-ph.CO]} \BibitemShut {NoStop}%
\bibitem [{\citenamefont {Lemke}\ \emph {et~al.}(2025)\citenamefont {Lemke}, \citenamefont {Mitridate},\ and\ \citenamefont {Gersbach}}]{Lemke:2024cdu}%
  \BibitemOpen
  \bibfield  {author} {\bibinfo {author} {\bibfnamefont {A.-M.}\ \bibnamefont {Lemke}}, \bibinfo {author} {\bibfnamefont {A.}~\bibnamefont {Mitridate}}, \ and\ \bibinfo {author} {\bibfnamefont {K.~A.}\ \bibnamefont {Gersbach}},\ }\href {\doibase 10.1103/PhysRevD.111.063068} {\bibfield  {journal} {\bibinfo  {journal} {Phys. Rev. D}\ }\textbf {\bibinfo {volume} {111}},\ \bibinfo {pages} {063068} (\bibinfo {year} {2025})},\ \Eprint {http://arxiv.org/abs/2407.08705} {arXiv:2407.08705 [astro-ph.HE]} \BibitemShut {NoStop}%
\bibitem [{\citenamefont {Yang}\ \emph {et~al.}(2025)\citenamefont {Yang}, \citenamefont {Guo}, \citenamefont {Cao}, \citenamefont {Shao},\ and\ \citenamefont {Yuan}}]{Yang:2024mqz}%
  \BibitemOpen
  \bibfield  {author} {\bibinfo {author} {\bibfnamefont {Q.}~\bibnamefont {Yang}}, \bibinfo {author} {\bibfnamefont {X.}~\bibnamefont {Guo}}, \bibinfo {author} {\bibfnamefont {Z.}~\bibnamefont {Cao}}, \bibinfo {author} {\bibfnamefont {X.}~\bibnamefont {Shao}}, \ and\ \bibinfo {author} {\bibfnamefont {X.}~\bibnamefont {Yuan}},\ }\href {\doibase 10.3847/1538-4357/aded02} {\bibfield  {journal} {\bibinfo  {journal} {Astrophys. J.}\ }\textbf {\bibinfo {volume} {989}},\ \bibinfo {pages} {157} (\bibinfo {year} {2025})},\ \Eprint {http://arxiv.org/abs/2408.05043} {arXiv:2408.05043 [astro-ph.CO]} \BibitemShut {NoStop}%
\bibitem [{\citenamefont {Olmez}\ \emph {et~al.}(2012)\citenamefont {Olmez}, \citenamefont {Mandic},\ and\ \citenamefont {Siemens}}]{Olmez:2011cg}%
  \BibitemOpen
  \bibfield  {author} {\bibinfo {author} {\bibfnamefont {S.}~\bibnamefont {Olmez}}, \bibinfo {author} {\bibfnamefont {V.}~\bibnamefont {Mandic}}, \ and\ \bibinfo {author} {\bibfnamefont {X.}~\bibnamefont {Siemens}},\ }\href {\doibase 10.1088/1475-7516/2012/07/009} {\bibfield  {journal} {\bibinfo  {journal} {JCAP}\ }\textbf {\bibinfo {volume} {07}},\ \bibinfo {pages} {009} (\bibinfo {year} {2012})},\ \Eprint {http://arxiv.org/abs/1106.5555} {arXiv:1106.5555 [astro-ph.CO]} \BibitemShut {NoStop}%
\bibitem [{\citenamefont {Kuroyanagi}\ \emph {et~al.}(2017)\citenamefont {Kuroyanagi}, \citenamefont {Takahashi}, \citenamefont {Yonemaru},\ and\ \citenamefont {Kumamoto}}]{Kuroyanagi:2016ugi}%
  \BibitemOpen
  \bibfield  {author} {\bibinfo {author} {\bibfnamefont {S.}~\bibnamefont {Kuroyanagi}}, \bibinfo {author} {\bibfnamefont {K.}~\bibnamefont {Takahashi}}, \bibinfo {author} {\bibfnamefont {N.}~\bibnamefont {Yonemaru}}, \ and\ \bibinfo {author} {\bibfnamefont {H.}~\bibnamefont {Kumamoto}},\ }\href {\doibase 10.1103/PhysRevD.95.043531} {\bibfield  {journal} {\bibinfo  {journal} {Phys. Rev. D}\ }\textbf {\bibinfo {volume} {95}},\ \bibinfo {pages} {043531} (\bibinfo {year} {2017})},\ \Eprint {http://arxiv.org/abs/1604.00332} {arXiv:1604.00332 [astro-ph.CO]} \BibitemShut {NoStop}%
\bibitem [{\citenamefont {Jenkins}\ and\ \citenamefont {Sakellariadou}(2018)}]{Jenkins:2018nty}%
  \BibitemOpen
  \bibfield  {author} {\bibinfo {author} {\bibfnamefont {A.~C.}\ \bibnamefont {Jenkins}}\ and\ \bibinfo {author} {\bibfnamefont {M.}~\bibnamefont {Sakellariadou}},\ }\href {\doibase 10.1103/PhysRevD.98.063509} {\bibfield  {journal} {\bibinfo  {journal} {Phys. Rev. D}\ }\textbf {\bibinfo {volume} {98}},\ \bibinfo {pages} {063509} (\bibinfo {year} {2018})},\ \Eprint {http://arxiv.org/abs/1802.06046} {arXiv:1802.06046 [astro-ph.CO]} \BibitemShut {NoStop}%
\bibitem [{\citenamefont {Geller}\ \emph {et~al.}(2018)\citenamefont {Geller}, \citenamefont {Hook}, \citenamefont {Sundrum},\ and\ \citenamefont {Tsai}}]{Geller:2018mwu}%
  \BibitemOpen
  \bibfield  {author} {\bibinfo {author} {\bibfnamefont {M.}~\bibnamefont {Geller}}, \bibinfo {author} {\bibfnamefont {A.}~\bibnamefont {Hook}}, \bibinfo {author} {\bibfnamefont {R.}~\bibnamefont {Sundrum}}, \ and\ \bibinfo {author} {\bibfnamefont {Y.}~\bibnamefont {Tsai}},\ }\href {\doibase 10.1103/PhysRevLett.121.201303} {\bibfield  {journal} {\bibinfo  {journal} {Phys. Rev. Lett.}\ }\textbf {\bibinfo {volume} {121}},\ \bibinfo {pages} {201303} (\bibinfo {year} {2018})},\ \Eprint {http://arxiv.org/abs/1803.10780} {arXiv:1803.10780 [hep-ph]} \BibitemShut {NoStop}%
\bibitem [{\citenamefont {Liu}\ \emph {et~al.}(2021)\citenamefont {Liu}, \citenamefont {Cai},\ and\ \citenamefont {Guo}}]{Liu:2020mru}%
  \BibitemOpen
  \bibfield  {author} {\bibinfo {author} {\bibfnamefont {J.}~\bibnamefont {Liu}}, \bibinfo {author} {\bibfnamefont {R.-G.}\ \bibnamefont {Cai}}, \ and\ \bibinfo {author} {\bibfnamefont {Z.-K.}\ \bibnamefont {Guo}},\ }\href {\doibase 10.1103/PhysRevLett.126.141303} {\bibfield  {journal} {\bibinfo  {journal} {Phys. Rev. Lett.}\ }\textbf {\bibinfo {volume} {126}},\ \bibinfo {pages} {141303} (\bibinfo {year} {2021})},\ \Eprint {http://arxiv.org/abs/2010.03225} {arXiv:2010.03225 [astro-ph.CO]} \BibitemShut {NoStop}%
\bibitem [{\citenamefont {Li}\ \emph {et~al.}(2022)\citenamefont {Li}, \citenamefont {Huang}, \citenamefont {Wang},\ and\ \citenamefont {Zhang}}]{Li:2021iva}%
  \BibitemOpen
  \bibfield  {author} {\bibinfo {author} {\bibfnamefont {Y.}~\bibnamefont {Li}}, \bibinfo {author} {\bibfnamefont {F.~P.}\ \bibnamefont {Huang}}, \bibinfo {author} {\bibfnamefont {X.}~\bibnamefont {Wang}}, \ and\ \bibinfo {author} {\bibfnamefont {X.}~\bibnamefont {Zhang}},\ }\href {\doibase 10.1103/PhysRevD.105.083527} {\bibfield  {journal} {\bibinfo  {journal} {Phys. Rev. D}\ }\textbf {\bibinfo {volume} {105}},\ \bibinfo {pages} {083527} (\bibinfo {year} {2022})},\ \Eprint {http://arxiv.org/abs/2112.01409} {arXiv:2112.01409 [astro-ph.CO]} \BibitemShut {NoStop}%
\bibitem [{\citenamefont {Ota}(2020)}]{Ota:2020vfn}%
  \BibitemOpen
  \bibfield  {author} {\bibinfo {author} {\bibfnamefont {A.}~\bibnamefont {Ota}},\ }\href {\doibase 10.1103/PhysRevD.101.103511} {\bibfield  {journal} {\bibinfo  {journal} {Phys. Rev. D}\ }\textbf {\bibinfo {volume} {101}},\ \bibinfo {pages} {103511} (\bibinfo {year} {2020})},\ \Eprint {http://arxiv.org/abs/2001.00409} {arXiv:2001.00409 [astro-ph.CO]} \BibitemShut {NoStop}%
\bibitem [{\citenamefont {Zhang}\ \emph {et~al.}(2022)\citenamefont {Zhang}, \citenamefont {Zhou},\ and\ \citenamefont {Chang}}]{Zhang:2022dgx}%
  \BibitemOpen
  \bibfield  {author} {\bibinfo {author} {\bibfnamefont {X.}~\bibnamefont {Zhang}}, \bibinfo {author} {\bibfnamefont {J.-Z.}\ \bibnamefont {Zhou}}, \ and\ \bibinfo {author} {\bibfnamefont {Z.}~\bibnamefont {Chang}},\ }\href {\doibase 10.1140/epjc/s10052-022-10742-x} {\bibfield  {journal} {\bibinfo  {journal} {Eur. Phys. J. C}\ }\textbf {\bibinfo {volume} {82}},\ \bibinfo {pages} {781} (\bibinfo {year} {2022})},\ \Eprint {http://arxiv.org/abs/2208.12948} {arXiv:2208.12948 [astro-ph.CO]} \BibitemShut {NoStop}%
\bibitem [{\citenamefont {Chen}\ and\ \citenamefont {Ota}(2022)}]{Chen:2022qec}%
  \BibitemOpen
  \bibfield  {author} {\bibinfo {author} {\bibfnamefont {C.}~\bibnamefont {Chen}}\ and\ \bibinfo {author} {\bibfnamefont {A.}~\bibnamefont {Ota}},\ }\href {\doibase 10.1103/PhysRevD.106.063507} {\bibfield  {journal} {\bibinfo  {journal} {Phys. Rev. D}\ }\textbf {\bibinfo {volume} {106}},\ \bibinfo {pages} {063507} (\bibinfo {year} {2022})},\ \Eprint {http://arxiv.org/abs/2205.07810} {arXiv:2205.07810 [astro-ph.CO]} \BibitemShut {NoStop}%
\bibitem [{\citenamefont {Kuang}\ \emph {et~al.}(2023)\citenamefont {Kuang}, \citenamefont {Zhou}, \citenamefont {Wu},\ and\ \citenamefont {Chang}}]{Kuang:2023urj}%
  \BibitemOpen
  \bibfield  {author} {\bibinfo {author} {\bibfnamefont {Y.-T.}\ \bibnamefont {Kuang}}, \bibinfo {author} {\bibfnamefont {J.-Z.}\ \bibnamefont {Zhou}}, \bibinfo {author} {\bibfnamefont {D.}~\bibnamefont {Wu}}, \ and\ \bibinfo {author} {\bibfnamefont {Z.}~\bibnamefont {Chang}},\ }\href@noop {} {\  (\bibinfo {year} {2023})},\ \Eprint {http://arxiv.org/abs/2309.06676} {arXiv:2309.06676 [astro-ph.CO]} \BibitemShut {NoStop}%
\bibitem [{\citenamefont {Li}\ \emph {et~al.}(2023)\citenamefont {Li}, \citenamefont {Wang}, \citenamefont {Zhao},\ and\ \citenamefont {Kohri}}]{Li:2023qua}%
  \BibitemOpen
  \bibfield  {author} {\bibinfo {author} {\bibfnamefont {J.-P.}\ \bibnamefont {Li}}, \bibinfo {author} {\bibfnamefont {S.}~\bibnamefont {Wang}}, \bibinfo {author} {\bibfnamefont {Z.-C.}\ \bibnamefont {Zhao}}, \ and\ \bibinfo {author} {\bibfnamefont {K.}~\bibnamefont {Kohri}},\ }\href {\doibase 10.1088/1475-7516/2023/10/056} {\bibfield  {journal} {\bibinfo  {journal} {JCAP}\ }\textbf {\bibinfo {volume} {10}},\ \bibinfo {pages} {056} (\bibinfo {year} {2023})},\ \Eprint {http://arxiv.org/abs/2305.19950} {arXiv:2305.19950 [astro-ph.CO]} \BibitemShut {NoStop}%
\bibitem [{\citenamefont {Yu}\ and\ \citenamefont {Wang}(2024)}]{Yu:2023jrs}%
  \BibitemOpen
  \bibfield  {author} {\bibinfo {author} {\bibfnamefont {Y.-H.}\ \bibnamefont {Yu}}\ and\ \bibinfo {author} {\bibfnamefont {S.}~\bibnamefont {Wang}},\ }\href {\doibase 10.1103/PhysRevD.109.083501} {\bibfield  {journal} {\bibinfo  {journal} {Phys. Rev. D}\ }\textbf {\bibinfo {volume} {109}},\ \bibinfo {pages} {083501} (\bibinfo {year} {2024})},\ \Eprint {http://arxiv.org/abs/2310.14606} {arXiv:2310.14606 [astro-ph.CO]} \BibitemShut {NoStop}%
\bibitem [{\citenamefont {Mukherjee}\ \emph {et~al.}(2026)\citenamefont {Mukherjee}, \citenamefont {Ragavendra},\ and\ \citenamefont {Sethi}}]{Mukherjee:2025dcv}%
  \BibitemOpen
  \bibfield  {author} {\bibinfo {author} {\bibfnamefont {D.}~\bibnamefont {Mukherjee}}, \bibinfo {author} {\bibfnamefont {H.~V.}\ \bibnamefont {Ragavendra}}, \ and\ \bibinfo {author} {\bibfnamefont {S.~K.}\ \bibnamefont {Sethi}},\ }\href {\doibase 10.1103/qd1s-9fxl} {\bibfield  {journal} {\bibinfo  {journal} {Phys. Rev. D}\ }\textbf {\bibinfo {volume} {113}},\ \bibinfo {pages} {023533} (\bibinfo {year} {2026})},\ \Eprint {http://arxiv.org/abs/2506.23798} {arXiv:2506.23798 [astro-ph.CO]} \BibitemShut {NoStop}%
\bibitem [{\citenamefont {Luo}\ \emph {et~al.}(2025)\citenamefont {Luo}, \citenamefont {Yu}, \citenamefont {Li},\ and\ \citenamefont {Wang}}]{Luo:2025lgr}%
  \BibitemOpen
  \bibfield  {author} {\bibinfo {author} {\bibfnamefont {D.}~\bibnamefont {Luo}}, \bibinfo {author} {\bibfnamefont {Y.-H.}\ \bibnamefont {Yu}}, \bibinfo {author} {\bibfnamefont {J.-P.}\ \bibnamefont {Li}}, \ and\ \bibinfo {author} {\bibfnamefont {S.}~\bibnamefont {Wang}},\ }\href {\doibase 10.1088/1475-7516/2025/04/085} {\bibfield  {journal} {\bibinfo  {journal} {JCAP}\ }\textbf {\bibinfo {volume} {04}},\ \bibinfo {pages} {085} (\bibinfo {year} {2025})},\ \Eprint {http://arxiv.org/abs/2501.02965} {arXiv:2501.02965 [astro-ph.CO]} \BibitemShut {NoStop}%
\bibitem [{\citenamefont {Li}\ \emph {et~al.}(2025{\natexlab{a}})\citenamefont {Li}, \citenamefont {Wang}, \citenamefont {Zhao},\ and\ \citenamefont {Kohri}}]{Li:2025met}%
  \BibitemOpen
  \bibfield  {author} {\bibinfo {author} {\bibfnamefont {J.-P.}\ \bibnamefont {Li}}, \bibinfo {author} {\bibfnamefont {S.}~\bibnamefont {Wang}}, \bibinfo {author} {\bibfnamefont {Z.-C.}\ \bibnamefont {Zhao}}, \ and\ \bibinfo {author} {\bibfnamefont {K.}~\bibnamefont {Kohri}},\ }\href@noop {} {\  (\bibinfo {year} {2025}{\natexlab{a}})},\ \Eprint {http://arxiv.org/abs/2505.16820} {arXiv:2505.16820 [astro-ph.CO]} \BibitemShut {NoStop}%
\bibitem [{\citenamefont {Li}\ \emph {et~al.}(2024{\natexlab{a}})\citenamefont {Li}, \citenamefont {Wang}, \citenamefont {Zhao},\ and\ \citenamefont {Kohri}}]{Li:2024zwx}%
  \BibitemOpen
  \bibfield  {author} {\bibinfo {author} {\bibfnamefont {J.-P.}\ \bibnamefont {Li}}, \bibinfo {author} {\bibfnamefont {S.}~\bibnamefont {Wang}}, \bibinfo {author} {\bibfnamefont {Z.-C.}\ \bibnamefont {Zhao}}, \ and\ \bibinfo {author} {\bibfnamefont {K.}~\bibnamefont {Kohri}},\ }\href {\doibase 10.1088/1475-7516/2024/05/109} {\bibfield  {journal} {\bibinfo  {journal} {JCAP}\ }\textbf {\bibinfo {volume} {05}},\ \bibinfo {pages} {109} (\bibinfo {year} {2024}{\natexlab{a}})},\ \Eprint {http://arxiv.org/abs/2403.00238} {arXiv:2403.00238 [astro-ph.CO]} \BibitemShut {NoStop}%
\bibitem [{\citenamefont {Vilenkin}(1981)}]{Vilenkin:1981bx}%
  \BibitemOpen
  \bibfield  {author} {\bibinfo {author} {\bibfnamefont {A.}~\bibnamefont {Vilenkin}},\ }\href {\doibase 10.1016/0370-2693(81)91144-8} {\bibfield  {journal} {\bibinfo  {journal} {Phys. Lett. B}\ }\textbf {\bibinfo {volume} {107}},\ \bibinfo {pages} {47} (\bibinfo {year} {1981})}\BibitemShut {NoStop}%
\bibitem [{\citenamefont {Hogan}\ and\ \citenamefont {Rees}(1984)}]{Hogan:1984is}%
  \BibitemOpen
  \bibfield  {author} {\bibinfo {author} {\bibfnamefont {C.~J.}\ \bibnamefont {Hogan}}\ and\ \bibinfo {author} {\bibfnamefont {M.~J.}\ \bibnamefont {Rees}},\ }\href {\doibase 10.1038/311109a0} {\bibfield  {journal} {\bibinfo  {journal} {Nature}\ }\textbf {\bibinfo {volume} {311}},\ \bibinfo {pages} {109} (\bibinfo {year} {1984})}\BibitemShut {NoStop}%
\bibitem [{\citenamefont {Vachaspati}\ and\ \citenamefont {Vilenkin}(1985)}]{Vachaspati:1984gt}%
  \BibitemOpen
  \bibfield  {author} {\bibinfo {author} {\bibfnamefont {T.}~\bibnamefont {Vachaspati}}\ and\ \bibinfo {author} {\bibfnamefont {A.}~\bibnamefont {Vilenkin}},\ }\href {\doibase 10.1103/PhysRevD.31.3052} {\bibfield  {journal} {\bibinfo  {journal} {Phys. Rev. D}\ }\textbf {\bibinfo {volume} {31}},\ \bibinfo {pages} {3052} (\bibinfo {year} {1985})}\BibitemShut {NoStop}%
\bibitem [{\citenamefont {Damour}\ and\ \citenamefont {Vilenkin}(2005)}]{Damour:2004kw}%
  \BibitemOpen
  \bibfield  {author} {\bibinfo {author} {\bibfnamefont {T.}~\bibnamefont {Damour}}\ and\ \bibinfo {author} {\bibfnamefont {A.}~\bibnamefont {Vilenkin}},\ }\href {\doibase 10.1103/PhysRevD.71.063510} {\bibfield  {journal} {\bibinfo  {journal} {Phys. Rev. D}\ }\textbf {\bibinfo {volume} {71}},\ \bibinfo {pages} {063510} (\bibinfo {year} {2005})},\ \Eprint {http://arxiv.org/abs/hep-th/0410222} {arXiv:hep-th/0410222} \BibitemShut {NoStop}%
\bibitem [{\citenamefont {Blanco-Pillado}\ and\ \citenamefont {Olum}(2017)}]{Blanco-Pillado:2017oxo}%
  \BibitemOpen
  \bibfield  {author} {\bibinfo {author} {\bibfnamefont {J.~J.}\ \bibnamefont {Blanco-Pillado}}\ and\ \bibinfo {author} {\bibfnamefont {K.~D.}\ \bibnamefont {Olum}},\ }\href {\doibase 10.1103/PhysRevD.96.104046} {\bibfield  {journal} {\bibinfo  {journal} {Phys. Rev. D}\ }\textbf {\bibinfo {volume} {96}},\ \bibinfo {pages} {104046} (\bibinfo {year} {2017})},\ \Eprint {http://arxiv.org/abs/1709.02693} {arXiv:1709.02693 [astro-ph.CO]} \BibitemShut {NoStop}%
\bibitem [{\citenamefont {Ringeval}\ and\ \citenamefont {Suyama}(2017)}]{Ringeval:2017eww}%
  \BibitemOpen
  \bibfield  {author} {\bibinfo {author} {\bibfnamefont {C.}~\bibnamefont {Ringeval}}\ and\ \bibinfo {author} {\bibfnamefont {T.}~\bibnamefont {Suyama}},\ }\href {\doibase 10.1088/1475-7516/2017/12/027} {\bibfield  {journal} {\bibinfo  {journal} {JCAP}\ }\textbf {\bibinfo {volume} {12}},\ \bibinfo {pages} {027} (\bibinfo {year} {2017})},\ \Eprint {http://arxiv.org/abs/1709.03845} {arXiv:1709.03845 [astro-ph.CO]} \BibitemShut {NoStop}%
\bibitem [{\citenamefont {Baeza-Ballesteros}\ \emph {et~al.}(2024)\citenamefont {Baeza-Ballesteros}, \citenamefont {Copeland}, \citenamefont {Figueroa},\ and\ \citenamefont {Lizarraga}}]{Baeza-Ballesteros:2023say}%
  \BibitemOpen
  \bibfield  {author} {\bibinfo {author} {\bibfnamefont {J.}~\bibnamefont {Baeza-Ballesteros}}, \bibinfo {author} {\bibfnamefont {E.~J.}\ \bibnamefont {Copeland}}, \bibinfo {author} {\bibfnamefont {D.~G.}\ \bibnamefont {Figueroa}}, \ and\ \bibinfo {author} {\bibfnamefont {J.}~\bibnamefont {Lizarraga}},\ }\href {\doibase 10.1103/PhysRevD.110.043522} {\bibfield  {journal} {\bibinfo  {journal} {Phys. Rev. D}\ }\textbf {\bibinfo {volume} {110}},\ \bibinfo {pages} {043522} (\bibinfo {year} {2024})},\ \Eprint {http://arxiv.org/abs/2308.08456} {arXiv:2308.08456 [astro-ph.CO]} \BibitemShut {NoStop}%
\bibitem [{\citenamefont {Fu}\ \emph {et~al.}(2023)\citenamefont {Fu}, \citenamefont {Ghoshal},\ and\ \citenamefont {King}}]{Fu:2023nrn}%
  \BibitemOpen
  \bibfield  {author} {\bibinfo {author} {\bibfnamefont {B.}~\bibnamefont {Fu}}, \bibinfo {author} {\bibfnamefont {A.}~\bibnamefont {Ghoshal}}, \ and\ \bibinfo {author} {\bibfnamefont {S.~F.}\ \bibnamefont {King}},\ }\href {\doibase 10.1007/JHEP11(2023)071} {\bibfield  {journal} {\bibinfo  {journal} {JHEP}\ }\textbf {\bibinfo {volume} {11}},\ \bibinfo {pages} {071} (\bibinfo {year} {2023})},\ \Eprint {http://arxiv.org/abs/2306.07334} {arXiv:2306.07334 [hep-ph]} \BibitemShut {NoStop}%
\bibitem [{\citenamefont {Qiu}\ and\ \citenamefont {Yu}(2023)}]{Qiu:2023wbs}%
  \BibitemOpen
  \bibfield  {author} {\bibinfo {author} {\bibfnamefont {Z.-Y.}\ \bibnamefont {Qiu}}\ and\ \bibinfo {author} {\bibfnamefont {Z.-H.}\ \bibnamefont {Yu}},\ }\href {\doibase 10.1088/1674-1137/acd9bf} {\bibfield  {journal} {\bibinfo  {journal} {Chin. Phys. C}\ }\textbf {\bibinfo {volume} {47}},\ \bibinfo {pages} {085104} (\bibinfo {year} {2023})},\ \Eprint {http://arxiv.org/abs/2304.02506} {arXiv:2304.02506 [hep-ph]} \BibitemShut {NoStop}%
\bibitem [{\citenamefont {Li}\ \emph {et~al.}(2025{\natexlab{b}})\citenamefont {Li}, \citenamefont {Yu},\ and\ \citenamefont {Pan}}]{Li:2024iua}%
  \BibitemOpen
  \bibfield  {author} {\bibinfo {author} {\bibfnamefont {M.}~\bibnamefont {Li}}, \bibinfo {author} {\bibfnamefont {J.}~\bibnamefont {Yu}}, \ and\ \bibinfo {author} {\bibfnamefont {Z.}~\bibnamefont {Pan}},\ }\href {\doibase 10.1103/PhysRevD.111.023009} {\bibfield  {journal} {\bibinfo  {journal} {Phys. Rev. D}\ }\textbf {\bibinfo {volume} {111}},\ \bibinfo {pages} {023009} (\bibinfo {year} {2025}{\natexlab{b}})},\ \Eprint {http://arxiv.org/abs/2403.01846} {arXiv:2403.01846 [gr-qc]} \BibitemShut {NoStop}%
\bibitem [{\citenamefont {Hu}\ and\ \citenamefont {Kamada}(2025)}]{Hu:2025sxv}%
  \BibitemOpen
  \bibfield  {author} {\bibinfo {author} {\bibfnamefont {Y.}~\bibnamefont {Hu}}\ and\ \bibinfo {author} {\bibfnamefont {K.}~\bibnamefont {Kamada}},\ }\href {\doibase 10.1088/1475-7516/2025/04/044} {\bibfield  {journal} {\bibinfo  {journal} {JCAP}\ }\textbf {\bibinfo {volume} {04}},\ \bibinfo {pages} {044} (\bibinfo {year} {2025})},\ \Eprint {http://arxiv.org/abs/2501.18380} {arXiv:2501.18380 [astro-ph.CO]} \BibitemShut {NoStop}%
\bibitem [{\citenamefont {Hogan}(1986)}]{Hogan:1986dsh}%
  \BibitemOpen
  \bibfield  {author} {\bibinfo {author} {\bibfnamefont {C.~J.}\ \bibnamefont {Hogan}},\ }\href {\doibase 10.1093/mnras/218.4.629} {\bibfield  {journal} {\bibinfo  {journal} {Mon. Not. Roy. Astron. Soc.}\ }\textbf {\bibinfo {volume} {218}},\ \bibinfo {pages} {629} (\bibinfo {year} {1986})}\BibitemShut {NoStop}%
\bibitem [{\citenamefont {Barni}\ \emph {et~al.}(2026)\citenamefont {Barni}, \citenamefont {Blasi}, \citenamefont {Madge},\ and\ \citenamefont {Vanvlasselaer}}]{Barni:2025gnm}%
  \BibitemOpen
  \bibfield  {author} {\bibinfo {author} {\bibfnamefont {G.}~\bibnamefont {Barni}}, \bibinfo {author} {\bibfnamefont {S.}~\bibnamefont {Blasi}}, \bibinfo {author} {\bibfnamefont {E.}~\bibnamefont {Madge}}, \ and\ \bibinfo {author} {\bibfnamefont {M.}~\bibnamefont {Vanvlasselaer}},\ }\href {\doibase 10.1088/1475-7516/2026/03/066} {\bibfield  {journal} {\bibinfo  {journal} {JCAP}\ }\textbf {\bibinfo {volume} {03}},\ \bibinfo {pages} {066} (\bibinfo {year} {2026})},\ \Eprint {http://arxiv.org/abs/2510.21439} {arXiv:2510.21439 [hep-ph]} \BibitemShut {NoStop}%
\bibitem [{\citenamefont {Chatrchyan}\ \emph {et~al.}(2026)\citenamefont {Chatrchyan}, \citenamefont {Marsh},\ and\ \citenamefont {Nikolis}}]{Chatrchyan:2025wop}%
  \BibitemOpen
  \bibfield  {author} {\bibinfo {author} {\bibfnamefont {A.}~\bibnamefont {Chatrchyan}}, \bibinfo {author} {\bibfnamefont {M.~C.~D.}\ \bibnamefont {Marsh}}, \ and\ \bibinfo {author} {\bibfnamefont {C.}~\bibnamefont {Nikolis}},\ }\href {\doibase 10.1103/2v2f-1jvz} {\bibfield  {journal} {\bibinfo  {journal} {Phys. Rev. Lett.}\ }\textbf {\bibinfo {volume} {136}},\ \bibinfo {pages} {041005} (\bibinfo {year} {2026})},\ \Eprint {http://arxiv.org/abs/2507.01191} {arXiv:2507.01191 [hep-ph]} \BibitemShut {NoStop}%
\bibitem [{\citenamefont {Inomata}\ \emph {et~al.}(2025)\citenamefont {Inomata}, \citenamefont {Kamionkowski}, \citenamefont {Kasai},\ and\ \citenamefont {Shakya}}]{Inomata:2024rkt}%
  \BibitemOpen
  \bibfield  {author} {\bibinfo {author} {\bibfnamefont {K.}~\bibnamefont {Inomata}}, \bibinfo {author} {\bibfnamefont {M.}~\bibnamefont {Kamionkowski}}, \bibinfo {author} {\bibfnamefont {K.}~\bibnamefont {Kasai}}, \ and\ \bibinfo {author} {\bibfnamefont {B.}~\bibnamefont {Shakya}},\ }\href {\doibase 10.1103/k4s5-8zqy} {\bibfield  {journal} {\bibinfo  {journal} {Phys. Rev. D}\ }\textbf {\bibinfo {volume} {112}},\ \bibinfo {pages} {083523} (\bibinfo {year} {2025})},\ \Eprint {http://arxiv.org/abs/2412.17912} {arXiv:2412.17912 [astro-ph.CO]} \BibitemShut {NoStop}%
\bibitem [{\citenamefont {Shao}\ \emph {et~al.}(2025)\citenamefont {Shao}, \citenamefont {Mao},\ and\ \citenamefont {Huang}}]{Shao:2024dxt}%
  \BibitemOpen
  \bibfield  {author} {\bibinfo {author} {\bibfnamefont {J.}~\bibnamefont {Shao}}, \bibinfo {author} {\bibfnamefont {H.}~\bibnamefont {Mao}}, \ and\ \bibinfo {author} {\bibfnamefont {M.}~\bibnamefont {Huang}},\ }\href {\doibase 10.1103/PhysRevD.111.023052} {\bibfield  {journal} {\bibinfo  {journal} {Phys. Rev. D}\ }\textbf {\bibinfo {volume} {111}},\ \bibinfo {pages} {023052} (\bibinfo {year} {2025})},\ \Eprint {http://arxiv.org/abs/2410.06780} {arXiv:2410.06780 [hep-ph]} \BibitemShut {NoStop}%
\bibitem [{\citenamefont {Tian}\ \emph {et~al.}(2025)\citenamefont {Tian}, \citenamefont {Wang},\ and\ \citenamefont {Bal{\'a}zs}}]{Tian:2024ysd}%
  \BibitemOpen
  \bibfield  {author} {\bibinfo {author} {\bibfnamefont {C.}~\bibnamefont {Tian}}, \bibinfo {author} {\bibfnamefont {X.}~\bibnamefont {Wang}}, \ and\ \bibinfo {author} {\bibfnamefont {C.}~\bibnamefont {Bal{\'a}zs}},\ }\href {\doibase 10.1140/epjc/s10052-025-14826-2} {\bibfield  {journal} {\bibinfo  {journal} {Eur. Phys. J. C}\ }\textbf {\bibinfo {volume} {85}},\ \bibinfo {pages} {1091} (\bibinfo {year} {2025})},\ \Eprint {http://arxiv.org/abs/2409.14505} {arXiv:2409.14505 [hep-ph]} \BibitemShut {NoStop}%
\bibitem [{\citenamefont {Niemi}\ and\ \citenamefont {Tenkanen}(2025)}]{Niemi:2024vzw}%
  \BibitemOpen
  \bibfield  {author} {\bibinfo {author} {\bibfnamefont {L.}~\bibnamefont {Niemi}}\ and\ \bibinfo {author} {\bibfnamefont {T.~V.~I.}\ \bibnamefont {Tenkanen}},\ }\href {\doibase 10.1103/PhysRevD.111.075034} {\bibfield  {journal} {\bibinfo  {journal} {Phys. Rev. D}\ }\textbf {\bibinfo {volume} {111}},\ \bibinfo {pages} {075034} (\bibinfo {year} {2025})},\ \Eprint {http://arxiv.org/abs/2408.15912} {arXiv:2408.15912 [hep-ph]} \BibitemShut {NoStop}%
\bibitem [{\citenamefont {Jiang}\ \emph {et~al.}(2023)\citenamefont {Jiang}, \citenamefont {Huang},\ and\ \citenamefont {Li}}]{Jiang:2023nkj}%
  \BibitemOpen
  \bibfield  {author} {\bibinfo {author} {\bibfnamefont {S.}~\bibnamefont {Jiang}}, \bibinfo {author} {\bibfnamefont {F.~P.}\ \bibnamefont {Huang}}, \ and\ \bibinfo {author} {\bibfnamefont {C.~S.}\ \bibnamefont {Li}},\ }\href {\doibase 10.1103/PhysRevD.108.063508} {\bibfield  {journal} {\bibinfo  {journal} {Phys. Rev. D}\ }\textbf {\bibinfo {volume} {108}},\ \bibinfo {pages} {063508} (\bibinfo {year} {2023})},\ \Eprint {http://arxiv.org/abs/2305.02218} {arXiv:2305.02218 [hep-ph]} \BibitemShut {NoStop}%
\bibitem [{\citenamefont {Cai}\ \emph {et~al.}(2023)\citenamefont {Cai}, \citenamefont {Wang},\ and\ \citenamefont {Yuwen}}]{Cai:2023guc}%
  \BibitemOpen
  \bibfield  {author} {\bibinfo {author} {\bibfnamefont {R.-G.}\ \bibnamefont {Cai}}, \bibinfo {author} {\bibfnamefont {S.-J.}\ \bibnamefont {Wang}}, \ and\ \bibinfo {author} {\bibfnamefont {Z.-Y.}\ \bibnamefont {Yuwen}},\ }\href {\doibase 10.1103/PhysRevD.108.L021502} {\bibfield  {journal} {\bibinfo  {journal} {Phys. Rev. D}\ }\textbf {\bibinfo {volume} {108}},\ \bibinfo {pages} {L021502} (\bibinfo {year} {2023})},\ \Eprint {http://arxiv.org/abs/2305.00074} {arXiv:2305.00074 [gr-qc]} \BibitemShut {NoStop}%
\bibitem [{\citenamefont {Liu}\ \emph {et~al.}(2023)\citenamefont {Liu}, \citenamefont {Bian}, \citenamefont {Cai}, \citenamefont {Guo},\ and\ \citenamefont {Wang}}]{Liu:2022lvz}%
  \BibitemOpen
  \bibfield  {author} {\bibinfo {author} {\bibfnamefont {J.}~\bibnamefont {Liu}}, \bibinfo {author} {\bibfnamefont {L.}~\bibnamefont {Bian}}, \bibinfo {author} {\bibfnamefont {R.-G.}\ \bibnamefont {Cai}}, \bibinfo {author} {\bibfnamefont {Z.-K.}\ \bibnamefont {Guo}}, \ and\ \bibinfo {author} {\bibfnamefont {S.-J.}\ \bibnamefont {Wang}},\ }\href {\doibase 10.1103/PhysRevLett.130.051001} {\bibfield  {journal} {\bibinfo  {journal} {Phys. Rev. Lett.}\ }\textbf {\bibinfo {volume} {130}},\ \bibinfo {pages} {051001} (\bibinfo {year} {2023})},\ \Eprint {http://arxiv.org/abs/2208.14086} {arXiv:2208.14086 [astro-ph.CO]} \BibitemShut {NoStop}%
\bibitem [{\citenamefont {Ananda}\ \emph {et~al.}(2007)\citenamefont {Ananda}, \citenamefont {Clarkson},\ and\ \citenamefont {Wands}}]{Ananda:2006af}%
  \BibitemOpen
  \bibfield  {author} {\bibinfo {author} {\bibfnamefont {K.~N.}\ \bibnamefont {Ananda}}, \bibinfo {author} {\bibfnamefont {C.}~\bibnamefont {Clarkson}}, \ and\ \bibinfo {author} {\bibfnamefont {D.}~\bibnamefont {Wands}},\ }\href {\doibase 10.1103/PhysRevD.75.123518} {\bibfield  {journal} {\bibinfo  {journal} {Phys. Rev. D}\ }\textbf {\bibinfo {volume} {75}},\ \bibinfo {pages} {123518} (\bibinfo {year} {2007})},\ \Eprint {http://arxiv.org/abs/gr-qc/0612013} {arXiv:gr-qc/0612013} \BibitemShut {NoStop}%
\bibitem [{\citenamefont {Baumann}\ \emph {et~al.}(2007)\citenamefont {Baumann}, \citenamefont {Steinhardt}, \citenamefont {Takahashi},\ and\ \citenamefont {Ichiki}}]{Baumann:2007zm}%
  \BibitemOpen
  \bibfield  {author} {\bibinfo {author} {\bibfnamefont {D.}~\bibnamefont {Baumann}}, \bibinfo {author} {\bibfnamefont {P.~J.}\ \bibnamefont {Steinhardt}}, \bibinfo {author} {\bibfnamefont {K.}~\bibnamefont {Takahashi}}, \ and\ \bibinfo {author} {\bibfnamefont {K.}~\bibnamefont {Ichiki}},\ }\href {\doibase 10.1103/PhysRevD.76.084019} {\bibfield  {journal} {\bibinfo  {journal} {Phys. Rev. D}\ }\textbf {\bibinfo {volume} {76}},\ \bibinfo {pages} {084019} (\bibinfo {year} {2007})},\ \Eprint {http://arxiv.org/abs/hep-th/0703290} {arXiv:hep-th/0703290} \BibitemShut {NoStop}%
\bibitem [{\citenamefont {Espinosa}\ \emph {et~al.}(2018)\citenamefont {Espinosa}, \citenamefont {Racco},\ and\ \citenamefont {Riotto}}]{Espinosa:2018eve}%
  \BibitemOpen
  \bibfield  {author} {\bibinfo {author} {\bibfnamefont {J.~R.}\ \bibnamefont {Espinosa}}, \bibinfo {author} {\bibfnamefont {D.}~\bibnamefont {Racco}}, \ and\ \bibinfo {author} {\bibfnamefont {A.}~\bibnamefont {Riotto}},\ }\href {\doibase 10.1088/1475-7516/2018/09/012} {\bibfield  {journal} {\bibinfo  {journal} {JCAP}\ }\textbf {\bibinfo {volume} {09}},\ \bibinfo {pages} {012} (\bibinfo {year} {2018})},\ \Eprint {http://arxiv.org/abs/1804.07732} {arXiv:1804.07732 [hep-ph]} \BibitemShut {NoStop}%
\bibitem [{\citenamefont {Kohri}\ and\ \citenamefont {Terada}(2018)}]{Kohri:2018awv}%
  \BibitemOpen
  \bibfield  {author} {\bibinfo {author} {\bibfnamefont {K.}~\bibnamefont {Kohri}}\ and\ \bibinfo {author} {\bibfnamefont {T.}~\bibnamefont {Terada}},\ }\href {\doibase 10.1103/PhysRevD.97.123532} {\bibfield  {journal} {\bibinfo  {journal} {Phys. Rev. D}\ }\textbf {\bibinfo {volume} {97}},\ \bibinfo {pages} {123532} (\bibinfo {year} {2018})},\ \Eprint {http://arxiv.org/abs/1804.08577} {arXiv:1804.08577 [gr-qc]} \BibitemShut {NoStop}%
\bibitem [{\citenamefont {Sasaki}\ \emph {et~al.}(2018)\citenamefont {Sasaki}, \citenamefont {Suyama}, \citenamefont {Tanaka},\ and\ \citenamefont {Yokoyama}}]{Sasaki:2018dmp}%
  \BibitemOpen
  \bibfield  {author} {\bibinfo {author} {\bibfnamefont {M.}~\bibnamefont {Sasaki}}, \bibinfo {author} {\bibfnamefont {T.}~\bibnamefont {Suyama}}, \bibinfo {author} {\bibfnamefont {T.}~\bibnamefont {Tanaka}}, \ and\ \bibinfo {author} {\bibfnamefont {S.}~\bibnamefont {Yokoyama}},\ }\href {\doibase 10.1088/1361-6382/aaa7b4} {\bibfield  {journal} {\bibinfo  {journal} {Class. Quant. Grav.}\ }\textbf {\bibinfo {volume} {35}},\ \bibinfo {pages} {063001} (\bibinfo {year} {2018})},\ \Eprint {http://arxiv.org/abs/1801.05235} {arXiv:1801.05235 [astro-ph.CO]} \BibitemShut {NoStop}%
\bibitem [{\citenamefont {Dom{\`e}nech}(2021)}]{Domenech:2021ztg}%
  \BibitemOpen
  \bibfield  {author} {\bibinfo {author} {\bibfnamefont {G.}~\bibnamefont {Dom{\`e}nech}},\ }\href {\doibase 10.3390/universe7110398} {\bibfield  {journal} {\bibinfo  {journal} {Universe}\ }\textbf {\bibinfo {volume} {7}},\ \bibinfo {pages} {398} (\bibinfo {year} {2021})},\ \Eprint {http://arxiv.org/abs/2109.01398} {arXiv:2109.01398 [gr-qc]} \BibitemShut {NoStop}%
\bibitem [{\citenamefont {Yuan}\ and\ \citenamefont {Huang}(2021)}]{Yuan:2021qgz}%
  \BibitemOpen
  \bibfield  {author} {\bibinfo {author} {\bibfnamefont {C.}~\bibnamefont {Yuan}}\ and\ \bibinfo {author} {\bibfnamefont {Q.-G.}\ \bibnamefont {Huang}},\ }\href {\doibase 10.1016/j.isci.2021.102860} {\bibfield  {journal} {\bibinfo  {journal} {iScience}\ }\textbf {\bibinfo {volume} {24}},\ \bibinfo {pages} {102860} (\bibinfo {year} {2021})},\ \Eprint {http://arxiv.org/abs/2103.04739} {arXiv:2103.04739 [astro-ph.GA]} \BibitemShut {NoStop}%
\bibitem [{\citenamefont {Dom{\`e}nech}\ \emph {et~al.}(2024)\citenamefont {Dom{\`e}nech}, \citenamefont {Pi}, \citenamefont {Wang},\ and\ \citenamefont {Wang}}]{Domenech:2024rks}%
  \BibitemOpen
  \bibfield  {author} {\bibinfo {author} {\bibfnamefont {G.}~\bibnamefont {Dom{\`e}nech}}, \bibinfo {author} {\bibfnamefont {S.}~\bibnamefont {Pi}}, \bibinfo {author} {\bibfnamefont {A.}~\bibnamefont {Wang}}, \ and\ \bibinfo {author} {\bibfnamefont {J.}~\bibnamefont {Wang}},\ }\href {\doibase 10.1088/1475-7516/2024/08/054} {\bibfield  {journal} {\bibinfo  {journal} {JCAP}\ }\textbf {\bibinfo {volume} {08}},\ \bibinfo {pages} {054} (\bibinfo {year} {2024})},\ \Eprint {http://arxiv.org/abs/2402.18965} {arXiv:2402.18965 [astro-ph.CO]} \BibitemShut {NoStop}%
\bibitem [{\citenamefont {Wang}\ \emph {et~al.}(2024{\natexlab{a}})\citenamefont {Wang}, \citenamefont {Zhao},\ and\ \citenamefont {Zhu}}]{Wang:2023sij}%
  \BibitemOpen
  \bibfield  {author} {\bibinfo {author} {\bibfnamefont {S.}~\bibnamefont {Wang}}, \bibinfo {author} {\bibfnamefont {Z.-C.}\ \bibnamefont {Zhao}}, \ and\ \bibinfo {author} {\bibfnamefont {Q.-H.}\ \bibnamefont {Zhu}},\ }\href {\doibase 10.1103/PhysRevResearch.6.013207} {\bibfield  {journal} {\bibinfo  {journal} {Phys. Rev. Res.}\ }\textbf {\bibinfo {volume} {6}},\ \bibinfo {pages} {013207} (\bibinfo {year} {2024}{\natexlab{a}})},\ \Eprint {http://arxiv.org/abs/2307.03095} {arXiv:2307.03095 [astro-ph.CO]} \BibitemShut {NoStop}%
\bibitem [{\citenamefont {Chen}\ \emph {et~al.}(2024)\citenamefont {Chen}, \citenamefont {Li}, \citenamefont {Liu},\ and\ \citenamefont {Yi}}]{Chen:2024fir}%
  \BibitemOpen
  \bibfield  {author} {\bibinfo {author} {\bibfnamefont {Z.-C.}\ \bibnamefont {Chen}}, \bibinfo {author} {\bibfnamefont {J.}~\bibnamefont {Li}}, \bibinfo {author} {\bibfnamefont {L.}~\bibnamefont {Liu}}, \ and\ \bibinfo {author} {\bibfnamefont {Z.}~\bibnamefont {Yi}},\ }\href {\doibase 10.1103/PhysRevD.109.L101302} {\bibfield  {journal} {\bibinfo  {journal} {Phys. Rev. D}\ }\textbf {\bibinfo {volume} {109}},\ \bibinfo {pages} {L101302} (\bibinfo {year} {2024})},\ \Eprint {http://arxiv.org/abs/2401.09818} {arXiv:2401.09818 [gr-qc]} \BibitemShut {NoStop}%
\bibitem [{\citenamefont {Chen}\ \emph {et~al.}(2020)\citenamefont {Chen}, \citenamefont {Yuan},\ and\ \citenamefont {Huang}}]{Chen:2019xse}%
  \BibitemOpen
  \bibfield  {author} {\bibinfo {author} {\bibfnamefont {Z.-C.}\ \bibnamefont {Chen}}, \bibinfo {author} {\bibfnamefont {C.}~\bibnamefont {Yuan}}, \ and\ \bibinfo {author} {\bibfnamefont {Q.-G.}\ \bibnamefont {Huang}},\ }\href {\doibase 10.1103/PhysRevLett.124.251101} {\bibfield  {journal} {\bibinfo  {journal} {Phys. Rev. Lett.}\ }\textbf {\bibinfo {volume} {124}},\ \bibinfo {pages} {25} (\bibinfo {year} {2020})},\ \Eprint {http://arxiv.org/abs/1910.12239} {arXiv:1910.12239 [astro-ph.CO]} \BibitemShut {NoStop}%
\bibitem [{\citenamefont {Yuan}\ \emph {et~al.}(2019)\citenamefont {Yuan}, \citenamefont {Chen},\ and\ \citenamefont {Huang}}]{Yuan:2019udt}%
  \BibitemOpen
  \bibfield  {author} {\bibinfo {author} {\bibfnamefont {C.}~\bibnamefont {Yuan}}, \bibinfo {author} {\bibfnamefont {Z.-C.}\ \bibnamefont {Chen}}, \ and\ \bibinfo {author} {\bibfnamefont {Q.-G.}\ \bibnamefont {Huang}},\ }\href {\doibase 10.1103/PhysRevD.100.081301} {\bibfield  {journal} {\bibinfo  {journal} {Phys. Rev. D}\ }\textbf {\bibinfo {volume} {100}},\ \bibinfo {pages} {8} (\bibinfo {year} {2019})},\ \Eprint {http://arxiv.org/abs/1906.11549} {arXiv:1906.11549 [astro-ph.CO]} \BibitemShut {NoStop}%
\bibitem [{\citenamefont {Aghanim}\ \emph {et~al.}(2020)\citenamefont {Aghanim} \emph {et~al.}}]{Planck:2018vyg}%
  \BibitemOpen
  \bibfield  {author} {\bibinfo {author} {\bibfnamefont {N.}~\bibnamefont {Aghanim}} \emph {et~al.} (\bibinfo {collaboration} {Planck}),\ }\href {\doibase 10.1051/0004-6361/201833910} {\bibfield  {journal} {\bibinfo  {journal} {Astron. Astrophys.}\ }\textbf {\bibinfo {volume} {641}},\ \bibinfo {pages} {A6} (\bibinfo {year} {2020})},\ \bibinfo {note} {[Erratum: Astron.Astrophys. 652, C4 (2021)]},\ \Eprint {http://arxiv.org/abs/1807.06209} {arXiv:1807.06209 [astro-ph.CO]} \BibitemShut {NoStop}%
\bibitem [{\citenamefont {Nakama}\ and\ \citenamefont {Suyama}(2015)}]{Nakama:2015nea}%
  \BibitemOpen
  \bibfield  {author} {\bibinfo {author} {\bibfnamefont {T.}~\bibnamefont {Nakama}}\ and\ \bibinfo {author} {\bibfnamefont {T.}~\bibnamefont {Suyama}},\ }\href {\doibase 10.1103/PhysRevD.92.121304} {\bibfield  {journal} {\bibinfo  {journal} {Phys. Rev. D}\ }\textbf {\bibinfo {volume} {92}},\ \bibinfo {pages} {121304} (\bibinfo {year} {2015})},\ \Eprint {http://arxiv.org/abs/1506.05228} {arXiv:1506.05228 [gr-qc]} \BibitemShut {NoStop}%
\bibitem [{\citenamefont {Afzal}\ \emph {et~al.}(2023)\citenamefont {Afzal} \emph {et~al.}}]{NANOGrav:2023hvm}%
  \BibitemOpen
  \bibfield  {author} {\bibinfo {author} {\bibfnamefont {A.}~\bibnamefont {Afzal}} \emph {et~al.} (\bibinfo {collaboration} {NANOGrav}),\ }\href {\doibase 10.3847/2041-8213/acdc91} {\bibfield  {journal} {\bibinfo  {journal} {Astrophys. J. Lett.}\ }\textbf {\bibinfo {volume} {951}} (\bibinfo {year} {2023}),\ 10.3847/2041-8213/acdc91},\ \Eprint {http://arxiv.org/abs/2306.16219} {arXiv:2306.16219 [astro-ph.HE]} \BibitemShut {NoStop}%
\bibitem [{\citenamefont {Zhu}\ \emph {et~al.}(2024)\citenamefont {Zhu}, \citenamefont {Zhao}, \citenamefont {Wang},\ and\ \citenamefont {Zhang}}]{Zhu:2023gmx}%
  \BibitemOpen
  \bibfield  {author} {\bibinfo {author} {\bibfnamefont {Q.-H.}\ \bibnamefont {Zhu}}, \bibinfo {author} {\bibfnamefont {Z.-C.}\ \bibnamefont {Zhao}}, \bibinfo {author} {\bibfnamefont {S.}~\bibnamefont {Wang}}, \ and\ \bibinfo {author} {\bibfnamefont {X.}~\bibnamefont {Zhang}},\ }\href {\doibase 10.1088/1674-1137/ad79d5} {\bibfield  {journal} {\bibinfo  {journal} {Chin. Phys. C}\ }\textbf {\bibinfo {volume} {48}},\ \bibinfo {pages} {125105} (\bibinfo {year} {2024})},\ \Eprint {http://arxiv.org/abs/2307.13574} {arXiv:2307.13574 [astro-ph.CO]} \BibitemShut {NoStop}%
\bibitem [{\citenamefont {Harigaya}\ \emph {et~al.}(2023)\citenamefont {Harigaya}, \citenamefont {Inomata},\ and\ \citenamefont {Terada}}]{Harigaya:2023pmw}%
  \BibitemOpen
  \bibfield  {author} {\bibinfo {author} {\bibfnamefont {K.}~\bibnamefont {Harigaya}}, \bibinfo {author} {\bibfnamefont {K.}~\bibnamefont {Inomata}}, \ and\ \bibinfo {author} {\bibfnamefont {T.}~\bibnamefont {Terada}},\ }\href {\doibase 10.1103/PhysRevD.108.123538} {\bibfield  {journal} {\bibinfo  {journal} {Phys. Rev. D}\ }\textbf {\bibinfo {volume} {108}},\ \bibinfo {pages} {123538} (\bibinfo {year} {2023})},\ \Eprint {http://arxiv.org/abs/2309.00228} {arXiv:2309.00228 [astro-ph.CO]} \BibitemShut {NoStop}%
\bibitem [{\citenamefont {De~Luca}\ \emph {et~al.}(2023)\citenamefont {De~Luca}, \citenamefont {Kehagias},\ and\ \citenamefont {Riotto}}]{DeLuca:2023tun}%
  \BibitemOpen
  \bibfield  {author} {\bibinfo {author} {\bibfnamefont {V.}~\bibnamefont {De~Luca}}, \bibinfo {author} {\bibfnamefont {A.}~\bibnamefont {Kehagias}}, \ and\ \bibinfo {author} {\bibfnamefont {A.}~\bibnamefont {Riotto}},\ }\href {\doibase 10.1103/PhysRevD.108.063531} {\bibfield  {journal} {\bibinfo  {journal} {Phys. Rev. D}\ }\textbf {\bibinfo {volume} {108}},\ \bibinfo {pages} {063531} (\bibinfo {year} {2023})},\ \Eprint {http://arxiv.org/abs/2307.13633} {arXiv:2307.13633 [astro-ph.CO]} \BibitemShut {NoStop}%
\bibitem [{\citenamefont {Choudhury}\ \emph {et~al.}(2024)\citenamefont {Choudhury}, \citenamefont {Dey}, \citenamefont {Karde}, \citenamefont {Panda},\ and\ \citenamefont {Sami}}]{Choudhury:2023fwk}%
  \BibitemOpen
  \bibfield  {author} {\bibinfo {author} {\bibfnamefont {S.}~\bibnamefont {Choudhury}}, \bibinfo {author} {\bibfnamefont {K.}~\bibnamefont {Dey}}, \bibinfo {author} {\bibfnamefont {A.}~\bibnamefont {Karde}}, \bibinfo {author} {\bibfnamefont {S.}~\bibnamefont {Panda}}, \ and\ \bibinfo {author} {\bibfnamefont {M.}~\bibnamefont {Sami}},\ }\href {\doibase 10.1016/j.physletb.2024.138925} {\bibfield  {journal} {\bibinfo  {journal} {Phys. Lett. B}\ }\textbf {\bibinfo {volume} {856}},\ \bibinfo {pages} {138925} (\bibinfo {year} {2024})},\ \Eprint {http://arxiv.org/abs/2310.11034} {arXiv:2310.11034 [astro-ph.CO]} \BibitemShut {NoStop}%
\bibitem [{\citenamefont {Wang}\ \emph {et~al.}(2025)\citenamefont {Wang}, \citenamefont {Ma},\ and\ \citenamefont {Cai}}]{Wang:2024nmd}%
  \BibitemOpen
  \bibfield  {author} {\bibinfo {author} {\bibfnamefont {X.}~\bibnamefont {Wang}}, \bibinfo {author} {\bibfnamefont {X.-H.}\ \bibnamefont {Ma}}, \ and\ \bibinfo {author} {\bibfnamefont {Y.-F.}\ \bibnamefont {Cai}},\ }\href {\doibase 10.1142/S0218271825500270} {\bibfield  {journal} {\bibinfo  {journal} {Int. J. Mod. Phys. D}\ }\textbf {\bibinfo {volume} {34}},\ \bibinfo {pages} {2550027} (\bibinfo {year} {2025})},\ \Eprint {http://arxiv.org/abs/2412.19631} {arXiv:2412.19631 [astro-ph.CO]} \BibitemShut {NoStop}%
\bibitem [{\citenamefont {Choudhury}\ \emph {et~al.}(2025)\citenamefont {Choudhury}, \citenamefont {Dey}, \citenamefont {Ganguly}, \citenamefont {Karde}, \citenamefont {Singh},\ and\ \citenamefont {Tiwari}}]{Choudhury:2024kjj}%
  \BibitemOpen
  \bibfield  {author} {\bibinfo {author} {\bibfnamefont {S.}~\bibnamefont {Choudhury}}, \bibinfo {author} {\bibfnamefont {K.}~\bibnamefont {Dey}}, \bibinfo {author} {\bibfnamefont {S.}~\bibnamefont {Ganguly}}, \bibinfo {author} {\bibfnamefont {A.}~\bibnamefont {Karde}}, \bibinfo {author} {\bibfnamefont {S.~K.}\ \bibnamefont {Singh}}, \ and\ \bibinfo {author} {\bibfnamefont {P.}~\bibnamefont {Tiwari}},\ }\href {\doibase 10.1140/epjc/s10052-025-14176-z} {\bibfield  {journal} {\bibinfo  {journal} {Eur. Phys. J. C}\ }\textbf {\bibinfo {volume} {85}},\ \bibinfo {pages} {472} (\bibinfo {year} {2025})},\ \Eprint {http://arxiv.org/abs/2409.18983} {arXiv:2409.18983 [astro-ph.CO]} \BibitemShut {NoStop}%
\bibitem [{\citenamefont {Yu}\ and\ \citenamefont {Wang}(2025)}]{Yu:2024xmz}%
  \BibitemOpen
  \bibfield  {author} {\bibinfo {author} {\bibfnamefont {Y.-H.}\ \bibnamefont {Yu}}\ and\ \bibinfo {author} {\bibfnamefont {S.}~\bibnamefont {Wang}},\ }\href {\doibase 10.1007/s11433-024-2499-9} {\bibfield  {journal} {\bibinfo  {journal} {Sci. China Phys. Mech. Astron.}\ }\textbf {\bibinfo {volume} {68}},\ \bibinfo {pages} {210412} (\bibinfo {year} {2025})},\ \Eprint {http://arxiv.org/abs/2405.02960} {arXiv:2405.02960 [astro-ph.CO]} \BibitemShut {NoStop}%
\bibitem [{\citenamefont {Suyama}\ and\ \citenamefont {Yokoyama}(2011)}]{Suyama:2011pu}%
  \BibitemOpen
  \bibfield  {author} {\bibinfo {author} {\bibfnamefont {T.}~\bibnamefont {Suyama}}\ and\ \bibinfo {author} {\bibfnamefont {J.}~\bibnamefont {Yokoyama}},\ }\href {\doibase 10.1103/PhysRevD.84.083511} {\bibfield  {journal} {\bibinfo  {journal} {Phys. Rev. D}\ }\textbf {\bibinfo {volume} {84}},\ \bibinfo {pages} {083511} (\bibinfo {year} {2011})},\ \Eprint {http://arxiv.org/abs/1106.5983} {arXiv:1106.5983 [astro-ph.CO]} \BibitemShut {NoStop}%
\bibitem [{\citenamefont {Cai}\ \emph {et~al.}(2018)\citenamefont {Cai}, \citenamefont {Tong}, \citenamefont {Wang},\ and\ \citenamefont {Yan}}]{Cai:2018tuh}%
  \BibitemOpen
  \bibfield  {author} {\bibinfo {author} {\bibfnamefont {Y.-F.}\ \bibnamefont {Cai}}, \bibinfo {author} {\bibfnamefont {X.}~\bibnamefont {Tong}}, \bibinfo {author} {\bibfnamefont {D.-G.}\ \bibnamefont {Wang}}, \ and\ \bibinfo {author} {\bibfnamefont {S.-F.}\ \bibnamefont {Yan}},\ }\href {\doibase 10.1103/PhysRevLett.121.081306} {\bibfield  {journal} {\bibinfo  {journal} {Phys. Rev. Lett.}\ }\textbf {\bibinfo {volume} {121}},\ \bibinfo {pages} {081306} (\bibinfo {year} {2018})},\ \Eprint {http://arxiv.org/abs/1805.03639} {arXiv:1805.03639 [astro-ph.CO]} \BibitemShut {NoStop}%
\bibitem [{\citenamefont {Cai}\ \emph {et~al.}(2019{\natexlab{a}})\citenamefont {Cai}, \citenamefont {Chen}, \citenamefont {Tong}, \citenamefont {Wang},\ and\ \citenamefont {Yan}}]{Cai:2019jah}%
  \BibitemOpen
  \bibfield  {author} {\bibinfo {author} {\bibfnamefont {Y.-F.}\ \bibnamefont {Cai}}, \bibinfo {author} {\bibfnamefont {C.}~\bibnamefont {Chen}}, \bibinfo {author} {\bibfnamefont {X.}~\bibnamefont {Tong}}, \bibinfo {author} {\bibfnamefont {D.-G.}\ \bibnamefont {Wang}}, \ and\ \bibinfo {author} {\bibfnamefont {S.-F.}\ \bibnamefont {Yan}},\ }\href {\doibase 10.1103/PhysRevD.100.043518} {\bibfield  {journal} {\bibinfo  {journal} {Phys. Rev. D}\ }\textbf {\bibinfo {volume} {100}},\ \bibinfo {pages} {043518} (\bibinfo {year} {2019}{\natexlab{a}})},\ \Eprint {http://arxiv.org/abs/1902.08187} {arXiv:1902.08187 [astro-ph.CO]} \BibitemShut {NoStop}%
\bibitem [{\citenamefont {Cai}\ \emph {et~al.}(2020)\citenamefont {Cai}, \citenamefont {Guo}, \citenamefont {Liu}, \citenamefont {Liu},\ and\ \citenamefont {Yang}}]{Cai:2019bmk}%
  \BibitemOpen
  \bibfield  {author} {\bibinfo {author} {\bibfnamefont {R.-G.}\ \bibnamefont {Cai}}, \bibinfo {author} {\bibfnamefont {Z.-K.}\ \bibnamefont {Guo}}, \bibinfo {author} {\bibfnamefont {J.}~\bibnamefont {Liu}}, \bibinfo {author} {\bibfnamefont {L.}~\bibnamefont {Liu}}, \ and\ \bibinfo {author} {\bibfnamefont {X.-Y.}\ \bibnamefont {Yang}},\ }\href {\doibase 10.1088/1475-7516/2020/06/013} {\bibfield  {journal} {\bibinfo  {journal} {JCAP}\ }\textbf {\bibinfo {volume} {06}},\ \bibinfo {pages} {013} (\bibinfo {year} {2020})},\ \Eprint {http://arxiv.org/abs/1912.10437} {arXiv:1912.10437 [astro-ph.CO]} \BibitemShut {NoStop}%
\bibitem [{\citenamefont {{\"O}zsoy}\ and\ \citenamefont {Tasinato}(2020)}]{Ozsoy:2019lyy}%
  \BibitemOpen
  \bibfield  {author} {\bibinfo {author} {\bibfnamefont {O.}~\bibnamefont {{\"O}zsoy}}\ and\ \bibinfo {author} {\bibfnamefont {G.}~\bibnamefont {Tasinato}},\ }\href {\doibase 10.1088/1475-7516/2020/04/048} {\bibfield  {journal} {\bibinfo  {journal} {JCAP}\ }\textbf {\bibinfo {volume} {04}},\ \bibinfo {pages} {048} (\bibinfo {year} {2020})},\ \Eprint {http://arxiv.org/abs/1912.01061} {arXiv:1912.01061 [astro-ph.CO]} \BibitemShut {NoStop}%
\bibitem [{\citenamefont {Cai}\ \emph {et~al.}(2021)\citenamefont {Cai}, \citenamefont {Jiang}, \citenamefont {Sasaki}, \citenamefont {Vardanyan},\ and\ \citenamefont {Zhou}}]{Cai:2021yvq}%
  \BibitemOpen
  \bibfield  {author} {\bibinfo {author} {\bibfnamefont {Y.-F.}\ \bibnamefont {Cai}}, \bibinfo {author} {\bibfnamefont {J.}~\bibnamefont {Jiang}}, \bibinfo {author} {\bibfnamefont {M.}~\bibnamefont {Sasaki}}, \bibinfo {author} {\bibfnamefont {V.}~\bibnamefont {Vardanyan}}, \ and\ \bibinfo {author} {\bibfnamefont {Z.}~\bibnamefont {Zhou}},\ }\href {\doibase 10.1103/PhysRevLett.127.251301} {\bibfield  {journal} {\bibinfo  {journal} {Phys. Rev. Lett.}\ }\textbf {\bibinfo {volume} {127}},\ \bibinfo {pages} {251301} (\bibinfo {year} {2021})},\ \Eprint {http://arxiv.org/abs/2105.12554} {arXiv:2105.12554 [astro-ph.CO]} \BibitemShut {NoStop}%
\bibitem [{\citenamefont {Maldacena}(2003)}]{Maldacena:2002vr}%
  \BibitemOpen
  \bibfield  {author} {\bibinfo {author} {\bibfnamefont {J.~M.}\ \bibnamefont {Maldacena}},\ }\href {\doibase 10.1088/1126-6708/2003/05/013} {\bibfield  {journal} {\bibinfo  {journal} {JHEP}\ }\textbf {\bibinfo {volume} {05}},\ \bibinfo {pages} {013} (\bibinfo {year} {2003})},\ \Eprint {http://arxiv.org/abs/astro-ph/0210603} {arXiv:astro-ph/0210603} \BibitemShut {NoStop}%
\bibitem [{\citenamefont {Byrnes}\ and\ \citenamefont {Choi}(2010)}]{Byrnes:2010em}%
  \BibitemOpen
  \bibfield  {author} {\bibinfo {author} {\bibfnamefont {C.~T.}\ \bibnamefont {Byrnes}}\ and\ \bibinfo {author} {\bibfnamefont {K.-Y.}\ \bibnamefont {Choi}},\ }\href {\doibase 10.1155/2010/724525} {\bibfield  {journal} {\bibinfo  {journal} {Adv. Astron.}\ }\textbf {\bibinfo {volume} {2010}},\ \bibinfo {pages} {724525} (\bibinfo {year} {2010})},\ \Eprint {http://arxiv.org/abs/1002.3110} {arXiv:1002.3110 [astro-ph.CO]} \BibitemShut {NoStop}%
\bibitem [{\citenamefont {Dimastrogiovanni}\ \emph {et~al.}(2010)\citenamefont {Dimastrogiovanni}, \citenamefont {Bartolo}, \citenamefont {Matarrese},\ and\ \citenamefont {Riotto}}]{Dimastrogiovanni:2010sm}%
  \BibitemOpen
  \bibfield  {author} {\bibinfo {author} {\bibfnamefont {E.}~\bibnamefont {Dimastrogiovanni}}, \bibinfo {author} {\bibfnamefont {N.}~\bibnamefont {Bartolo}}, \bibinfo {author} {\bibfnamefont {S.}~\bibnamefont {Matarrese}}, \ and\ \bibinfo {author} {\bibfnamefont {A.}~\bibnamefont {Riotto}},\ }\href {\doibase 10.1155/2010/752670} {\bibfield  {journal} {\bibinfo  {journal} {Adv. Astron.}\ }\textbf {\bibinfo {volume} {2010}},\ \bibinfo {pages} {752670} (\bibinfo {year} {2010})},\ \Eprint {http://arxiv.org/abs/1001.4049} {arXiv:1001.4049 [astro-ph.CO]} \BibitemShut {NoStop}%
\bibitem [{\citenamefont {Adshead}\ \emph {et~al.}(2012)\citenamefont {Adshead}, \citenamefont {Dvorkin}, \citenamefont {Hu},\ and\ \citenamefont {Lim}}]{Adshead:2011jq}%
  \BibitemOpen
  \bibfield  {author} {\bibinfo {author} {\bibfnamefont {P.}~\bibnamefont {Adshead}}, \bibinfo {author} {\bibfnamefont {C.}~\bibnamefont {Dvorkin}}, \bibinfo {author} {\bibfnamefont {W.}~\bibnamefont {Hu}}, \ and\ \bibinfo {author} {\bibfnamefont {E.~A.}\ \bibnamefont {Lim}},\ }\href {\doibase 10.1103/PhysRevD.85.023531} {\bibfield  {journal} {\bibinfo  {journal} {Phys. Rev. D}\ }\textbf {\bibinfo {volume} {85}},\ \bibinfo {pages} {023531} (\bibinfo {year} {2012})},\ \Eprint {http://arxiv.org/abs/1110.3050} {arXiv:1110.3050 [astro-ph.CO]} \BibitemShut {NoStop}%
\bibitem [{\citenamefont {Cai}\ \emph {et~al.}(2019{\natexlab{b}})\citenamefont {Cai}, \citenamefont {Pi},\ and\ \citenamefont {Sasaki}}]{Cai:2018dig}%
  \BibitemOpen
  \bibfield  {author} {\bibinfo {author} {\bibfnamefont {R.-g.}\ \bibnamefont {Cai}}, \bibinfo {author} {\bibfnamefont {S.}~\bibnamefont {Pi}}, \ and\ \bibinfo {author} {\bibfnamefont {M.}~\bibnamefont {Sasaki}},\ }\href {\doibase 10.1103/PhysRevLett.122.201101} {\bibfield  {journal} {\bibinfo  {journal} {Phys. Rev. Lett.}\ }\textbf {\bibinfo {volume} {122}},\ \bibinfo {pages} {201101} (\bibinfo {year} {2019}{\natexlab{b}})},\ \Eprint {http://arxiv.org/abs/1810.11000} {arXiv:1810.11000 [astro-ph.CO]} \BibitemShut {NoStop}%
\bibitem [{\citenamefont {Franciolini}\ \emph {et~al.}(2023)\citenamefont {Franciolini}, \citenamefont {Iovino}, \citenamefont {Vaskonen},\ and\ \citenamefont {Veermae}}]{Franciolini:2023pbf}%
  \BibitemOpen
  \bibfield  {author} {\bibinfo {author} {\bibfnamefont {G.}~\bibnamefont {Franciolini}}, \bibinfo {author} {\bibfnamefont {A.}~\bibnamefont {Iovino}, \bibfnamefont {Junior.}}, \bibinfo {author} {\bibfnamefont {V.}~\bibnamefont {Vaskonen}}, \ and\ \bibinfo {author} {\bibfnamefont {H.}~\bibnamefont {Veermae}},\ }\href {\doibase 10.1103/PhysRevLett.131.201401} {\bibfield  {journal} {\bibinfo  {journal} {Phys. Rev. Lett.}\ }\textbf {\bibinfo {volume} {131}},\ \bibinfo {pages} {201401} (\bibinfo {year} {2023})},\ \Eprint {http://arxiv.org/abs/2306.17149} {arXiv:2306.17149 [astro-ph.CO]} \BibitemShut {NoStop}%
\bibitem [{\citenamefont {Gorji}\ \emph {et~al.}(2023)\citenamefont {Gorji}, \citenamefont {Sasaki},\ and\ \citenamefont {Suyama}}]{Gorji:2023sil}%
  \BibitemOpen
  \bibfield  {author} {\bibinfo {author} {\bibfnamefont {M.~A.}\ \bibnamefont {Gorji}}, \bibinfo {author} {\bibfnamefont {M.}~\bibnamefont {Sasaki}}, \ and\ \bibinfo {author} {\bibfnamefont {T.}~\bibnamefont {Suyama}},\ }\href {\doibase 10.1016/j.physletb.2023.138214} {\bibfield  {journal} {\bibinfo  {journal} {Phys. Lett. B}\ }\textbf {\bibinfo {volume} {846}},\ \bibinfo {pages} {138214} (\bibinfo {year} {2023})},\ \Eprint {http://arxiv.org/abs/2307.13109} {arXiv:2307.13109 [astro-ph.CO]} \BibitemShut {NoStop}%
\bibitem [{\citenamefont {Zhao}\ \emph {et~al.}(2025)\citenamefont {Zhao}, \citenamefont {Wang}, \citenamefont {Li},\ and\ \citenamefont {Kohri}}]{Zhao:2024gan}%
  \BibitemOpen
  \bibfield  {author} {\bibinfo {author} {\bibfnamefont {Z.-C.}\ \bibnamefont {Zhao}}, \bibinfo {author} {\bibfnamefont {S.}~\bibnamefont {Wang}}, \bibinfo {author} {\bibfnamefont {J.-P.}\ \bibnamefont {Li}}, \ and\ \bibinfo {author} {\bibfnamefont {K.}~\bibnamefont {Kohri}},\ }\href {\doibase 10.1140/epjc/s10052-025-15115-8} {\bibfield  {journal} {\bibinfo  {journal} {Eur. Phys. J. C}\ }\textbf {\bibinfo {volume} {85}},\ \bibinfo {pages} {1406} (\bibinfo {year} {2025})},\ \Eprint {http://arxiv.org/abs/2412.02500} {arXiv:2412.02500 [astro-ph.CO]} \BibitemShut {NoStop}%
\bibitem [{\citenamefont {Li}\ \emph {et~al.}(2024{\natexlab{b}})\citenamefont {Li}, \citenamefont {Wang}, \citenamefont {Zhao},\ and\ \citenamefont {Kohri}}]{Li:2023xtl}%
  \BibitemOpen
  \bibfield  {author} {\bibinfo {author} {\bibfnamefont {J.-P.}\ \bibnamefont {Li}}, \bibinfo {author} {\bibfnamefont {S.}~\bibnamefont {Wang}}, \bibinfo {author} {\bibfnamefont {Z.-C.}\ \bibnamefont {Zhao}}, \ and\ \bibinfo {author} {\bibfnamefont {K.}~\bibnamefont {Kohri}},\ }\href {\doibase 10.1088/1475-7516/2024/06/039} {\bibfield  {journal} {\bibinfo  {journal} {JCAP}\ }\textbf {\bibinfo {volume} {06}},\ \bibinfo {pages} {039} (\bibinfo {year} {2024}{\natexlab{b}})},\ \Eprint {http://arxiv.org/abs/2309.07792} {arXiv:2309.07792 [astro-ph.CO]} \BibitemShut {NoStop}%
\bibitem [{\citenamefont {Wang}\ \emph {et~al.}(2024{\natexlab{b}})\citenamefont {Wang}, \citenamefont {Zhao}, \citenamefont {Li},\ and\ \citenamefont {Zhu}}]{Wang:2023ost}%
  \BibitemOpen
  \bibfield  {author} {\bibinfo {author} {\bibfnamefont {S.}~\bibnamefont {Wang}}, \bibinfo {author} {\bibfnamefont {Z.-C.}\ \bibnamefont {Zhao}}, \bibinfo {author} {\bibfnamefont {J.-P.}\ \bibnamefont {Li}}, \ and\ \bibinfo {author} {\bibfnamefont {Q.-H.}\ \bibnamefont {Zhu}},\ }\href {\doibase 10.1103/PhysRevResearch.6.L012060} {\bibfield  {journal} {\bibinfo  {journal} {Phys. Rev. Res.}\ }\textbf {\bibinfo {volume} {6}},\ \bibinfo {pages} {L012060} (\bibinfo {year} {2024}{\natexlab{b}})},\ \Eprint {http://arxiv.org/abs/2307.00572} {arXiv:2307.00572 [astro-ph.CO]} \BibitemShut {NoStop}%
\bibitem [{\citenamefont {Ackerman}\ \emph {et~al.}(2007)\citenamefont {Ackerman}, \citenamefont {Carroll},\ and\ \citenamefont {Wise}}]{Ackerman:2007nb}%
  \BibitemOpen
  \bibfield  {author} {\bibinfo {author} {\bibfnamefont {L.}~\bibnamefont {Ackerman}}, \bibinfo {author} {\bibfnamefont {S.~M.}\ \bibnamefont {Carroll}}, \ and\ \bibinfo {author} {\bibfnamefont {M.~B.}\ \bibnamefont {Wise}},\ }\href {\doibase 10.1103/PhysRevD.75.083502} {\bibfield  {journal} {\bibinfo  {journal} {Phys. Rev. D}\ }\textbf {\bibinfo {volume} {75}},\ \bibinfo {pages} {083502} (\bibinfo {year} {2007})},\ \bibinfo {note} {[Erratum: Phys.Rev.D 80, 069901 (2009)]},\ \Eprint {http://arxiv.org/abs/astro-ph/0701357} {arXiv:astro-ph/0701357} \BibitemShut {NoStop}%
\bibitem [{\citenamefont {Yokoyama}\ and\ \citenamefont {Soda}(2008)}]{Yokoyama:2008xw}%
  \BibitemOpen
  \bibfield  {author} {\bibinfo {author} {\bibfnamefont {S.}~\bibnamefont {Yokoyama}}\ and\ \bibinfo {author} {\bibfnamefont {J.}~\bibnamefont {Soda}},\ }\href {\doibase 10.1088/1475-7516/2008/08/005} {\bibfield  {journal} {\bibinfo  {journal} {JCAP}\ }\textbf {\bibinfo {volume} {08}},\ \bibinfo {pages} {005} (\bibinfo {year} {2008})},\ \Eprint {http://arxiv.org/abs/0805.4265} {arXiv:0805.4265 [astro-ph]} \BibitemShut {NoStop}%
\bibitem [{\citenamefont {Dulaney}\ and\ \citenamefont {Gresham}(2010)}]{Dulaney:2010sq}%
  \BibitemOpen
  \bibfield  {author} {\bibinfo {author} {\bibfnamefont {T.~R.}\ \bibnamefont {Dulaney}}\ and\ \bibinfo {author} {\bibfnamefont {M.~I.}\ \bibnamefont {Gresham}},\ }\href {\doibase 10.1103/PhysRevD.81.103532} {\bibfield  {journal} {\bibinfo  {journal} {Phys. Rev. D}\ }\textbf {\bibinfo {volume} {81}},\ \bibinfo {pages} {103532} (\bibinfo {year} {2010})},\ \Eprint {http://arxiv.org/abs/1001.2301} {arXiv:1001.2301 [astro-ph.CO]} \BibitemShut {NoStop}%
\bibitem [{\citenamefont {Maleknejad}\ \emph {et~al.}(2013)\citenamefont {Maleknejad}, \citenamefont {Sheikh-Jabbari},\ and\ \citenamefont {Soda}}]{Maleknejad:2012fw}%
  \BibitemOpen
  \bibfield  {author} {\bibinfo {author} {\bibfnamefont {A.}~\bibnamefont {Maleknejad}}, \bibinfo {author} {\bibfnamefont {M.~M.}\ \bibnamefont {Sheikh-Jabbari}}, \ and\ \bibinfo {author} {\bibfnamefont {J.}~\bibnamefont {Soda}},\ }\href {\doibase 10.1016/j.physrep.2013.03.003} {\bibfield  {journal} {\bibinfo  {journal} {Phys. Rept.}\ }\textbf {\bibinfo {volume} {528}},\ \bibinfo {pages} {161} (\bibinfo {year} {2013})},\ \Eprint {http://arxiv.org/abs/1212.2921} {arXiv:1212.2921 [hep-th]} \BibitemShut {NoStop}%
\bibitem [{\citenamefont {Soda}(2012)}]{Soda:2012zm}%
  \BibitemOpen
  \bibfield  {author} {\bibinfo {author} {\bibfnamefont {J.}~\bibnamefont {Soda}},\ }\href {\doibase 10.1088/0264-9381/29/8/083001} {\bibfield  {journal} {\bibinfo  {journal} {Class. Quant. Grav.}\ }\textbf {\bibinfo {volume} {29}},\ \bibinfo {pages} {083001} (\bibinfo {year} {2012})},\ \Eprint {http://arxiv.org/abs/1201.6434} {arXiv:1201.6434 [hep-th]} \BibitemShut {NoStop}%
\bibitem [{\citenamefont {Chen}\ \emph {et~al.}(2025)\citenamefont {Chen}, \citenamefont {An},\ and\ \citenamefont {Shu}}]{Chen:2025qyv}%
  \BibitemOpen
  \bibfield  {author} {\bibinfo {author} {\bibfnamefont {C.-B.}\ \bibnamefont {Chen}}, \bibinfo {author} {\bibfnamefont {B.-X.}\ \bibnamefont {An}}, \ and\ \bibinfo {author} {\bibfnamefont {F.-W.}\ \bibnamefont {Shu}},\ }\href {\doibase 10.1103/rv6r-y6m7} {\bibfield  {journal} {\bibinfo  {journal} {Phys. Rev. D}\ }\textbf {\bibinfo {volume} {112}},\ \bibinfo {pages} {10} (\bibinfo {year} {2025})},\ \Eprint {http://arxiv.org/abs/2507.16807} {arXiv:2507.16807 [astro-ph.CO]} \BibitemShut {NoStop}%
\bibitem [{\citenamefont {Chang}\ \emph {et~al.}(2013)\citenamefont {Chang}, \citenamefont {Li}, \citenamefont {Li},\ and\ \citenamefont {Wang}}]{Chang:2013xwa}%
  \BibitemOpen
  \bibfield  {author} {\bibinfo {author} {\bibfnamefont {Z.}~\bibnamefont {Chang}}, \bibinfo {author} {\bibfnamefont {M.-H.}\ \bibnamefont {Li}}, \bibinfo {author} {\bibfnamefont {X.}~\bibnamefont {Li}}, \ and\ \bibinfo {author} {\bibfnamefont {S.}~\bibnamefont {Wang}},\ }\href {\doibase 10.1140/epjc/s10052-013-2459-x} {\bibfield  {journal} {\bibinfo  {journal} {Eur. Phys. J. C}\ }\textbf {\bibinfo {volume} {73}},\ \bibinfo {pages} {2459} (\bibinfo {year} {2013})},\ \Eprint {http://arxiv.org/abs/1303.1593} {arXiv:1303.1593 [astro-ph.CO]} \BibitemShut {NoStop}%
\bibitem [{\citenamefont {Chang}\ \emph {et~al.}(2015)\citenamefont {Chang}, \citenamefont {Li},\ and\ \citenamefont {Wang}}]{Chang:2013lxa}%
  \BibitemOpen
  \bibfield  {author} {\bibinfo {author} {\bibfnamefont {Z.}~\bibnamefont {Chang}}, \bibinfo {author} {\bibfnamefont {X.}~\bibnamefont {Li}}, \ and\ \bibinfo {author} {\bibfnamefont {S.}~\bibnamefont {Wang}},\ }\href {\doibase 10.1088/1674-1137/39/5/055101} {\bibfield  {journal} {\bibinfo  {journal} {Chin. Phys. C}\ }\textbf {\bibinfo {volume} {39}},\ \bibinfo {pages} {055101} (\bibinfo {year} {2015})},\ \Eprint {http://arxiv.org/abs/1307.4542} {arXiv:1307.4542 [astro-ph.CO]} \BibitemShut {NoStop}%
\bibitem [{\citenamefont {Jain}\ and\ \citenamefont {Rath}(2015)}]{Jain:2014cpa}%
  \BibitemOpen
  \bibfield  {author} {\bibinfo {author} {\bibfnamefont {P.}~\bibnamefont {Jain}}\ and\ \bibinfo {author} {\bibfnamefont {P.~K.}\ \bibnamefont {Rath}},\ }\href {\doibase 10.1140/epjc/s10052-015-3333-9} {\bibfield  {journal} {\bibinfo  {journal} {Eur. Phys. J. C}\ }\textbf {\bibinfo {volume} {75}},\ \bibinfo {pages} {113} (\bibinfo {year} {2015})},\ \Eprint {http://arxiv.org/abs/1407.1714} {arXiv:1407.1714 [astro-ph.CO]} \BibitemShut {NoStop}%
\bibitem [{\citenamefont {Li}\ \emph {et~al.}(2015)\citenamefont {Li}, \citenamefont {Wang},\ and\ \citenamefont {Chang}}]{Li:2015sja}%
  \BibitemOpen
  \bibfield  {author} {\bibinfo {author} {\bibfnamefont {X.}~\bibnamefont {Li}}, \bibinfo {author} {\bibfnamefont {S.}~\bibnamefont {Wang}}, \ and\ \bibinfo {author} {\bibfnamefont {Z.}~\bibnamefont {Chang}},\ }\href {\doibase 10.1140/epjc/s10052-015-3468-8} {\bibfield  {journal} {\bibinfo  {journal} {Eur. Phys. J. C}\ }\textbf {\bibinfo {volume} {75}},\ \bibinfo {pages} {260} (\bibinfo {year} {2015})},\ \Eprint {http://arxiv.org/abs/1502.02256} {arXiv:1502.02256 [gr-qc]} \BibitemShut {NoStop}%
\bibitem [{\citenamefont {Rath}\ \emph {et~al.}(2015)\citenamefont {Rath}, \citenamefont {Aluri},\ and\ \citenamefont {Jain}}]{Rath:2014cka}%
  \BibitemOpen
  \bibfield  {author} {\bibinfo {author} {\bibfnamefont {P.~K.}\ \bibnamefont {Rath}}, \bibinfo {author} {\bibfnamefont {P.~K.}\ \bibnamefont {Aluri}}, \ and\ \bibinfo {author} {\bibfnamefont {P.}~\bibnamefont {Jain}},\ }\href {\doibase 10.1103/PhysRevD.91.023515} {\bibfield  {journal} {\bibinfo  {journal} {Phys. Rev. D}\ }\textbf {\bibinfo {volume} {91}},\ \bibinfo {pages} {023515} (\bibinfo {year} {2015})},\ \Eprint {http://arxiv.org/abs/1403.2567} {arXiv:1403.2567 [astro-ph.CO]} \BibitemShut {NoStop}%
\bibitem [{\citenamefont {Chang}\ \emph {et~al.}(2018)\citenamefont {Chang}, \citenamefont {Rath}, \citenamefont {Sang},\ and\ \citenamefont {Zhao}}]{Chang:2018msh}%
  \BibitemOpen
  \bibfield  {author} {\bibinfo {author} {\bibfnamefont {Z.}~\bibnamefont {Chang}}, \bibinfo {author} {\bibfnamefont {P.~K.}\ \bibnamefont {Rath}}, \bibinfo {author} {\bibfnamefont {Y.}~\bibnamefont {Sang}}, \ and\ \bibinfo {author} {\bibfnamefont {D.}~\bibnamefont {Zhao}},\ }\href {\doibase 10.1088/1674-4527/18/3/29} {\bibfield  {journal} {\bibinfo  {journal} {Res. Astron. Astrophys.}\ }\textbf {\bibinfo {volume} {18}},\ \bibinfo {pages} {029} (\bibinfo {year} {2018})},\ \Eprint {http://arxiv.org/abs/1801.02773} {arXiv:1801.02773 [astro-ph.CO]} \BibitemShut {NoStop}%
\bibitem [{\citenamefont {Bennett}\ \emph {et~al.}(2011)\citenamefont {Bennett} \emph {et~al.}}]{Bennett:2010jb}%
  \BibitemOpen
  \bibfield  {author} {\bibinfo {author} {\bibfnamefont {C.~L.}\ \bibnamefont {Bennett}} \emph {et~al.},\ }\href {\doibase 10.1088/0067-0049/192/2/17} {\bibfield  {journal} {\bibinfo  {journal} {Astrophys. J. Suppl.}\ }\textbf {\bibinfo {volume} {192}},\ \bibinfo {pages} {17} (\bibinfo {year} {2011})},\ \Eprint {http://arxiv.org/abs/1001.4758} {arXiv:1001.4758 [astro-ph.CO]} \BibitemShut {NoStop}%
\bibitem [{\citenamefont {Ade}\ \emph {et~al.}(2016)\citenamefont {Ade} \emph {et~al.}}]{Planck:2015igc}%
  \BibitemOpen
  \bibfield  {author} {\bibinfo {author} {\bibfnamefont {P.~A.~R.}\ \bibnamefont {Ade}} \emph {et~al.} (\bibinfo {collaboration} {Planck}),\ }\href {\doibase 10.1051/0004-6361/201526681} {\bibfield  {journal} {\bibinfo  {journal} {Astron. Astrophys.}\ }\textbf {\bibinfo {volume} {594}},\ \bibinfo {pages} {A16} (\bibinfo {year} {2016})},\ \Eprint {http://arxiv.org/abs/1506.07135} {arXiv:1506.07135 [astro-ph.CO]} \BibitemShut {NoStop}%
\bibitem [{\citenamefont {Akrami}\ \emph {et~al.}(2020)\citenamefont {Akrami} \emph {et~al.}}]{Planck:2019evm}%
  \BibitemOpen
  \bibfield  {author} {\bibinfo {author} {\bibfnamefont {Y.}~\bibnamefont {Akrami}} \emph {et~al.} (\bibinfo {collaboration} {Planck}),\ }\href {\doibase 10.1051/0004-6361/201935201} {\bibfield  {journal} {\bibinfo  {journal} {Astron. Astrophys.}\ }\textbf {\bibinfo {volume} {641}},\ \bibinfo {pages} {A7} (\bibinfo {year} {2020})},\ \Eprint {http://arxiv.org/abs/1906.02552} {arXiv:1906.02552 [astro-ph.CO]} \BibitemShut {NoStop}%
\bibitem [{\citenamefont {Eisenstein}\ \emph {et~al.}(2005)\citenamefont {Eisenstein} \emph {et~al.}}]{SDSS:2005xqv}%
  \BibitemOpen
  \bibfield  {author} {\bibinfo {author} {\bibfnamefont {D.~J.}\ \bibnamefont {Eisenstein}} \emph {et~al.} (\bibinfo {collaboration} {SDSS}),\ }\href {\doibase 10.1086/466512} {\bibfield  {journal} {\bibinfo  {journal} {Astrophys. J.}\ }\textbf {\bibinfo {volume} {633}},\ \bibinfo {pages} {560} (\bibinfo {year} {2005})},\ \Eprint {http://arxiv.org/abs/astro-ph/0501171} {arXiv:astro-ph/0501171} \BibitemShut {NoStop}%
\bibitem [{\citenamefont {Bautista}\ \emph {et~al.}(2020)\citenamefont {Bautista} \emph {et~al.}}]{eBOSS:2020lta}%
  \BibitemOpen
  \bibfield  {author} {\bibinfo {author} {\bibfnamefont {J.~E.}\ \bibnamefont {Bautista}} \emph {et~al.} (\bibinfo {collaboration} {eBOSS}),\ }\href {\doibase 10.1093/mnras/staa2800} {\bibfield  {journal} {\bibinfo  {journal} {Mon. Not. Roy. Astron. Soc.}\ }\textbf {\bibinfo {volume} {500}},\ \bibinfo {pages} {736} (\bibinfo {year} {2020})},\ \Eprint {http://arxiv.org/abs/2007.08993} {arXiv:2007.08993 [astro-ph.CO]} \BibitemShut {NoStop}%
\bibitem [{\citenamefont {Hou}\ \emph {et~al.}(2020)\citenamefont {Hou} \emph {et~al.}}]{eBOSS:2020gbb}%
  \BibitemOpen
  \bibfield  {author} {\bibinfo {author} {\bibfnamefont {J.}~\bibnamefont {Hou}} \emph {et~al.} (\bibinfo {collaboration} {eBOSS}),\ }\href {\doibase 10.1093/mnras/staa3234} {\bibfield  {journal} {\bibinfo  {journal} {Mon. Not. Roy. Astron. Soc.}\ }\textbf {\bibinfo {volume} {500}},\ \bibinfo {pages} {1201} (\bibinfo {year} {2020})},\ \Eprint {http://arxiv.org/abs/2007.08998} {arXiv:2007.08998 [astro-ph.CO]} \BibitemShut {NoStop}%
\bibitem [{\citenamefont {Andrade}\ \emph {et~al.}(2025)\citenamefont {Andrade} \emph {et~al.}}]{DESI:2025qqy}%
  \BibitemOpen
  \bibfield  {author} {\bibinfo {author} {\bibfnamefont {U.}~\bibnamefont {Andrade}} \emph {et~al.} (\bibinfo {collaboration} {DESI}),\ }\href {\doibase 10.1103/kdys-w8vl} {\bibfield  {journal} {\bibinfo  {journal} {Phys. Rev. D}\ }\textbf {\bibinfo {volume} {112}},\ \bibinfo {pages} {083512} (\bibinfo {year} {2025})},\ \Eprint {http://arxiv.org/abs/2503.14742} {arXiv:2503.14742 [astro-ph.CO]} \BibitemShut {NoStop}%
\bibitem [{\citenamefont {Abdul~Karim}\ \emph {et~al.}(2025)\citenamefont {Abdul~Karim} \emph {et~al.}}]{DESI:2025zgx}%
  \BibitemOpen
  \bibfield  {author} {\bibinfo {author} {\bibfnamefont {M.}~\bibnamefont {Abdul~Karim}} \emph {et~al.} (\bibinfo {collaboration} {DESI}),\ }\href {\doibase 10.1103/tr6y-kpc6} {\bibfield  {journal} {\bibinfo  {journal} {Phys. Rev. D}\ }\textbf {\bibinfo {volume} {112}},\ \bibinfo {pages} {083515} (\bibinfo {year} {2025})},\ \Eprint {http://arxiv.org/abs/2503.14738} {arXiv:2503.14738 [astro-ph.CO]} \BibitemShut {NoStop}%
\bibitem [{\citenamefont {Abbott}\ \emph {et~al.}(2021{\natexlab{a}})\citenamefont {Abbott} \emph {et~al.}}]{KAGRA:2021mth}%
  \BibitemOpen
  \bibfield  {author} {\bibinfo {author} {\bibfnamefont {R.}~\bibnamefont {Abbott}} \emph {et~al.} (\bibinfo {collaboration} {KAGRA, Virgo, LIGO Scientific}),\ }\href {\doibase 10.1103/PhysRevD.104.022005} {\bibfield  {journal} {\bibinfo  {journal} {Phys. Rev. D}\ }\textbf {\bibinfo {volume} {104}},\ \bibinfo {pages} {022005} (\bibinfo {year} {2021}{\natexlab{a}})},\ \Eprint {http://arxiv.org/abs/2103.08520} {arXiv:2103.08520 [gr-qc]} \BibitemShut {NoStop}%
\bibitem [{\citenamefont {Abbott}\ \emph {et~al.}(2021{\natexlab{b}})\citenamefont {Abbott} \emph {et~al.}}]{KAGRA:2021kbb}%
  \BibitemOpen
  \bibfield  {author} {\bibinfo {author} {\bibfnamefont {R.}~\bibnamefont {Abbott}} \emph {et~al.} (\bibinfo {collaboration} {KAGRA, Virgo, LIGO Scientific}),\ }\href {\doibase 10.1103/PhysRevD.104.022004} {\bibfield  {journal} {\bibinfo  {journal} {Phys. Rev. D}\ }\textbf {\bibinfo {volume} {104}},\ \bibinfo {pages} {022004} (\bibinfo {year} {2021}{\natexlab{b}})},\ \Eprint {http://arxiv.org/abs/2101.12130} {arXiv:2101.12130 [gr-qc]} \BibitemShut {NoStop}%
\bibitem [{\citenamefont {Abbott}\ \emph {et~al.}(2019)\citenamefont {Abbott} \emph {et~al.}}]{LIGOScientific:2019vic}%
  \BibitemOpen
  \bibfield  {author} {\bibinfo {author} {\bibfnamefont {B.~P.}\ \bibnamefont {Abbott}} \emph {et~al.} (\bibinfo {collaboration} {LIGO Scientific, Virgo}),\ }\href {\doibase 10.1103/PhysRevD.100.061101} {\bibfield  {journal} {\bibinfo  {journal} {Phys. Rev. D}\ }\textbf {\bibinfo {volume} {100}},\ \bibinfo {pages} {061101} (\bibinfo {year} {2019})},\ \Eprint {http://arxiv.org/abs/1903.02886} {arXiv:1903.02886 [gr-qc]} \BibitemShut {NoStop}%
\bibitem [{\citenamefont {Mingarelli}\ \emph {et~al.}(2013)\citenamefont {Mingarelli}, \citenamefont {Sidery}, \citenamefont {Mandel},\ and\ \citenamefont {Vecchio}}]{Mingarelli:2013dsa}%
  \BibitemOpen
  \bibfield  {author} {\bibinfo {author} {\bibfnamefont {C.~M.~F.}\ \bibnamefont {Mingarelli}}, \bibinfo {author} {\bibfnamefont {T.}~\bibnamefont {Sidery}}, \bibinfo {author} {\bibfnamefont {I.}~\bibnamefont {Mandel}}, \ and\ \bibinfo {author} {\bibfnamefont {A.}~\bibnamefont {Vecchio}},\ }\href {\doibase 10.1103/PhysRevD.88.062005} {\bibfield  {journal} {\bibinfo  {journal} {Phys. Rev. D}\ }\textbf {\bibinfo {volume} {88}},\ \bibinfo {pages} {062005} (\bibinfo {year} {2013})},\ \Eprint {http://arxiv.org/abs/1306.5394} {arXiv:1306.5394 [astro-ph.HE]} \BibitemShut {NoStop}%
\bibitem [{\citenamefont {Taylor}\ and\ \citenamefont {Gair}(2013)}]{Taylor:2013esa}%
  \BibitemOpen
  \bibfield  {author} {\bibinfo {author} {\bibfnamefont {S.~R.}\ \bibnamefont {Taylor}}\ and\ \bibinfo {author} {\bibfnamefont {J.~R.}\ \bibnamefont {Gair}},\ }\href {\doibase 10.1103/PhysRevD.88.084001} {\bibfield  {journal} {\bibinfo  {journal} {Phys. Rev. D}\ }\textbf {\bibinfo {volume} {88}},\ \bibinfo {pages} {084001} (\bibinfo {year} {2013})},\ \Eprint {http://arxiv.org/abs/1306.5395} {arXiv:1306.5395 [gr-qc]} \BibitemShut {NoStop}%
\bibitem [{\citenamefont {Gair}\ \emph {et~al.}(2014)\citenamefont {Gair}, \citenamefont {Romano}, \citenamefont {Taylor},\ and\ \citenamefont {Mingarelli}}]{Gair:2014rwa}%
  \BibitemOpen
  \bibfield  {author} {\bibinfo {author} {\bibfnamefont {J.}~\bibnamefont {Gair}}, \bibinfo {author} {\bibfnamefont {J.~D.}\ \bibnamefont {Romano}}, \bibinfo {author} {\bibfnamefont {S.}~\bibnamefont {Taylor}}, \ and\ \bibinfo {author} {\bibfnamefont {C.~M.~F.}\ \bibnamefont {Mingarelli}},\ }\href {\doibase 10.1103/PhysRevD.90.082001} {\bibfield  {journal} {\bibinfo  {journal} {Phys. Rev. D}\ }\textbf {\bibinfo {volume} {90}},\ \bibinfo {pages} {082001} (\bibinfo {year} {2014})},\ \Eprint {http://arxiv.org/abs/1406.4664} {arXiv:1406.4664 [gr-qc]} \BibitemShut {NoStop}%
\bibitem [{\citenamefont {Romano}\ and\ \citenamefont {Cornish}(2017)}]{Romano:2016dpx}%
  \BibitemOpen
  \bibfield  {author} {\bibinfo {author} {\bibfnamefont {J.~D.}\ \bibnamefont {Romano}}\ and\ \bibinfo {author} {\bibfnamefont {N.~J.}\ \bibnamefont {Cornish}},\ }\href {\doibase 10.1007/s41114-017-0004-1} {\bibfield  {journal} {\bibinfo  {journal} {Living Rev. Rel.}\ }\textbf {\bibinfo {volume} {20}},\ \bibinfo {pages} {2} (\bibinfo {year} {2017})},\ \Eprint {http://arxiv.org/abs/1608.06889} {arXiv:1608.06889 [gr-qc]} \BibitemShut {NoStop}%
\bibitem [{\citenamefont {Pol}\ \emph {et~al.}(2022)\citenamefont {Pol}, \citenamefont {Taylor},\ and\ \citenamefont {Romano}}]{Pol:2022sjn}%
  \BibitemOpen
  \bibfield  {author} {\bibinfo {author} {\bibfnamefont {N.}~\bibnamefont {Pol}}, \bibinfo {author} {\bibfnamefont {S.~R.}\ \bibnamefont {Taylor}}, \ and\ \bibinfo {author} {\bibfnamefont {J.~D.}\ \bibnamefont {Romano}},\ }\href {\doibase 10.3847/1538-4357/ac9836} {\bibfield  {journal} {\bibinfo  {journal} {Astrophys. J.}\ }\textbf {\bibinfo {volume} {940}},\ \bibinfo {pages} {173} (\bibinfo {year} {2022})},\ \Eprint {http://arxiv.org/abs/2206.09936} {arXiv:2206.09936 [astro-ph.HE]} \BibitemShut {NoStop}%
\bibitem [{\citenamefont {Wang}\ and\ \citenamefont {Zhao}(2024)}]{Wang:2023div}%
  \BibitemOpen
  \bibfield  {author} {\bibinfo {author} {\bibfnamefont {S.}~\bibnamefont {Wang}}\ and\ \bibinfo {author} {\bibfnamefont {Z.-C.}\ \bibnamefont {Zhao}},\ }\href {\doibase 10.1103/PhysRevD.109.L061502} {\bibfield  {journal} {\bibinfo  {journal} {Phys. Rev. D}\ }\textbf {\bibinfo {volume} {109}},\ \bibinfo {pages} {L061502} (\bibinfo {year} {2024})},\ \Eprint {http://arxiv.org/abs/2307.04680} {arXiv:2307.04680 [astro-ph.HE]} \BibitemShut {NoStop}%
\bibitem [{\citenamefont {Chen}\ \emph {et~al.}(2026)\citenamefont {Chen} \emph {et~al.}}]{Chen:2026mid}%
  \BibitemOpen
  \bibfield  {author} {\bibinfo {author} {\bibfnamefont {Y.}~\bibnamefont {Chen}} \emph {et~al.},\ }\href {\doibase 10.1103/czxp-zrd6} {\bibfield  {journal} {\bibinfo  {journal} {Phys. Rev. D}\ }\textbf {\bibinfo {volume} {113}},\ \bibinfo {pages} {043042} (\bibinfo {year} {2026})},\ \Eprint {http://arxiv.org/abs/2602.11529} {arXiv:2602.11529 [astro-ph.HE]} \BibitemShut {NoStop}%
\bibitem [{\citenamefont {Cusin}\ and\ \citenamefont {Tasinato}(2022)}]{Cusin:2022cbb}%
  \BibitemOpen
  \bibfield  {author} {\bibinfo {author} {\bibfnamefont {G.}~\bibnamefont {Cusin}}\ and\ \bibinfo {author} {\bibfnamefont {G.}~\bibnamefont {Tasinato}},\ }\href {\doibase 10.1088/1475-7516/2022/08/036} {\bibfield  {journal} {\bibinfo  {journal} {JCAP}\ }\textbf {\bibinfo {volume} {08}},\ \bibinfo {pages} {036} (\bibinfo {year} {2022})},\ \Eprint {http://arxiv.org/abs/2201.10464} {arXiv:2201.10464 [astro-ph.CO]} \BibitemShut {NoStop}%
\bibitem [{\citenamefont {Chowdhury}\ \emph {et~al.}(2023)\citenamefont {Chowdhury}, \citenamefont {Tasinato},\ and\ \citenamefont {Zavala}}]{Chowdhury:2022pnv}%
  \BibitemOpen
  \bibfield  {author} {\bibinfo {author} {\bibfnamefont {D.}~\bibnamefont {Chowdhury}}, \bibinfo {author} {\bibfnamefont {G.}~\bibnamefont {Tasinato}}, \ and\ \bibinfo {author} {\bibfnamefont {I.}~\bibnamefont {Zavala}},\ }\href {\doibase 10.1103/PhysRevD.107.083516} {\bibfield  {journal} {\bibinfo  {journal} {Phys. Rev. D}\ }\textbf {\bibinfo {volume} {107}},\ \bibinfo {pages} {083516} (\bibinfo {year} {2023})},\ \Eprint {http://arxiv.org/abs/2209.05770} {arXiv:2209.05770 [gr-qc]} \BibitemShut {NoStop}%
\bibitem [{\citenamefont {Tasinato}(2023)}]{Tasinato:2023zcg}%
  \BibitemOpen
  \bibfield  {author} {\bibinfo {author} {\bibfnamefont {G.}~\bibnamefont {Tasinato}},\ }\href {\doibase 10.1103/PhysRevD.108.103521} {\bibfield  {journal} {\bibinfo  {journal} {Phys. Rev. D}\ }\textbf {\bibinfo {volume} {108}},\ \bibinfo {pages} {103521} (\bibinfo {year} {2023})},\ \Eprint {http://arxiv.org/abs/2309.00403} {arXiv:2309.00403 [gr-qc]} \BibitemShut {NoStop}%
\bibitem [{\citenamefont {Heisenberg}\ \emph {et~al.}(2025)\citenamefont {Heisenberg}, \citenamefont {Inchausp{\'e}},\ and\ \citenamefont {Maibach}}]{Heisenberg:2024var}%
  \BibitemOpen
  \bibfield  {author} {\bibinfo {author} {\bibfnamefont {L.}~\bibnamefont {Heisenberg}}, \bibinfo {author} {\bibfnamefont {H.}~\bibnamefont {Inchausp{\'e}}}, \ and\ \bibinfo {author} {\bibfnamefont {D.}~\bibnamefont {Maibach}},\ }\href {\doibase 10.1088/1475-7516/2025/01/044} {\bibfield  {journal} {\bibinfo  {journal} {JCAP}\ }\textbf {\bibinfo {volume} {01}},\ \bibinfo {pages} {044} (\bibinfo {year} {2025})},\ \Eprint {http://arxiv.org/abs/2401.14849} {arXiv:2401.14849 [gr-qc]} \BibitemShut {NoStop}%
\bibitem [{\citenamefont {Mentasti}\ \emph {et~al.}(2026)\citenamefont {Mentasti}, \citenamefont {Contaldi},\ and\ \citenamefont {Peloso}}]{Mentasti:2025ywl}%
  \BibitemOpen
  \bibfield  {author} {\bibinfo {author} {\bibfnamefont {G.}~\bibnamefont {Mentasti}}, \bibinfo {author} {\bibfnamefont {C.~R.}\ \bibnamefont {Contaldi}}, \ and\ \bibinfo {author} {\bibfnamefont {M.}~\bibnamefont {Peloso}},\ }\href {\doibase 10.1088/1475-7516/2026/02/068} {\bibfield  {journal} {\bibinfo  {journal} {JCAP}\ }\textbf {\bibinfo {volume} {02}},\ \bibinfo {pages} {068} (\bibinfo {year} {2026})},\ \Eprint {http://arxiv.org/abs/2507.16901} {arXiv:2507.16901 [astro-ph.CO]} \BibitemShut {NoStop}%
\bibitem [{\citenamefont {Maggiore}(2018)}]{Maggiore:2018sht}%
  \BibitemOpen
  \bibfield  {author} {\bibinfo {author} {\bibfnamefont {M.}~\bibnamefont {Maggiore}},\ }\href@noop {} {\emph {\bibinfo {title} {{Gravitational Waves. Vol. 2: Astrophysics and Cosmology}}}}\ (\bibinfo  {publisher} {Oxford University Press},\ \bibinfo {year} {2018})\BibitemShut {NoStop}%
\bibitem [{\citenamefont {Detweiler}(1979)}]{Detweiler:1979wn}%
  \BibitemOpen
  \bibfield  {author} {\bibinfo {author} {\bibfnamefont {S.~L.}\ \bibnamefont {Detweiler}},\ }\href {\doibase 10.1086/157593} {\bibfield  {journal} {\bibinfo  {journal} {Astrophys. J.}\ }\textbf {\bibinfo {volume} {234}},\ \bibinfo {pages} {1100} (\bibinfo {year} {1979})}\BibitemShut {NoStop}%
\bibitem [{\citenamefont {Hellings}\ and\ \citenamefont {Downs}(1983)}]{Hellings:1983fr}%
  \BibitemOpen
  \bibfield  {author} {\bibinfo {author} {\bibfnamefont {R.~w.}\ \bibnamefont {Hellings}}\ and\ \bibinfo {author} {\bibfnamefont {G.~s.}\ \bibnamefont {Downs}},\ }\href {\doibase 10.1086/183954} {\bibfield  {journal} {\bibinfo  {journal} {Astrophys. J. Lett.}\ }\textbf {\bibinfo {volume} {265}},\ \bibinfo {pages} {L39} (\bibinfo {year} {1983})}\BibitemShut {NoStop}%
\bibitem [{\citenamefont {Vallisneri}\ \emph {et~al.}(2025)\citenamefont {Vallisneri}, \citenamefont {Meyers}, \citenamefont {Wright}, \citenamefont {Johnson}, \citenamefont {Baier},\ and\ \citenamefont {van Haasteren}}]{Vallisneri_nanograv_discovery_2025}%
  \BibitemOpen
  \bibfield  {author} {\bibinfo {author} {\bibfnamefont {M.}~\bibnamefont {Vallisneri}}, \bibinfo {author} {\bibfnamefont {P.~M.}\ \bibnamefont {Meyers}}, \bibinfo {author} {\bibfnamefont {D.}~\bibnamefont {Wright}}, \bibinfo {author} {\bibfnamefont {A.~D.}\ \bibnamefont {Johnson}}, \bibinfo {author} {\bibfnamefont {J.~G.}\ \bibnamefont {Baier}}, \ and\ \bibinfo {author} {\bibfnamefont {R.}~\bibnamefont {van Haasteren}},\ }\href {https://github.com/nanograv/discovery} {\enquote {\bibinfo {title} {{nanograv/discovery}},}\ } (\bibinfo {year} {2025})\BibitemShut {NoStop}%
\bibitem [{\citenamefont {Agazie}\ \emph {et~al.}(2023{\natexlab{e}})\citenamefont {Agazie} \emph {et~al.}}]{NANOGrav:2023hde}%
  \BibitemOpen
  \bibfield  {author} {\bibinfo {author} {\bibfnamefont {G.}~\bibnamefont {Agazie}} \emph {et~al.} (\bibinfo {collaboration} {NANOGrav}),\ }\href {\doibase 10.3847/2041-8213/acda9a} {\bibfield  {journal} {\bibinfo  {journal} {Astrophys. J. Lett.}\ }\textbf {\bibinfo {volume} {951}},\ \bibinfo {pages} {L9} (\bibinfo {year} {2023}{\natexlab{e}})},\ \Eprint {http://arxiv.org/abs/2306.16217} {arXiv:2306.16217 [astro-ph.HE]} \BibitemShut {NoStop}%
\bibitem [{\citenamefont {Pi}\ and\ \citenamefont {Sasaki}(2020)}]{Pi:2020otn}%
  \BibitemOpen
  \bibfield  {author} {\bibinfo {author} {\bibfnamefont {S.}~\bibnamefont {Pi}}\ and\ \bibinfo {author} {\bibfnamefont {M.}~\bibnamefont {Sasaki}},\ }\href {\doibase 10.1088/1475-7516/2020/09/037} {\bibfield  {journal} {\bibinfo  {journal} {JCAP}\ }\textbf {\bibinfo {volume} {09}},\ \bibinfo {pages} {037} (\bibinfo {year} {2020})},\ \Eprint {http://arxiv.org/abs/2005.12306} {arXiv:2005.12306 [gr-qc]} \BibitemShut {NoStop}%
\end{thebibliography}%
\end{document}